\documentclass[reprint,
superscriptaddress,
 amsmath,amssymb,
 aps,
pra,
floatfix,
]{revtex4-2}

\usepackage{lineno}  

\usepackage[dvipsnames]{xcolor}
\usepackage{graphicx}
\usepackage{dcolumn}
\usepackage{bm}
\usepackage{amsmath,accents,amssymb}
\usepackage{subcaption}
\usepackage{graphicx}
\usepackage{placeins}
\usepackage{comment}
\usepackage{xcolor}
\usepackage{booktabs} 
\usepackage{rotating}

\newcommand\R{\mbox{Re}}
\newcommand\Ca{\mbox{Ca}}
\newcommand\Ka{\mbox{Ka}}
\newcommand\Ha{\mbox{Ha}}
\newcommand\N{\mbox{N}}
\newcommand\Sr{\mbox{S}}
\newcommand\Rm{\mbox{R\textsubscript{m}}}

\usepackage{bm} 

\begin{document}
\preprint{APS/123-QED}

\title{Integral modelling and Reinforcement Learning control of 3D liquid metal coating on a moving substrate}

\author{Fabio Pino}
 \email{Currently at Department of Applied Mathematics and Theoretical Physics (DAMTP), University of Cambridge, Cambridge, United Kingdom, correspondence at fp448@cam.ac.uk}
\affiliation{The von Karman Institute for Fluid Dynamics, EA Department, Sint Genesius Rode, Belgium}
\affiliation{Transfers, Interfaces and Processes (TIPs), Université libre de Bruxelles, 1050 Brussels, Belgium}
\author{Edoardo Fracchia}
\affiliation{The von Karman Institute for Fluid Dynamics, EA Department, Sint Genesius Rode, Belgium}  
\affiliation{Department of Mechanical and Aerospace Engineering, Politecnico di Torino, Torino, Italy} 
 \email{Currently at Department of Mechanical and Aerospace Engineering, Politecnico di Torino, Torino, Italy}
\author{Benoit Scheid}%
\affiliation{Transfers, Interfaces and Processes (TIPs), Université libre de Bruxelles, 1050 Brussels, Belgium}%
\author{Miguel A. Mendez}%
\affiliation{The von Karman Institute for Fluid Dynamics, EA Department, Sint Genesius Rode, Belgium}

\affiliation{Aero-Thermo-Mechanics Laboratory, Université Libre de Bruxelles, Elsene, Brussels, Belgium}

\affiliation{Aerospace Engineering Research Group, Universidad Carlos III de Madrid, Leganés, Spain}

\date{\today}

\begin{abstract}
Metallic coatings are used to enhance the durability of metal surfaces by protecting them from corrosion. These protective layers are typically deposited in a fluid state via a liquid film. Controlling instabilities in the liquid film is crucial to achieving uniform, high-quality coatings. This study explores the possibility of controlling liquid films on a moving substrate using a combination of gas jets and electromagnetic actuators. To model the 3D liquid film, we extend existing integral models to incorporate the effects of electromagnetic actuators. The control strategy was developed within a reinforcement learning framework, in which the Proximal Policy Optimisation (PPO) algorithm interacts with the liquid film via pneumatic and electromagnetic actuators to optimise a reward function that accounts for instability-wave amplitude through a trial-and-error process. The PPO identified an optimal control law that reduced interface instabilities via a novel mechanism: gas jets push crests, and electromagnets raise troughs via the Lorentz force.
\end{abstract}

\maketitle

\section{Introduction} \label{sec:intro}
The interaction between liquid films and external magnetic fields is a relevant physical phenomenon with many engineering applications, including the analysis of magnetic soap films \cite{lalli2024thin}, the evaluation of magnetic wiping solutions in dip coating processes \cite{dumont2011new,pino2024multi}, the control of mould flow during continuous steel casting \cite{lehman1994fluid}, and the design of liquid metal films for plasma confinement and heat removal inside nuclear fusion reactors \cite{zakharov2003magnetic,narula2005study}.

The complexity of the magnetohydrodynamic MHD equations, describing the liquid and electromagnetic dynamics, poses significant challenges for numerical simulations \cite{morley2004progress}, often requiring customised software implementations with prohibitive computational costs \cite{lunz2019flow,smolentsev2021status,munipalli2003development}. To avoid these limitations, \citet{smolentsev2005open} derived a simplified MHD model based on the long-wave assumption. Other authors used reduced-dimensional models with a single evolution equation for film thickness, based on depth-averaged equations and a gradient expansion of the velocity profile \citep{lalli2024thin,lunz2019flow,seric2014interfacial}.

Reduced-dimensional models are commonly used to capture the relevant features of liquid film dynamics by employing simplified governing equations, thereby reducing computational costs. \citet{shkadov1968wave} and \citet{kapitza1949wave} developed a 2D Integral Boundary Layer (IBL) model for moderate Reynolds number flows in terms of film thickness and flow rate, which outperformed the single-equation model in capturing the film dynamics. 

A two-dimensional IBL model describing a liquid film flowing over a moving substrate under external pressure and shear-stress distributions at the free surface was first introduced by \citet{mendez2021dynamics}. The formulation was later extended to three dimensions to account for spanwise effects \citep{ivanova2023evolution}, and its predictions were validated against Direct Numerical Simulations (DNS) of the full governing equations \citep{barreiro2021dynamics}. This 3D IBL framework was subsequently employed to reproduce the dynamics of liquid films in the hot-dip galvanising process with impinging gas jets.

The hot-dip galvanising process consists in coating a metal strip with a thin layer of zinc to protect it against corrosion. The zinc layer is applied by immersing the metal sheet in molten zinc and then withdrawing it vertically. As the sheet is withdrawn, it entrains a thin layer of liquid zinc, which then solidifies into a protective coating. The thickness of the final coating is controlled by impinging gas jets that remove the excess liquid coming from the bath \cite{buchlin1997modelling,gosset2019experimental}. For withdrawal velocities above 2~m/s, the gas jets begin to oscillate, inducing an instability in the liquid film known as undulation \cite{gosset2019experimental,mendez2015research,mendez2019experimental}. This instability manifests as nearly two-dimensional travelling waves \cite{barreiro2021dynamics,barreiro2023damping,BarreiroVillaverde2024}, resulting in an uneven coating layer and reduced surface quality.

Developing active control laws to suppress undulation waves is essential for achieving uniform coatings and reducing material and energy waste. Various strategies have been proposed for liquid-film systems, such as modulating electric fields \cite{wray2022electrostatic} or controlling mass injection at the solid boundary \cite{thompson2016stabilising}. A particularly attractive alternative is the use of blowing gas jets and electromagnetic actuators. Electromagnetic actuation, in particular, has recently shown promise for controlling liquid metal films in plasma-confinement applications. Experiments with Galinstan films flowing over inclined surfaces under transverse magnetic fields demonstrated effective stabilisation in both open- and closed-loop configurations using film-height feedback \cite{mirhoseini2017passive,mirhoseini2016resistive}, highlighting its potential for undulation control in coating flows.

Optimal control strategies for liquid-film flows have been explored using linear stability analysis \cite{thompson2016stabilising}, Linear Quadratic Regulator (LQR) techniques \cite{holroyd2024linear}, optimal-control formulations \cite{boujo2019pancake,tomlin2019optimal}, and Model Predictive Control (MPC) methods \cite{wray2022electrostatic}. However, applying such model-based approaches to hot-dip galvanising remains challenging due to strong noise, substantial measurement and actuation uncertainties, and unmodelled effects—such as high Reynolds numbers, thermal gradients, oxidation, and substrate vibration, that extend beyond the range of simplified models. In contrast, Reinforcement Learning (RL) \cite{viquerat2022review} offers a model-free framework for developing robust feedback laws by directly interacting with the process and optimising a reward function through trial and error. RL is inherently resilient to modelling errors and uncertainties, and has shown strong performance in fluid-dynamics applications \cite{vu2022online,zheng2021reinforcement,ran2021reinforcement,pino2023comparative}, including the stabilisation of two-dimensional falling films using only local pressure actuation and limited film-thickness observations \cite{belus2019exploiting}.

This study introduces a novel three-dimensional IBL model describing the dynamics of a liquid zinc film flowing over a vertically moving substrate under the influence of an external magnetic field and impinging gas jets. Building on the formulation of \citet{ivanova2023evolution}, the model incorporates Maxwell stresses at the free surface and Lorentz forces within the bulk. It is further employed to design optimal feedback control laws for gas jets and electromagnetic actuators to mitigate undulation instability. The resulting control strategy is derived using the Reinforcement Learning Proximal Policy Optimisation (PPO) algorithm \citep{schulman2017proximal}.

The remainder of this article is organised as follows. Section~\ref{sec:prob_definition} defines the problem and introduces the relevant physical quantities. Section~\ref{sec:int_model} outlines the reference quantities employed to nondimensionalize the governing equations. The modelling of gas jets and electromagnetic actuators is described in Section~\ref{subsec:modelling_actuators}. Section~\ref{sec:num_methods} details the numerical methods used to solve the reduced-dimensional model equations. The results on wave dynamics and control performance are presented in Section~\ref{sec:results}. Finally, Section~\ref{sec:conclusion} provides concluding remarks and future perspectives.

\section{Problem description and physical quantities} \label{sec:prob_definition}
We consider a 3D liquid film of thickness $h$ flowing over a substrate moving against gravity $g$ at constant speed $U_p=2.5\,$m/s in contact with air, under the action of external gas jets and electromagnetic actuators.

\begin{figure}
    \centering
	\includegraphics[width=\linewidth]{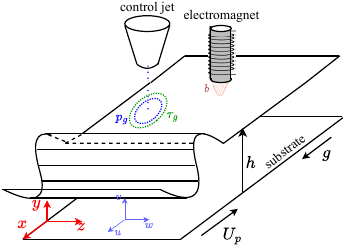}
	\caption{Scheme of the liquid film flowing over a substrate moving against gravity under the effect of a gas jet and electromagnetic actuators.}
	\label{fig:problem_scheme_general}
\end{figure}

For clarity, the physical quantities related to the liquid are written without a subscript, while those related to the air are denoted by the subscript "g". We consider a liquid film of molten zinc with density $\rho=6570\,\rm{kg/m}^3$, dynamic viscosity $\mu=3.5\,\rm{mPa\,s}$, kinematic viscosity $\nu=\mu/\rho$, surface tension $\sigma=700\,\rm{mN/m}$, electrical conductivity $\sigma_M=2.7\,\rm{MS}/\rm{m}$ (evaluated at $450^{\circ}\,$C \cite{giordanengo2000new}), magnetic permeability $\mu_{\rm M}=1.26\,\mu\rm{N/A}^2$, magnetic diffusivity $\eta_M=1/(\mu_M\sigma_M)=0.3\,\rm{m}^2/\rm{s}$ and  magnetic susceptibility $\chi=(\mu_{\rm M}/\mu_{\rm M0} - 1) = 2.7\times 10^{-3}$ where $\mu_{M0}$ is the magnetic permeability of the void. The air has magnetic permeability $\mu_{\rm Mg}=1.26\;\mu{\rm N/A}^2$ and magnetic susceptibility $\chi_{\rm g}=(\mu_{\rm Mg}/\mu_{\rm M0} - 1) = 3.73\times 10^{-7}$.

For later convenience in the derivation of the reduced-dimensional model, we express the magnetic permeability of the air $\mu_{\rm Mg}$ as a function of the magnetic permeability $\mu_M$ and susceptibility $\chi$ of the liquid and the magnetic susceptibility of the air $\chi_g$ through the following relation:
\begin{equation}
    \mu_{\rm Mg} = \frac{\chi_{\rm g} + 1}{\chi + 1}\mu_{\rm M}\,.
\end{equation}

Figure~\ref{fig:problem_scheme_general} illustrates the coordinate system $\{\mathbf{O}$; $x,y,z\}$ with the $x$ axis aligned with the gravity vector $g$ pointing downward, the $y$ axis pointing along the normal direction of the wall towards the free surface, and the $z$ axis spanning the direction of flow, with origin $\mathcal{O}$ on the substrate. Because of the problem's translational invariance, specifying a precise origin for the $ x$-axis is unnecessary. 

A local Cartesian coordinate system is defined on the free surface $y-h(x,z,t)=0$ tangent plane as $\{\mathbf{O}$; $\mathbf{n},\mathbf{t}_x,\mathbf{t}_z\}$, with $\mathbf{n}$ representing the normal vector pointing towards the air, and $\mathbf{t}_x $ and $\mathbf{t}_z$ representing the tangential directions along $x$ and $z$ respectively. These vectors are expressed in ($\mathcal{O}$; $x,y,z$) with the following relations:
\begin{subequations}
\label{eq:local_ref_frame}
\begin{gather}
    \mathbf{n} = \frac{(-\partial_x h, 1, -\partial_z h)^T}{\sqrt{(\partial_x h)^2 + (\partial_z h)^2 + 1}},\\
\mathbf{t}_x = \frac{(1,\partial_x h, 0)^T}{\sqrt{(\partial_xh)^2 + 1}}, \qquad     
  \mathbf{t}_z = \frac{(0,\partial_z h, 1)^T}{\sqrt{(\partial_zh)^2 + 1}},
\end{gather}
\end{subequations} having used the shorthand notation $\partial h/\partial x = \partial_xh$ for partial derivatives, and $T$ denotes the transpose of a vector or a matrix.

The film dynamics is described by the velocity field $\mathbf{u}=(u,v,w)^T$, and the pressure field $p$. The integral formulation of the film dynamics is defined in terms of film thickness $h$ and streamwise $q_x$ and spanwise $q_z$ flow rates per unit width, defined as:
\begin{equation}
\label{eq:flow_rate_def}
    q_x=\int_{0}^{h}u\, dy,\qquad\qquad q_z=\int_{0}^{h}w\, dy.
\end{equation}

Following the modelling approach proposed by \citet{gosset2007jet} and \citet{mendez2021dynamics}, the effect of the gas jets at the free-surface is modelled by a pressure $p_{\rm g}(x,z,t)$ and a shear stress $\bm{\tau}_{\rm g}(x,z,t) = (\tau_{{\rm g},x}, \tau_{{\rm g},z})^T$ distributions. 

The influence of the electromagnetic actuator is represented by an external magnetic field $\mathbf{b}_{\rm ext} = (0, -b, 0)^T$, where $b \geq 0$. Because of the low magnetic susceptibility of both liquid zinc $\chi$ and air $\chi_{\rm g}$, the magnetic fields in the liquid $\mathbf{b}$ and in the gas $\mathbf{b}_{\rm g}$, are approximated as being equal to $\mathbf{b}_{\rm ext}$, reading:
\begin{subequations}
\begin{gather}
\mathbf{b} = (1 + \chi)\mathbf{b}_{\rm ext} \approx (0,-b,0)^T,\\
\mathbf{b}_{\rm g} = (1 + \chi_{\rm g})\mathbf{b}_{\rm ext} \approx (0,-b,0)^T.
\end{gather}
\end{subequations}

The relative motion of the liquid zinc with respect to $\mathbf{b}$ induces a current in the bulk, denoted as $\mathbf{j}$ given by Faraday's law of induction. Neglecting the electric potential difference, as discussed in \cite{Dumont2011d}, the induction law reads:
\begin{equation}
    \mathbf{j} = \sigma_M(\mathbf{u}\times\mathbf{b}) = (-\sigma_Mbw,0,-\sigma_Mbu)^T.
    \label{current_density}
\end{equation} 

Details on the modelling of $p_{\rm g}$, $\bm{\tau}_{\rm g}$  , and $b$ are provided in Section~\ref{subsec:modelling_actuators}.

\section{Integral Modelling with Magnetic Actuators} \label{sec:int_model}
\begin{table}
\small\addtolength{\tabcolsep}{15pt}
\renewcommand{\arraystretch}{1.2}
\centering
\begin{tabular}{ccc}
\toprule 
 & Definition & Value \\
\midrule
$u_{\rm ref}$ & $U_p$ & 2.5\;\rm{m/s} \\
$h_{\rm ref}$ & $\sqrt{(\nu U_p)/g}$ & 330\; $\mu$\rm{m} \\
$b_{\rm ref}$ & ${\rm sup}(|b(x,y)|)$ & 0.583\; \rm{T} \\
$p_{\rm ref}$ & $\rho g h_{\rm ref}$ & 102.9\; \rm{Pa} \\
$\tau_{\rm ref}$ & $\mu u_{\rm ref}/ h_{\rm ref}$ & 21.87\; $\rm{N/m}^2$\\
$j_{\rm ref}$ & $\sigma_MU_pb_{\rm ref}$ & 40\; $\rm{MA/m}^2$\\
\midrule
$\R$ & $\sqrt{U_p^3/{(\rm g}\nu)}$ & 1237\\
$\Ca$ & $(U_p\mu)/\sigma$ & 0.01 \\
$\varepsilon$ & $\Ca^{1/3}$ & 0.22\\ 
$\delta$ & $\R\Ca^{1/3}$ &  266 \\
$\Ka$ & $\sigma/(\rho {\rm g}^{1/3} \nu^{4/3})$& 15713\\
$\Ha$ & $\sqrt{{\sigma_M b_{\rm ref}^2U_p}{\rho {\rm g}}}$ & 6\\
$\Rm$ & $\sqrt{{U_p^2\nu}/{({\rm g}\eta_M^2})}$ & $2.3\times 10^{-3}$ \\
$\Sr$ & $(\chi - \chi_g)/(1+\chi_g)$ & $2.7\times 10^{-3}$ \\
$\N$ & $b_{\rm ref}^2\sigma_M\sqrt{\mu/(U_p\rho^3g)}$ & $2.9\times 10^{-2}$ \\
$\Lambda$ & $\varepsilon\,\Sr\,\Ha^2/\Rm$ & 9.3\\
\bottomrule
\end{tabular}
\caption{Definition and value of the reference quantities and the non-dimensional groups used in this analysis.}
\label{tab:scaling_quant}
\end{table}
This section introduces the reference quantities and nondimensional groups (Subsection~\ref{subsec:ref_quantities}), the steady-state flat film solution under a uniform magnetic field (Subsection~\ref{Steady}), the first-order boundary layer equations (Subsection~\ref{longWAVE}), and the derivation of the 3D IBL model (Subsection~\ref{subsec:integral_model}).

\subsection{Reference quantities and nondimensional groups}\label{subsec:ref_quantities}

The liquid film governing equations, reported in Appendix~\ref{sec_apx:governing_eq}, are scaled as in \citet{ivanova2023evolution}, with the addition of two dimensionless numbers accounting for the electromagnetic force. In the following, all dimensionless quantities are denoted with a hat $\hat{\cdot}$. These are obtained by dividing dimensional variables by their reference quantities, identified by the subscript `$\rm ref$'. 
The streamwise and spanwise velocities are scaled by the substrate velocity $U_p$, and the film thickness is scaled based on the thickness arising from the viscous-gravity balance in steady-state condition \citep{mendez2021dynamics}, namely
\begin{equation}
\label{eq:scaling_0}
    u_{\rm ref} = w_{\rm ref} = U_p \qquad{\rm and}\qquad h_{\rm ref} = \sqrt{(\nu U_p)/{\rm g}}.
\end{equation}

The streamwise $x_{\rm ref}$ and the spanwise $z_{\rm ref}$ spatial scales are based on the long wave assumption, which sets:
\begin{equation}
\label{eq:scaling_long_wave}
    \frac{h_{\rm ref}}{x_{\rm ref}}=\frac{h_{\rm ref}}{z_{\rm ref}}=\varepsilon,
\end{equation}
where $\varepsilon=\Ca^{1/3}\ll1$ is the \textit{film parameter}, with $\Ca={u_{\rm ref} \mu}/{\sigma}$ the capillary number. Based on \eqref{eq:scaling_0} and \eqref{eq:scaling_long_wave}, the remaining variables are scaled accordingly:
\begin{subequations}
\label{eq:ref_quantity_0}
\begin{gather}
y_{\rm ref} = h_{\rm ref},\quad v_{\rm ref} = \varepsilon u_{\rm ref},\quad t_{\rm ref}=h_{\rm ref}/(u_{\rm ref}\varepsilon), \\
p_{\rm ref} = \rho {\rm g} x_{\rm ref}, \quad \tau_{\rm ref}=\mu u_{\rm ref}/ h_{\rm ref},    \quad q_{\rm ref} = u_{\rm ref}h_{\rm ref}.
\end{gather}
\end{subequations} 

The scaling introduced in \citet{ivanova2023evolution} is expanded by the reference quantities for the magnetic field $b_{\rm ref}$ and the induced current $j_{\rm ref}$, introduced by the magnetohydrodynamic extension of the model; these are taken as  
\begin{subequations}
\label{eq:ref_quantity_1}
\begin{gather}
    b_{\rm ref} ={\rm sup}(|b|)\qquad{\rm and}\qquad  j_{\rm ref}=\sigma_MU_pb_{\rm ref}\,.
\end{gather}
\end{subequations}

The capillary number $\Ca$ introduced is given by:
\begin{equation}
    \Ca=\R^{2/3}\Ka,
\end{equation}
where $\Ka$ is the Kapitza number and $\R$ the Reynolds number, which are defined as:
\begin{equation}
\Ka = \frac{\sigma}{\rho\,g^{1/3}\,\nu^{4/3}}, \qquad \R = \frac{\rho u_{\rm ref}\,h_{\rm ref}}{\mu} = \sqrt{\frac{\rho\,U_p^3}{g\,\mu}}\,.
\end{equation}

In addition to the Reynolds, Kapitza and capillary numbers, the other non-dimensional groups arising from this scaling are the reduced Reynolds number $\delta = \varepsilon \R$, the Hartmann number,
\begin{equation}
    \Ha = b_{\rm ref} h_{\rm ref} \sqrt{\frac{\sigma_M}{\mu}} = \sqrt{\frac{\sigma_M b_{\rm ref}^2 U_p}{\rho g}},
\end{equation}
the magnetic Reynolds number $\Rm$, the susceptibility ratio $\Sr$, defined as:
\begin{equation}
    \Rm = \frac{u_{\rm ref}\,h_{\rm ref}}{\eta_M},\qquad\qquad \Sr = \frac{\chi - \chi_g}{1+\chi_g},
\end{equation}
and the Stuart number, defined as:
\begin{equation}
    \N=\frac{\Ha^2}{\R} = \varepsilon\Big(\frac{t_{\rm ref}}{t_{\rm ref,M}}\Big)^2,
\end{equation}
where $t_{\rm ref,M}$ is the time scale of the Loretz force, defined as:
\begin{equation}
\label{eq:time_scale_magn}
    t_{\rm ref,M} = \sqrt{\frac{\rho\,b\,x_{\rm ref}}{\sigma_M\,b_{\rm ref}^2\,u_{\rm ref}}}\,,
\end{equation}
where $bx_{\rm ref}$ is the spatial scale over which the electromagnetic actuators exert influence.

We also introduce the nondimensional parameter $\Lambda$
\begin{equation}
\Lambda = \varepsilon \Sr \Ha^2 / \Rm,
\end{equation}
to characterise the intensity of the normal component of the Maxwell stresses at the free surface.

Table~\ref{tab:scaling_quant} reports the values of the scaling quantities and nondimensional groups used in the control test cases described in Subsection~\ref{subsec:test_cases}. 

\subsection{Flat-film steady state solution}\label{Steady}

Before deriving the reduced-dimensional model, it is worth analysing the steady-state flat-film solution of the governing equations detailed in Appendix \ref{sec_apx:governing_eq}, excluding the influence of external gas-jet actuators. Assuming a stable ($\partial_t h=0$) and flat ($\partial_x=0\wedge\partial_z=0$) surface, without external shear stress ($\tau_{g,x}=\tau_{g,z}=0$) and no external pressure ($p_g=0$) distributions, and considering a uniform transversal magnetic field $b$, the flow field becomes parallel ($w=0$) and unidimensional ($v=0$). The Navier-Stokes equation reduces to the balance of viscous stresses, gravity, and Lorentz force along the streamwise direction $x$:
\begin{equation}
\label{eq:simp_eq_strady}
    \frac{\mu}{\rho}\partial_{yy}u + g - \frac{\sigma_M}{\rho}b^2u=0,
\end{equation}
with the boundary equations \eqref{eq:bc_strip} and \eqref{eq:dynamics_bs} reducing to:
\begin{equation}
\label{eq:bcs}
    u|_{y=0} = - U_p,\qquad\qquad \partial_{y}u|_{y=h} = 0.
\end{equation}

The non-dimensional steady-state velocity $\bar{u}(\hat{y})$ of \eqref{eq:simp_eq_strady} with \eqref{eq:bcs} reads:
\begin{equation}
\label{eq:general_sol_magn_stationary}
\begin{aligned}
    \bar{u}(\hat{y}) =& n_1\,exp(\Ha\;\hat{b}\;\hat{y}) +\\&+ n_2\,exp(-\Ha\;\hat{b}\;\hat{y}) + \frac{1}{\Ha^2\hat{b}^2},
\end{aligned}
\end{equation}
with integrating constants $n_1$ and $n_2$:
\begin{equation}
\label{eq:int_fun_111}
\begin{aligned}
    n_1 &= - \left(1 + n_2 + \frac{1}{\Ha^2\hat{b}^2}\right),\\
    n_2 &= - \frac{\,exp(\hat{h}\Ha\;\hat{b})\left(1 + \frac{1}{\Ha^2\hat{b}^2}\right)}{2\cosh{\left(\Ha\;\hat{b}\hat{h}\right)}}.
\end{aligned}
\end{equation}

Recasting \eqref{eq:general_sol_magn_stationary} with \eqref{eq:int_fun_111} in a more concise form, we obtain:
\begin{equation}
\label{eq:final_eq_velocity_magnetic}
\begin{split}
    \bar{u}(\hat{y}) =& \tanh{(\Bar{h}\Ha\;\hat{b})}\left(1 + \frac{1}{\Ha^2\hat{b}^2} \right)\sinh{(\hat{y}\Ha\;\hat{b})} \\&-\cosh{(\hat{y}\Ha\;\hat{b})}\left(1 + \frac{1}{\Ha^2\hat{b}^2}\right) + \frac{1}{\Ha^2\hat{b}^2},
\end{split}
\end{equation}
as in \citet{Dumont2011d}. Integrating \eqref{eq:final_eq_velocity_magnetic} over the steady-state flat film thickness $\bar{h}$ gives the steady-state flow rate $\Bar{q}_x$ in the streamwise direction:
\begin{equation}
    \label{eq:final_q_func_std_magn}
    \Bar{q}_x = \frac{\Bar{h}\Ha\;\hat{b}-\tanh{(\Bar{h}\Ha\;\hat{b})}\left(\Ha^2\hat{b}^2+1\right)}{\Ha^3\hat{b}^{3}}.
\end{equation}

Equations~\eqref{eq:final_eq_velocity_magnetic} and \eqref{eq:final_q_func_std_magn} describe a family of flat-film solutions that depend on the flat-film thickness $\Bar{h}$. This family of solutions and their asymptotic behaviours are discussed in Section~\ref{subsec:res_asym}.

\subsection{The long wave formulation}\label{longWAVE}

We assume that the distribution of external stresses at the free surface remains independent of the film dynamics, unless explicitly prescribed by a control law. Furthermore, advection of the magnetic field is neglected, and surface tension is assumed to be strong. These assumptions correspond to the following order-of-magnitude estimates for the magnetic Reynolds $\Rm$ and Capillary $\Ca$ numbers, reading:
\begin{equation}
    \Rm = O(\varepsilon), \qquad\qquad \Ca = O(\varepsilon^{3}),
\end{equation}
with the remaining nondimensional groups assumed to be $O(1)$ quantities.

The Integral Boundary Layer (IBL) model is derived from the first-order boundary-layer equations, obtained by applying the long-wave scaling introduced in Subsection~\ref{subsec:ref_quantities} to the magnetohydrodynamic (MHD) governing equations and boundary conditions reported in Appendix~\ref{sec_apx:governing_eq}. Retaining terms up to $O(\varepsilon)$ leads to the following set of equations:
\begin{subequations}
\label{eq:dimensionless-model}
\begin{gather}
\begin{split}
\partial_{\hat{x}} \hat{u} + \partial_{\hat{y}} \hat{v} + \partial_{\hat{z}} \hat{w} = 0, \label{eq:dimensionless-cont}    
\end{split}\\
\begin{split}
\delta \big( \partial_{\hat{t}} \hat{u} + \hat{u} \partial_{\hat{x}} \hat{u} + \hat{v} \partial_{\hat{y}} \hat{u} &+ \hat{w} \partial_{\hat{z}} \hat{u} \big) = -\partial_{\hat{x}} \hat{p_x} \\&+ \partial^2_{\hat{y}\hat{y}} \hat{u} + 1 - \Ha^{2}\hat{b}^2\hat{u}, \label{eq:dimensionless-xmom}
\end{split}\\
\begin{split}
\partial_{\hat{y}}\hat{p}=0,
\end{split}\\
\begin{split}
\delta \big( \partial_{\hat{t}} \hat{w} + \hat{u} \partial_{\hat{x}} \hat{w} + \hat{v} \partial_{\hat{y}} \hat{w} &+ \hat{w} \partial_{\hat{z}} \hat{w} \big) = -\partial_{\hat{z}} \hat{p_z} \\&+ \partial^2_{\hat{y}\hat{y}} \hat{w} - \Ha^{2}\hat{b}^2\hat{w}, \label{eq:dimensionless-zmom}    
\end{split}
\end{gather}
\end{subequations}
where $\hat{v}=O(\varepsilon)$.
The boundary conditions at the substrate $\hat{y}=0$ are given by:
\begin{equation}
\label{eq:bc_boundary_layer0}
    \hat{u} = -1,\qquad\qquad \hat{v}=0,\qquad\qquad \hat{w}=0, 
\end{equation} 
and at the free-surface $\hat{y}=\hat{h}$ are given by:
\begin{subequations}
\label{eq:bc_boundary_layer1}
	\begin{gather}
 	      \hat{v}  = \partial_{\hat{t}} \hat{h} + \hat{u} \partial_{\hat{x}} \hat{h} + \hat{w} \partial_{\hat{z}} \hat{h},\label{eq:kinematic_free_surface}\\
		\hat{p}  = \hat{p}_g - \frac{\varepsilon^3}{\Ca}(\partial_{\hat{x}\hat{x}} \hat{h} + \partial_{\hat{z}\hat{z}} \hat{h} ) - \frac{\Lambda}{2}\hat{b},  \label{eq:stress_norm_free_surf} \\
		\partial_{\hat{y}} \hat{u}  = \Xi, \qquad\qquad
		\partial_{\hat{y}} \hat{w}  = \Psi,
  \label{eq:stress_tang_free_surf}
	\end{gather}
\end{subequations} 
where $\Lambda\hat{b}/2$ represent the normal component of the Maxwell stresses, with $\Psi$ and $\Xi$ accounting for the gas-jet induced shear stress $\hat{\boldsymbol{\tau}}_g$ and the tangential component of the Maxwell stresses $\hat{\boldsymbol{\tau}}_M=(\hat{b}^2\partial_{\hat{x}}\hat{h},\hat{b}^2\partial_{\hat{z}}\hat{h})$ at the free surface, reading:
\begin{equation}
\label{eq:maxwell_stresses}
    \Psi = \hat{\tau}_{g,z} + \Lambda\,\hat{b}^2\partial_{\hat{z}}\hat{h},\qquad\Xi = \hat{\tau}_{g,x} + \Lambda\,\hat{b}^2\partial_{\hat{x}}\hat{h}.
\end{equation}

It is worth stressing that, besides the long-wave condition $\varepsilon \ll 1$, the validity of the boundary-layer approximation (and in particular the assumption $\partial_{\hat{y}}\hat{p}=0$) also requires that inertial effects remain moderate, which in turn demands the reduced Reynolds number $\delta = \varepsilon Re = O(1)$ \citep{kalliadasis2011falling}. Nevertheless, several studies have shown that integral (IBL) and weighted-integral (WIBL) models derived from \eqref{eq:dimensionless-model}, \eqref{eq:bc_boundary_layer0}, and \eqref{eq:bc_boundary_layer1} remain accurate well beyond this formal limit, at least in the absence of magnetic effects. For instance, \citet{Denner2018} validated a WIBL model up to $\delta = 110.3$ ($Re = 77$), \citet{mendez2021dynamics} obtained consistent results for a moving substrate at $\delta = 73.9$ and $Re = 478$, and \citet{BarreiroVillaverde2024} confirmed the same boundary-layer structure under jet-wiping conditions at $\delta \approx 76$. In the present configuration, the reduced Reynolds number reaches $\delta = 266$, which exceeds the asymptotic range but remains within one order of magnitude of previously validated cases, supporting the practical use of the boundary-layer approximation in this study.

\subsection{3D Integral boundary layer model}
\label{subsec:integral_model}
The integral boundary layer model is derived by integrating the boundary layer equations \eqref{eq:dimensionless-model} over the liquid film thickness ($\hat{h}$) using the Leibniz integral rule with the boundary conditions \eqref{eq:bc_boundary_layer0} and \eqref{eq:bc_boundary_layer1}. This results in a set of nonlinear hyperbolic partial differential equations in terms of $\hat{h}$, $\hat{q}_x$, and $\hat{q}_z$, reading:
\begin{eqnarray}
    \label{eq:model_gen}
    \partial_{\hat{t}} 	\begin{pmatrix} 
		\hat{h} \\
        \hat{q}_x \\
		\hat{q}_z  
	\end{pmatrix} + \nabla \cdot \mathbf{F} = \mathbf{s},
\end{eqnarray}
where $\mathbf{F}$ is the flux matrix defined as:
\begin{eqnarray}
\label{eq:flux_matrix}
	\mathbf{F} &=& 
	\begin{pmatrix} 
		F_{11} & F_{21} \\
        F_{12} & F_{22} \\
		F_{13} & F_{23} 
	\end{pmatrix}^T = 
	\begin{pmatrix} 
		\hat{q}_x & \hat{q}_z \\ 
        \int_0^{\hat{h}} \hat{u}^2 d\hat{y} & \int_0^{\hat{h}} \hat{u} \hat{w} d\hat{y} \\\int_0^{\hat{h}} \hat{u} \hat{w} d\hat{y} & \int_0^{\hat{h}} \hat{w}^2 d\hat{y} 
	\end{pmatrix}^T,
\end{eqnarray} and $\mathbf{s} = (s_1, s_2, s_3)^\text{T}$ is the vector containing the source terms, which is expressed as:
\begin{eqnarray*}
    \label{eq:sources}
\begin{aligned}
    \mathbf{s} =& \delta^{-1}
	\begin{pmatrix}
		0 \\
		 \hat{h} \Big( - \partial_{\hat{x}} \hat{p}_g + \partial_{\hat{x}\hat{x}\hat{x}} \hat{h} + \partial_{\hat{x}\hat{z}\hat{z}} \hat{h} - \frac{\Lambda}{2}\partial_{\hat{x}}\hat{b}\Big) \\
		 \hat{h} \Big( - \partial_{\hat{z}} \hat{p}_g + \partial_{\hat{z}\hat{z}\hat{z}} \hat{h} + \partial_{\hat{z}\hat{x}\hat{x}} \hat{h} - \frac{\Lambda}{2}\partial_{\hat{z}}\hat{b} \Big) 
	\end{pmatrix}+\\&+\delta^{-1}
 	\begin{pmatrix}
		0 \\ \Lambda\,\hat{b}^2\partial_{\hat{x}}\hat{h} + \hat{h} + \Delta\hat{\tau}_{x} - \Ha^2\hat{b}^2\hat{q}_x \\
	\Lambda\,\hat{b}^2\partial_{\hat{z}}\hat{h} + \Delta \hat{\tau}_{z} - \Ha^2\hat{b}^2\hat{q}_z
	\end{pmatrix},
\end{aligned}
\end{eqnarray*}
where $\Delta \hat{\tau}_{x} = \hat{\tau}_{g,x} + \hat{\tau}_{w,x}$ and $\Delta \hat{\tau}_{z} = \hat{\tau}_{g,z} + \hat{\tau}_{w,z}$ with the wall-shear stress vector components along the streamwise $\hat{\tau}_{w,x}$ and spanwise $\hat{\tau}_{w,z}$ directions given by:
\begin{equation}
\label{eq:def_wall_shear}
 \hat{\tau}_{w,x} = -\partial_{\hat{y}}u|_{\hat{y}=0}\,,\qquad\qquad \hat{\tau}_{w,z} = -\partial_{\hat{y}}w|_{\hat{y}=0}\,.
\end{equation}

The convective terms $F_{i,j}$ with $j > 1$, along with the wall-shear stress component, are expressed in terms of $\hat{h}$, $\hat{q}_x$, and $\hat{q}_z$ through \textit{closure relations}. These relations are based on the assumption of a self-similar velocity profile in streamwise and spanwise directions \citep{1965662, shkadov1967wave}. As discussed in the previous section, this assumption follows naturally from the boundary-layer structure implied by the long-wave formulation, although it formally requires moderate values of the reduced Reynolds number $\delta$. Here, we retain the traditional IBL closure approach on the same pragmatic grounds—namely, that the shape of these profiles reflects that of the leading-order steady-state solutions given in \eqref{eq:general_sol_magn_stationary}, here concisely rewritten as:

\begin{subequations}
\label{eq:param_velocity_profile}
\begin{gather}
    \hat{u}(\hat{y}) = c_{1}e^{\Upsilon\,\hat{y}} + c_{2}e^{-\Upsilon\,\hat{y}} + \frac{c_{3}}{\Upsilon^2},\\
    \hat{w}(\hat{y}) = c_{4}e^{\Upsilon\,\hat{y}} + c_{5}e^{-\Upsilon\,\hat{y}} + \frac{c_{6}}{\Upsilon^2},
\end{gather}
\end{subequations}
where $\Upsilon=\Ha\,\hat{b}$ and the coefficients $c_{1}, c_{2}, c_{3}, c_{4}, c_{5}, c_{6}$ depend on $\hat{h}, \hat{q}_x$, $\hat{q}_z$, $\Ha$, and $\hat{b}$. By solving a linear system of equations, which includes the non-slip condition at the substrate \eqref{eq:bc_boundary_layer0}, the tangential stress balance at the free surface \eqref{eq:stress_tang_free_surf}, and the flow rate definitions \eqref{eq:flow_rate_def}, the following expressions for these coefficients are derived:
\begin{subequations}
\label{eq:coeff_cs}
\begin{align}
    \begin{split}
        c_1 = \frac{-\hat{h} \Upsilon^2+\Xi e^{\hat{h} \Upsilon} (\hat{h} \Upsilon-1)-\hat{q}_x \Upsilon^2+\Xi}{\Upsilon \left(\hat{h} \Upsilon+e^{2 \hat{h} \Upsilon} (\hat{h} \Upsilon-1)+1\right)}\,,
    \end{split}\\
\begin{split}
    c_2 = \frac{e^{2 \hat{h} \Upsilon} \left(\Xi-\Upsilon^2 (\hat{h}+\hat{q}_x)\right)-\Xi e^{\hat{h} \Upsilon} (\hat{h} \Upsilon+1)}{\Upsilon \left(\hat{h} \Upsilon+e^{2 \hat{h} \Upsilon} (\hat{h} \Upsilon-1)+1\right)}\,,
\end{split}\\
\begin{split}
    c_3 = \frac{\Upsilon \left(\cosh (\hat{h} \Upsilon) \left(\hat{q}_x \Upsilon^2-\Xi\right)+\Upsilon \sinh (\hat{h} \Upsilon)+\Xi\right)}{\hat{h} \Upsilon \cosh (\hat{h} \Upsilon)-\sinh (\hat{h} \Upsilon)}\,,
\end{split}\\
\begin{split}
    c_4 = \frac{\Psi e^{\hat{h} \Upsilon} (\hat{h} \Upsilon-1)+\Psi-\hat{q}_z \Upsilon^2}{\Upsilon \left(\hat{h} \Upsilon+e^{2 \hat{h} \Upsilon} (\hat{h} \Upsilon-1)+1\right)}\,,
\end{split}\\
\begin{split}
    c_5 = \frac{e^{2 \hat{h} \Upsilon} \left(\Psi-\hat{q}_z \Upsilon^2\right)-\Psi e^{\hat{h} \Upsilon} (\hat{h} \Upsilon+1)}{\Upsilon \left(\hat{h} \Upsilon+e^{2 \hat{h} \Upsilon} (\hat{h} \Upsilon-1)+1\right)}\,,
\end{split}\\
\begin{split}
    c_6 = \frac{\Upsilon \left(\cosh (\hat{h} \Upsilon) \left(\hat{q}_z \Upsilon^2-\Psi\right)+\Psi\right)}{\hat{h} \Upsilon \cosh (\hat{h} \Upsilon)-\sinh (\hat{h} \Upsilon)}\,.
\end{split}
\end{align}
\end{subequations}

Introducing \eqref{eq:param_velocity_profile} with the coefficients \eqref{eq:coeff_cs} in the definition of the flux matrix \eqref{eq:flux_matrix} and the wall shear stress \eqref{eq:def_wall_shear} gives the following \textit{closure relations} for the convective flux terms:
\begin{subequations}
\label{eq:closure_flux}
\begin{equation}
\label{eq:closure_flux_1}
    \begin{split}
        &F_{12} = \Big[e^{2 \hat{h} \Upsilon} (\sinh (2 \hat{h} \Upsilon) (\Xi^2 (\hat{h}^2 \Upsilon^2+2)\\&+\Upsilon^4(\hat{h}-\hat{q}_x) (\hat{h}+3 \hat{q}_x)+2 \Upsilon^2 \Xi (\hat{h}+\hat{q}_x))\\&+2 \Upsilon^3 (\hat{h}^3 (-\Xi^2)+\hat{h} \Upsilon^2 (\hat{h}^2+2 \hat{h} \hat{q}_x+2 \hat{q}_x^2)\\&-2\hat{h} \Xi (\hat{h}+\hat{q}_x)+\hat{h}+2 \hat{q}_x)\\&+2 \Upsilon\cosh(2 \hat{h} \Upsilon) (\hat{h}\hat{q}_x^2\Upsilon^4-\Upsilon^2 (\hat{h}+2 \hat{q}_x)-2 \hat{h} \Xi^2)\\&-4\Xi\sinh(\hat{h} \Upsilon) (\Upsilon^2 (\hat{h}(\hat{h} \Upsilon^2 (\hat{h}+\hat{q}_x)-\hat{h} \Xi+1)\\&+\hat{q}_x)+\Xi)+4 \hat{h} \Upsilon \Xi \cosh (\hat{h} \Upsilon) (\Upsilon^2 (\hat{h}+\hat{q}_x)\\&+\Xi))\Big]/\Big[\Upsilon^3(\hat{h}\Upsilon+e^{2\hat{h} \Upsilon} (\hat{h}\Upsilon-1)+1)^2\Big]\,,
    \end{split}
\end{equation}
\begin{equation}
\label{eq:closure_flux_2}
    \begin{split}
        &F_{13} = \Big[e^{2 \hat{h} \Upsilon} (-2 \sinh (\hat{h} \Upsilon)(\Psi \Upsilon^2 (\hat{h}+\hat{q}_x) (\hat{h}^2 \Upsilon^2+1)\\&+\Xi (-2 \hat{h}^2 \Psi \Upsilon^2+\hat{h}^2 \hat{q}_z \Upsilon^4+2 \Psi+\hat{q}_z \Upsilon^2))\\&+\sinh (2 \hat{h} \Upsilon) (\Upsilon^2 \Xi (\hat{h}^2 \Psi+\hat{q}_z)+\Psi \Upsilon^2 (\hat{h}+\hat{q}_x)\\&+\hat{q}_z \Upsilon^4 (\hat{h}-3 \hat{q}_x)+2 \Psi \Xi)-2 \Upsilon^3 (\hat{h} (\hat{h}^2 \Psi \Xi+\hat{h} \Psi\\&-\hat{h} \hat{q}_z \Upsilon^2+\Psi \hat{q}_x-2 \hat{q}_x \hat{q}_z \Upsilon^2+\hat{q}_z \Xi)-\hat{q}_z)\\&+2 \hat{h} \Upsilon \cosh (\hat{h} \Upsilon) (\Psi \Upsilon^2 (\hat{h}+\hat{q}_x)+\Xi (2 \Psi+\hat{q}_z \Upsilon^2))\\&+2 \Upsilon \cosh (2 \hat{h} \Upsilon) (-2 \hat{h} \Psi \Xi+\hat{h} \hat{q}_x \hat{q}_z \Upsilon^4-\hat{q}_z \Upsilon^2))\Big]\\&/\Big[\Upsilon^3 (\hat{h} \Upsilon+e^{2 \hat{h} \Upsilon} (\hat{h} \Upsilon-1)+1)^2\Big]\,,
    \end{split}
\end{equation}
\begin{equation}
\label{eq:closure_flux_3}
    \begin{split}
        &F_{23} = \Big[e^{2 \hat{h} \Upsilon} (\sinh (2 \hat{h} \Upsilon) (\Psi \Upsilon^2 (\hat{h}^2 \Psi+2 \hat{q}_z)+2 \Psi^2-3 \hat{q}_z^2 \Upsilon^4)\\&-2 \hat{h} \Psi \Upsilon^3 (\hat{h}^2 \Psi+2 \hat{q}_z)+4 \Psi \sinh (\hat{h} \Upsilon) \left(\Upsilon^2 (\hat{h}^2 \Psi-\hat{q}_z\right)\\&-\hat{h}^2 \hat{q}_z \Upsilon^4-\Psi)+2 \hat{h} \Upsilon \cosh (2 \hat{h} \Upsilon) (\hat{q}_z^2\Upsilon^4-2 \Psi^2)\\&+4\hat{h}\Psi\Psi\Upsilon\cosh(\hat{h} \Upsilon) (\Psi+\hat{q}_z \Upsilon^2)+4 \hat{h} \hat{q}_z^2 \Upsilon^5)\Big]\\&/\Big[\Upsilon^3 (\hat{h} \Upsilon+e^{2 \hat{h} \Upsilon} (\hat{h} \Upsilon-1)+1)^2\Big],
    \end{split}
\end{equation}
\end{subequations}
with $F_{22} = F_{13}$ and for the wall shear stress term $\hat{\boldsymbol{\tau}}_w=(\hat{\tau}_{w,x},\hat{\tau}_{w,z})^T$, reading:
\begin{subequations}
\label{eq:wall_shear_stress_magn}
\begin{equation}
    \hat{\tau}_{w,x}^M = -\frac{2 e^{\hat{h} \Upsilon} \left(\sinh (\hat{h} \Upsilon) \left(\Upsilon^2 (\hat{h}+\hat{q}_x)-\Xi\right)+\hat{h} \Upsilon \Xi\right)}{\hat{h} \Upsilon+e^{2 \hat{h} \Upsilon} (\hat{h} \Upsilon-1)+1},
\end{equation}
\begin{equation}
    \hat{\tau}_{w,z}^M = \frac{\sinh (\hat{h} \Upsilon) \left(\Psi-\hat{q}_z \Upsilon^2\right)-\hat{h} \Psi \Upsilon}{\hat{h} \Upsilon \cosh (\hat{h} \Upsilon)-\sinh (\hat{h} \Upsilon)}.
\end{equation}
\end{subequations}
Equations~\eqref{eq:model_gen} with the closure relations \eqref{eq:closure_flux} and \eqref{eq:wall_shear_stress_magn} define the 3D IBL model with gas jets and electromagnetic actuators.

\section{Modelling control actuators} \label{subsec:modelling_actuators}
This section outlines the modelling of the gas jet and magnetic actuators, which appear in the reduced-dimensional model equations \eqref{eq:model_gen} and in the \textit{closure relations} \eqref{eq:closure_flux} and \eqref{eq:wall_shear_stress_magn} as pressure ($p_g$) and shear stress ($\boldsymbol\tau_g$) distributions at the free surface, and as a wall-normal magnetic field ($\hat{b}$). The pressure $p_g$ and shear stress $\boldsymbol\tau_g$ are modelled by adapting experimental correlations (Subsection~\ref{jet_actuators}), while the magnetic field $b$ is represented as an approximation of the field induced by an electromagnet (Subsection~\ref{subsec:model_magnet}).

\subsection{Modelling gas jet actuators}
\label{jet_actuators}

The pressure $\hat{p}_{\rm g}$ and shear stress $\hat{\boldsymbol{\tau}}_{\rm g} = (\hat{\tau}_{{\rm g},x}, \hat{\tau}_{{\rm g},z})^T$ distributions at the free surface are modelled using experimental correlations originally developed for circular gas jets impinging on a dry flat substrate \citep{beltaos1974impinging}. Although the present study considers two-dimensional (2D) gas jets, these correlations are convenient for the demonstrational purposes of this work and facilitate a straightforward adaptation from round to planar jet configurations.

We consider a 2D gas jet actuator with an equivalent nozzle diameter of $d = 1.5 , \mathrm{mm}$, positioned at a distance $H = 1.4 , \mathrm{cm}$ from the substrate. Since $H$ is much larger than the characteristic film thickness, the gas flow can be assumed to be decoupled from the liquid film dynamics, as shown in \citet{lacanette2006macroscopic} and \citet{gosset2019experimental}. Under this assumption, $\hat{p}_{\rm g}$ and $\hat{\boldsymbol{\tau}}_{\rm g}$ depend solely on the jet parameters and geometry, and are independent of the film evolution.

The correlations from \citet{beltaos1974impinging} define the pressure and shear stress distributions as:
\begin{equation}
p_g = p_s, f_p, \qquad\qquad
\tau_{g} = \max\left(0, \tau_{0m} f_{\tau}\right),
\end{equation}
where $p_s$ and $\tau_{0m}$ are the stagnation pressure and maximum wall shear stress, respectively, and $f_p$ and $f_{\tau}$ are functions of the radial coordinate $r = x - x_i$, measured from the jet impingement point $x_i$.

The characteristic scales $p_s$ and $\tau_{0m}$ are given by:
\begin{equation}
p_s = 50 \frac{\rho_g U_j^2}{2} \left(\frac{d}{H}\right)^2,
\;
\tau_{0m} = 0.16 \rho_g U_j^2 \left(\frac{d}{H}\right)^2,
\end{equation}
where $U_j$ is the jet exit velocity and $\rho_g$ the gas density.

The nondimensional spatial distributions read:
\begin{subequations}
\label{eq:corr_jet_3D_1}
\begin{align}
\label{eq:corr_jet_3D_pressure}
f_p &= \exp(-114\,\lambda^2), \\[3pt]
\label{eq:corr_jet_3D_shear}
f_{\tau} &= 
0.18\,\frac{1 - \exp(-114\,\lambda^2)}{\lambda} \notag\\
&\quad
- 9.43\,\lambda\,\exp(-114\,\lambda^2)
- \lambda^2,
\end{align}
\end{subequations}where $\lambda = r/H$.
Equation~\eqref{eq:corr_jet_3D_shear} represents a refined version of the original shear stress correlation, including an additional quadratic term for improved accuracy. To model 2D jets, the radial coordinate $r$ is replaced by the coordinate normal to the jet symmetry plane. In this transformation, the pressure distribution $p_g$ remains symmetric with respect to the stagnation line, while the shear stress $\tau_g$ becomes antisymmetric to reflect the change in flow direction across the impingement point, i.e.
\begin{equation}
    \tau_g(x) = \mathrm{sign}(x - x_i)\, \tau_{0m}\, f_\tau(|x - x_i|).
\end{equation}
This antisymmetric formulation ensures a physically consistent adaptation of the circular jet correlation to a planar jet configuration.

\subsection{Modelling electromagnetic actuators}
\label{subsec:model_magnet}

The magnetic field $\hat{b}$, induced by the 2D electromagnetic actuators, is obtained by simplifying the analytical expression for a magnetic field induced by a cylindrical solenoid perpendicular to the substrate with length $L_s$, radius $R_s$, and a metal core with magnetic permeability $\mu_m$. In a cylindrical reference frame $\{\mathbf{O};r_s,z_s\}$ centred at the midpoint of the solenoid, the magnetic field is given by complex analytical expressions derived by \citet{hampton2020closed} (see Appendix \ref{sec:approx_solenoid} for more details).

Assuming that the liquid film is positioned at a distance $z \ll R_s - L_s/2$ and considering a region near the centreline ($r_s \to 0$), the radial component of the magnetic field can be neglected and its axial component ($\hat{b}(\hat{x},\hat{t})$ in the reference frame of the liquid film) can be approximated as a Gaussian function in the $x$, with a time-dependent modulation described by the control function $\hat{b}_t(\hat{t})$. 

The 2D approximated magnetic field centred at $x_0$, reads:
\begin{equation}
\label{eq:approx_magnetic_field_11}
    \hat{b}(\hat{x}, \hat{t}) = \hat{b}_t(\hat{t})\,\exp\left(\frac{-(\hat{x}-\hat{x}_0)^2}{2\hat{\gamma}^2}\right),
\end{equation}
where  $\hat{\gamma}$ is the nondimensional standard deviation. The dimensional value of $\gamma$ defines the spatial scale at which the electromagnet affects the liquid film ($bx_{\rm ref} = \gamma x_{\rm ref}$). This value is crucial for calculating the characteristic timescale of the magnetic effect, as defined in \eqref{eq:time_scale_magn}.

The control function $\hat{b}_t$ and the standard deviations $\hat{\gamma}$ are expressed as a function of the geometrical and physical characteristics of the solenoid (detailed in Appendix~\ref{sec:approx_solenoid}), which read:
\begin{equation}
    \hat{b}_t=\frac{\mu_m I(t) n L_s}{4 R_s b_{\rm ref}\Ca^{1/3}},\quad\quad \hat{\gamma}=\frac{R_s}{2^{3/4} \sqrt{3} \sqrt[4]{17}h_{\rm ref}},
\end{equation}
where $n$ is the number of coils and $I(t)$ is time-varying current flowing through the system.

\section{Numerical Methods} \label{sec:num_methods}
The IBL model in \eqref{eq:model_gen} was implemented in numerical software using the Fourier pseudo-spectral method for spatial discretisation \citep{dutykh2016brief,canuto2007spectral}. This method transforms physical quantities from physical space to Fourier wave-number space, in which spatial derivatives are computed via Fourier differentiation. The liquid film height $\hat{h}$ and the flow rates $\hat{q}_{\hat{x}}$ and $\hat{q}_{\hat{z}}$ are approximated as linear combinations of $N \times M$ 2D plane waves, where $N, M \in \{2n \mid n \in \mathbb{N}^+\}$, expressed as:
\begin{subequations}
\label{eq:fourier_approx}
\begin{align}
    \hat{h} \approx & \sum_{k_x=-N/2}^{N/2-1} \sum_{k_z=-M/2}^{M/2-1} \tilde{h}^{[k_x,k_z]} e^{i(k_x \hat{x} + k_z \hat{z})}, \\
    \hat{q}_{\hat{x}} \approx & \sum_{k_x=-N/2}^{N/2-1} \sum_{k_z=-M/2}^{M/2-1} \tilde{q}_{\hat{x}}^{[k_x,k_z]} e^{i(k_x \hat{x} + k_z \hat{z})}, \\
    \hat{q}_{\hat{z}} \approx & \sum_{k_x=-N/2}^{N/2-1} \sum_{k_z=-M/2}^{M/2-1} \tilde{q}_{\hat{z}}^{[k_x,k_z]} e^{i(k_x \hat{x} + k_z \hat{z})},
\end{align}
\end{subequations}
where $k_x$ and $k_z$ are the nondimensional streamwise and spanwise wave numbers, and $\tilde{h}^{[k_x,k_z]}$, $\tilde{q}_{\hat{x}}^{[k_x,k_z]}$, and $\tilde{q}_{\hat{z}}^{[k_x,k_z]}$ are the Fourier coefficients. 

Substituting \eqref{eq:fourier_approx} into \eqref{eq:model_gen}, calculating the spatial derivative by differentiating \eqref{eq:fourier_approx} and subtracting the right-hand-side from the left-hand-side gives the residual of the spectral approximation. This is then projected onto a basis of Dirac delta functions $\delta_{\hat{x}_i,\hat{z}_j}$, defined on an equispaced grid (collocation points) where $\hat{x}_i = \frac{2\pi}{N} i$ and $\hat{z}_j = \frac{2\pi}{M} j$, with $i, j \in \mathbb{N}$ such that $i \leq N$ and $j \leq M$. Using the properties of the Dirac delta function, the inner product with the residual yields a discrete inverse Fourier transform (IDFT), which provides the values of the state variables at the grid points:
\begin{equation*}
\begin{split}
  & \langle \sum_{k_x=-N/2}^{N/2-1} \sum_{k_z=-M/2}^{M/2-1} \tilde{h}^{[k_x,k_z]} e^{i(k_x \hat{x} + k_z \hat{z})}, \delta_{\hat{x}_i, \hat{z}_j} \rangle = \\
  & \int_{\Omega} \sum_{k_x=-N/2}^{N/2-1} \sum_{k_z=-M/2}^{M/2-1} \tilde{h}^{[k_x,k_z]} e^{i(k_x \hat{x} + k_z \hat{z})} \delta_{\hat{x}_i, \hat{z}_j} \, d\hat{x} d\hat{z} = \\
  & \underbrace{\sum_{k_x=-N/2}^{N/2-1} \sum_{k_z=-M/2}^{M/2-1} \tilde{h}^{[k_x,k_z]} e^{i(k_x \hat{x}_i + k_z \hat{z}_i)}}_{\text{IDFT}} = \hat{h}(\hat{x}_i, \hat{z}_i).
\end{split}
\end{equation*}

Given the values of $\hat{h}$, $\hat{q}_x$, and $\hat{q}_z$ at the grid points for a specific time $\hat{t}$, the nonlinear terms are first computed in physical space and then transformed into Fourier wave number space using the Fast Fourier Transform (FFT). A 2D Finite Impulse Response (FIR) filter is applied to reduce aliasing errors, with a normalised cutoff wave number set to $0.8$ times the Nyquist frequency in both spatial directions. The filter is designed using a Hamming window, with the number of filter coefficients set to half the length of the input signal in each direction. The 2D filter is obtained by independently applying the 1D filters along the $x$ and $z$ directions. Finally, the derivatives are transformed back into physical space using the Inverse Discrete Fourier Transform (IDFT), and time integration is carried out using Euler’s method.

\section{Reinforcement Learning Control} 
\label{sec:reinforc_control}

The undulation control problem consists of finding the optimal weights $\mathbf{w}^*$ of a parametrised feedback control law $\pi_{\mathbf{w}}(\mathbf{s})$, which minimises an objective functional accounting for the amplitude of undulation waves. Based on $n_s$ observations of the liquid film thickness $\mathbf{s}=\mathbf{s}(\hat{h}(\hat{t};\mathbf{w}))\in\mathbb{R}^{n_{\rm s}}$ sampled at time $\hat{t}$, the parametric feedback control law $\pi_{\mathbf{w}}(\mathbf{s})$ gives the value of control actions $\mathbf{a}=\mathbf{a}(\hat{t})\in\mathbb{R}^{n_{\rm a}}$, which corresponds to the nozzle exit velocities $U_j$ or magnetic field intensities $b_j$ of the $n_{\rm a}$ actuators. The objective functional to minimise is given by the time integral of a \textit{reward function} $r(\hat{t}; \mathbf{w})$ over a liquid film simulation that lasts for a duration $T$. In this setting, the undulation control problem is expressed as:
\begin{equation}
\label{eq:opt_prob}
\min_{\mathbf{w}} \int_{0}^{T} r(\hat{t};\mathbf{w}) \, d\hat{t}.
\end{equation}

The reward function $r(\hat{t};\mathbf{w})$ provides an instantaneous performance measure of the feedback control law and it is defined as the space integral of a \textit{running cost} $\mathcal{L}(\hat{h}(\hat{t};\mathbf{w}))$, over the reward region $\Omega_r$, reading:
\begin{equation}
\label{eq:rew_def}
    r(\hat{t};\mathbf{w}) = \int_{\Omega_r} \mathcal{L}(\hat{h}(\hat{t};\mathbf{w})) \; d\hat{x} \, d\hat{z}.
\end{equation}

The running cost function $\mathcal{L}(\hat{h} (\hat{t};\mathbf{w}))$ quantifies the deviation of the liquid film thickness from its initial flat state and is defined as:
\begin{equation}
\label{eq:running_cost}
\mathcal{L}(\hat{h}(\hat{t};\mathbf{w})) = \exp\left(\text{std}\left(50(\hat{h}(\hat{t};\mathbf{w}) - \hat{h}_0)\right)\right) - 1,
\end{equation}
where $\hat{h}_0$ represents the initial thickness of the flat film at $\hat{t}=0$, $exp$ denotes the exponential function and $std$ denotes the standard deviation.

\subsection{Proximal Policy Optimization}
\label{subsec:proximal_policy}

The undulation control problem \eqref{eq:opt_prob} with \eqref{eq:rew_def} and \eqref{eq:running_cost}, constrained by the 3D IBL equations \eqref{eq:model_gen}, is solved using the Proximal Policy Optimisation (PPO) reinforcement learning algorithm \citep{schulman2017proximal} using the implementation provided by the Stable Baselines 3 library \citep{stable-baselines3}.

The PPO algorithm selects actions according to a stochastic control policy whose parameters are progressively updated during training. Given a set of liquid film observations $\mathbf{s}$, the control actions $\mathbf{a}$, representing the nozzle exit velocity and/or the magnetic field intensity, are sampled from a parameterised multivariate Gaussian distribution $\mathcal{N}$, reading:
\begin{equation}
    \mathbf{a}\sim\mathcal{N}\big(\mu_{\mathbf{w}}(\mathbf{s}),\, \sigma_{\mathbf{w}}(\mathbf{s})\big),
\end{equation}
where $\mu_{\mathbf{w}}(\mathbf{s})$ is the mean and $\sigma_{\mathbf{w}}(\mathbf{s})$ is the standard deviation. The value of $\mu_{\mathbf{w}}(\mathbf{s})$ and $\sigma_{\mathbf{w}}(\mathbf{s})$ is given by the function $\pi_{\mathbf{w}}(\mathbf{a}|\mathbf{s})$, which is represented by a neural network with weights $\mathbf{w}$ which consists of two hidden layers of 256 neurons each. 
This stochastic feedback control formulation introduces variability into the action-selection process, thereby encouraging exploration of alternative control strategies and improving overall control performance.

The PPO seeks the optimal parametrisation $\mathbf{w}^*$ through a trial-and-error process that runs 300 liquid film simulations in parallel, evaluates the instantaneous performance of the feedback control law, learns from accumulated experience, and then updates the parameterisation via gradient descent.
\begin{figure}
    \centering
	\includegraphics[width=.9\linewidth]{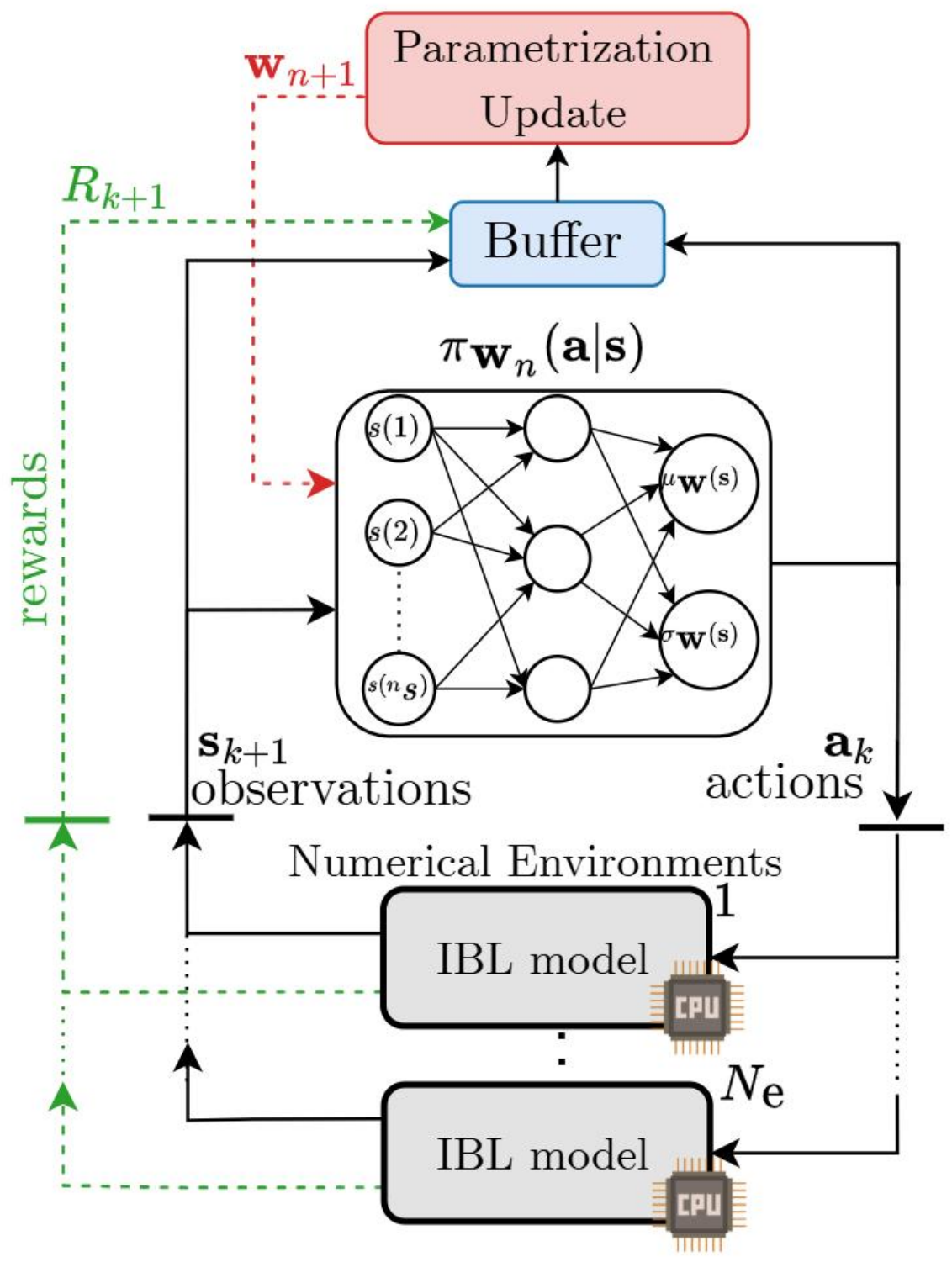}
	\caption{Schematic representation of the reinforcement learning process, in which actions $\mathbf{a}_k$ are selected according to the stochastic control policy $\pi_{\mathbf{w}_n}$ and passed to the numerical environments (grey boxes). These environments integrate the IBL equations and give updated liquid film thickness observations $\mathbf{s}_{k+1}$ and corresponding rewards $r_{k+1}$. The control policy parameters are then updated based on the rewards and observations stored in the buffer (blue box).}
    \label{fig:scheme_machine_learning_control}
\end{figure}

Figure~\ref{fig:scheme_machine_learning_control} shows the main steps of this process: actions are sampled from the Gaussian distribution $\pi_{\mathbf{w}_n}(\mathbf{a}_k|\mathbf{s}_{k-1})$, parameterized by the neural network weights $\mathbf{w}_n$ (white box). These actions are then passed on to $N_e=30$ liquid film simulations, known as \textit{numerical environments}, at time $\hat{t}_k = (k\,m\,-1)\Delta\hat{t}$, where $\Delta\hat{t}$ represents the integration time step. To accelerate the optimisation process, these numerical environments are executed in parallel across multiple CPUs (grey boxes), simulating shifted undulation waves to increase variability and improve the likelihood of learning a better parameterisation (see Subsection \ref{subsec:test_cases} for details).

The actions $\mathbf{a}_k$, which differ across environments, are held constant for $m=30$ integration time steps to prevent sudden changes that could hamper PPO learning. At the end of this period, each numerical environment outputs a vector of liquid film observations $\mathbf{s}_{k+1} = \mathbf{s}(\hat{h}(\hat{t}_{k+1}; \mathbf{w}_{n}))$ evaluated at the last timestep of the integration period $\hat{t}_{k+1}$, along with a total reward function $R_{k+1}$ given by the sum of the reward function at each time step, which reads:\begin{equation}
    R_{k+1}=\sum_{j=1}^{m} r(\hat{t}=(k\,m+j-1)\Delta\hat{t}; \mathbf{w}_{n}).
\end{equation}

The sequence of actions, observations, and rewards gathered during a simulation of duration $\hat{t}=T$ constitutes a \textit{trajectory}. The final time $T$ is defined as $T=\Delta\hat{t}(K,m - 1)$, where $K$ represents the algorithm's total number of interactions with the numerical environment per simulation. Trajectories from different simulations and numerical environments are stored in a buffer (blue box in Fig. \ref{fig:scheme_machine_learning_control}).

Leveraging the knowledge of liquid film dynamics acquired from stored trajectories, the PPO algorithm enhances control performance by updating the weights $\mathbf{w}_n$ every time $(k\cdot N_e\;\rm{mod}\;2048=1)$. This update is made by maximizing a clipped objective function, which entails the minimum of the probability ratio between the updated control law $\pi_{\mathbf{w}_{n+1}}(\mathbf{a}|\mathbf{s})$ and the old control law $\pi_{\mathbf{w}_n}(\mathbf{a}|\mathbf{s})$, along with a \textit{clipping function} $g=g(\iota, A)$ \cite{spinningup_ppo}. The maximisation problem is formulated as:
\begin{equation}
\label{eq:clip_obj}
\max_{\mathbf{w}_{n+1}} \, \mathbb{E}\left[\min\left(\frac{\pi_{\mathbf{w}_{n+1}}(\mathbf{a}|\mathbf{s})}{\pi_{{\mathbf{w}_{n}}}(\mathbf{a}|\mathbf{s})} A, g(\iota, A)\right)\right],
\end{equation}
where $A$ is the \textit{advantage function} and $\mathbb{E}$ denotes the empirical average over the trajectories in the buffer collected with the old control law.

The advantage function  $A$ estimates the benefit of taking an action $\mathbf{a}_{k}$ given an observation $\mathbf{s}_{k}$ and reads:
\begin{equation}
    A_k = -V(\mathbf{s}_k) + r_k + \gamma r_{k+1} + \gamma^2r_{k+2} + \cdots,
\end{equation}
where $\gamma \in [0,1]$ is the discount factor that determines the weight of future rewards, and $V(\mathbf{s}_k)$ represents the value function. This function calculates the expected sum of rewards starting from a given state $\mathbf{s}_k$, with actions chosen according to the control law $\pi_{\mathbf{w}_{n}}$ until the last interaction with the numerical environments ($k=K$), and it is defined as follows:
\begin{equation}
    V(\mathbf{s}_k) = \mathbb{E}\left[\sum_{j=k}^{K}\gamma^{j-k}r_{j}\right].
\end{equation}
The value function is approximated by a separate neural network with $256 \times 256$ neurons and is updated using the sum of rewards collected in the buffer.

The sign of the advantage function $A$ is used to prevent the algorithm from making too large updates in \eqref{eq:clip_obj}, which could result in ineffective control laws. To this end, the clipping function $g(\iota, A)$ defines a confidence region around the probability ratio $\pi_{\mathbf{w}_{n+1}}/\pi_{\mathbf{w}_{n}} = 1$, based on the sign of $A$, ensuring that successive control laws do not differ too much from one another. This function is defined as:
\begin{equation}
    g(\iota, A) = \begin{cases}
      (1 + \iota)A & \text{if } A \leq 0, \\
      (1 - \iota)A & \text{if } A > 0,
    \end{cases}
\end{equation}
where $\iota = 0.2$ is a hyperparameter that defines the clipping range.

\subsection{Control test cases}
\label{subsec:test_cases}
This section describes the numerical environment used to solve the control problem introduced in Section \ref{sec:reinforc_control} using the PPO algorithm presented in Subsection \ref{subsec:proximal_policy}. 

The IBL equations are discretised in space using the Fourier pseudo-spectral method (Subsection~\ref{sec:num_methods}) over a uniform grid of $180$ points along $\hat{x}$ and $88$ along $\hat{z}$. The discrete system is integrated in time via the explicit Euler method with a time step $\Delta \hat{t}= \Delta \hat{x}/80$ ($2\times 10^{-6}\,\rm{s}$). 

We consider an initially flat liquid film, with thickness and flow rates given by: 
\begin{eqnarray}
    \hat{h}_0 = 0.1 \, (34 \, \mu m),\quad\hat{q}_{x0} = \frac{1}{3}\hat{h}_0^3 - \hat{h}_0, \quad \hat{q}_{z0} = 0.
\end{eqnarray}
Given that the liquid film thickness is of order $O(10^{-1})$, the nondimensional group $\Lambda = 9.2$, $\hat{b} \in [0, 1]$ (see Table~\ref{tab:scaling_quant}), and the spatial derivatives of the film thickness are small, the Maxwell stress terms in the normal \eqref{eq:stress_norm_free_surf} and tangential \eqref{eq:stress_tang_free_surf} boundary conditions can be regarded as small quantities and are therefore neglected from our calculations for simplicity.

\begin{figure}
    \centering
	\includegraphics[width=\linewidth]{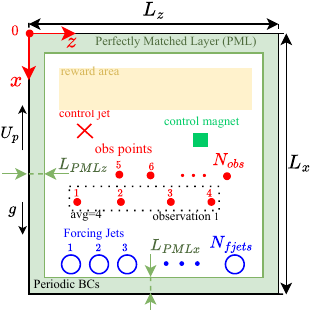}
	\caption{Rectangular domain with periodic boundary conditions used in the undulation control problem, showing the positions of the control jets (red crosses) and control magnets (blue squares), the observation points (red dots), the forcing jets (blue circles), and the reward area (light orange region), with the perfectly matched layer (green region).}
	\label{fig:scheme_spectral_env}
\end{figure}

The computational domain $\Omega$ is a rectangle of size $L_x = 45 \, (7.26\,\rm{cm})$ in the streamwise direction and $L_z = 22 \, (3.55\, \rm{cm})$ in the spanwise direction. Figure~\ref{fig:scheme_spectral_env} shows a schematic of the numerical environment where the domain $\Omega$ includes a set of control gas jets (red crosses) and electromagnets (green squares), the locations of the observation point of the liquid film thickness (indicated by red dots), a series of $N_{fj}=15$ forcing jets aligned along the $\hat{z}$ direction (depicted by blue circles) and a Perfectly Matched Layer (PML) (light green area) around the periodic boundaries (black contour line).

The PML is used to simulate open-flow conditions. This layer absorbs the incoming waves without reflection, preventing them from re-entering the domain through the periodic boundaries. A linear perfectly matched layer \citep{johnson2021notes,besse2022perfectly} is implemented with dimension $L_{PMLx} = L_x/6$ and $L_{PMLz} = L_z/6$. The PML is implemented by replacing the spatial derivative with the following expressions:
\begin{equation}
\label{eq:transformation_PRL}
    \partial_{\hat{x}}\rightarrow \Big(1 + \frac{i\sigma_x}{\omega}\Big)^{-1}\partial_{\hat{x}},\quad\quad\partial_{\hat{z}}\rightarrow \Big(1 + \frac{i\sigma_z}{\omega}\Big)^{-1}\partial_{\hat{z}},
\end{equation}
where $\omega$ is the angular frequency of the liquid film waves and $\sigma_x(\hat{x},\hat{z})$ and $\sigma_z(\hat{x},\hat{z})$ are positive functions which determine the strength of the absorbing layer, defined as:
\begin{subequations}
\begin{equation}
    \sigma_x(\hat{x},\hat{z})=
    \begin{cases}
      0, & \text{if}\ (\hat{x},\hat{z})\in\Omega_1 \\
      \hat{x}^3, & \text{if}\ (\hat{x},\hat{z})\in\Omega_{PML}
    \end{cases},
\end{equation}
\begin{equation}
    \sigma_z(\hat{x},\hat{z})=
    \begin{cases}
      0, & \text{if}\ (\hat{x},\hat{z})\in\Omega_1 \\
      \hat{z}^3, & \text{if}\ (\hat{x},\hat{z})\in\Omega_{PML}
    \end{cases}.
\end{equation}
\end{subequations}

The transformation in equation \eqref{eq:transformation_PRL} involves introducing a set of auxiliary equations, discretised both in space and time, similar to the IBL equations. For convenience, these equations are provided in Appendix \ref{appx:PML}.

Moving to undulation instability, this manifests in the liquid film as two-dimensional waves propagating streamwise, with minimal variation along $\hat{z}$. Numerical simulations from Large Eddies Simulations (LES) \citep{barreiro2021dynamics,BarreiroVillaverde2024} and 2D IBL model \citep{mendez2021dynamics} indicate that these waves are characterized by a nondimensional frequency in the range $[0.05, 0.2]$, a peak-to-peak amplitude ranging from $5\%$ to $10\%$ of the unperturbed film thickness, and wavelengths within the range $\lambda \in [25, 35] \, \text{mm}$.

These waves originate from the interaction between the wiping gas jet and the liquid film \citep{BarreiroVillaverde2024}. To emulate comparable undulation patterns over short distances, we introduce an artificial forcing mechanism that consists of numerical jets represented as source terms in the governing equations. These sources inject alternating positive and negative perturbations into the local flow rate, effectively seeding waves with amplitudes and wavelengths consistent with those reported in \citep{barreiro2021dynamics,BarreiroVillaverde2024}. The formulation of these sources follows the correlations defined for the control actuators (see Sec.~\ref{subsec:modelling_actuators}), but allows negative input velocities. Although such negative forcing is non-physical, the goal here is not to represent realistic jet actuation but rather to generate synthetic disturbances that reproduce the characteristic undulation instability within the computationally accessible domain. For these forcing jets, the velocity of the $j$-th jet is defined as
\begin{equation}
    U_{j,f}^i = \grave{A}_{fj}\cos(2\pi\hat{f}_{fj}\hat{t} + \phi_i),
\end{equation}
where $\phi_i\in[0.2\pi,0.5\pi]$ introduces mild spanwise variability. Moreover, to introduce greater variability into the simulation and promote the learning of robust control functions, the amplitude $\grave{A}_{fj}$ and the frequency $\hat{f}_{fj} $ are modelled as finite Fourier series $y_{fj}$ with randomly selected coefficients $a_j$ and $b_j$ \citep{filip2019smooth}, reading:
\begin{equation}
\label{eq:defyy}
\begin{split}
    y_{fj}(\hat{t}) = \sqrt{2}\sum_{j=1}^4\Big[a_j\cos{\left(2\pi j \hat{t}\right) + b_j\sin{(2\pi j\hat{t})}}\Big],
\end{split}
\end{equation}
where the coefficients $a_j$ and $b_j$, which are different for the amplitude and the frequency, are sampled from the normal distribution $\mathcal{N}(0,1/(y+1))$. Based on \eqref{eq:defyy}, the amplitude $\grave{A}_{fj} $ and frequency $\hat{f}_{fj}$ are rescaled in the range of interest as: 
\begin{subequations}
\begin{equation}
\grave{A}_{fj} = y_{fj} 42.55\,{\rm m/s} + 31.45\,{\rm m/s},    
\end{equation}
\begin{equation}
    \hat{f}_{fj} = y_{fj} 0.07 108\,{\rm Hz} + 93.3\, {\rm Hz}.
\end{equation}
\end{subequations}

\begin{figure}
    \centering
    \includegraphics[width=\linewidth]{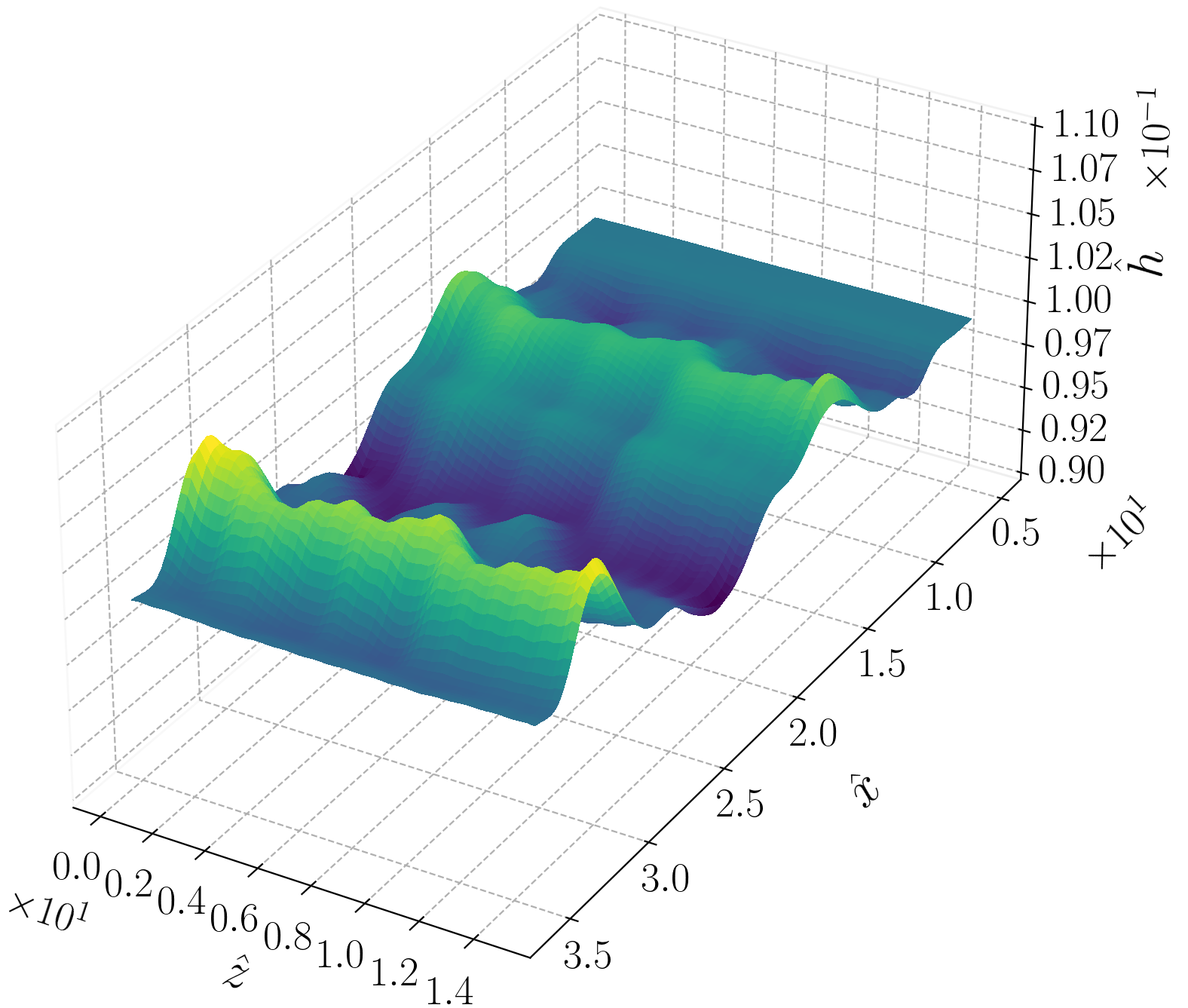}  
\caption{3D plot of the non-dimensional liquid film showing the propagation of the undulation instability waves generated by the forcing gas jets.}
    \label{undulation_3D_spectral}
\end{figure}

Figure~\ref{undulation_3D_spectral} shows the nondimensional uncontrolled liquid film with the undulation instability induced by the forcing jets propagating through the domain. The wavelength ranges from $25\,\text{mm}$ to $27\,\text{mm}$, in the range of most amplified waves identified in \cite{mendez2021dynamics}.

Regarding the control actuators, as introduced in subsection~\ref{subsec:modelling_actuators}, the magnetic field $\hat{b}_i$, is approximated as a Gaussian function centred at $(\hat{x}_{0,i}, \hat{z}_{0,i})$ with standard deviation where $\hat{\gamma} = 1\,(0.16 \, \rm{cm})$, and $b_{t,k} \in [0,1]$ is the control parameter. Although the standard deviation and the Hartmann number ($\Ha = 6$), do not lie on the Pareto front of optimal solutions \cite{pino2024multi}, the magnetic Gaussian roughly spans half the wavelength of the undulation instability, allowing the controller to direct the electromagnet's action to either a valley or a crest.

The pressure and shear distributions of the control and forcing jets are modelled using experimental correlations for 3D impinging circular jets (see Subsection~\ref{subsec:modelling_actuators}), with the velocity at the jet nozzle exit, $U_j \in [0,50] \, \text{m/s}$, used as the control parameter. To avoid sharp changes in the control action, this is smoothed with an exponential moving average, reading:
\begin{equation}
    (U_j)_{i+1,e} = (U_j)_{i,e} + 0.8((U_j)_{i+1,e}^{ML} - (U_j)_{i-1,e}),
\end{equation}
where $(U_j)_{i+1,e}$ is the action passed to the solver, $(U_j)_{i+1,e}^{ML}$ is the action selected by the PPO, and $(U_j)_{i-1,e}$ is the action passed to the solver at the previous interaction with the film.

Three actuator configurations were analysed: a standalone 2D gas jet, a standalone 2D electromagnet, and a combination of both actuators. For these, two control policies were considered. The first, referred to as the \emph{harmonic policy}, prescribes the actuation signal in a sinusoidal form,
\[
\hat{u}_c(t_k) = \acute{A}(\mathbf{s}_k)\,\sin\!\big(2\pi \acute{f}(\mathbf{s}_k)\, t + \acute{\phi}(\mathbf{s}_k)\big),
\]
where $\hat{u}_c$ denotes the actuator command, corresponding to either the jet velocity $U_j$ or the magnetic field intensity $\hat{b}_t$, and $\mathbf{s}_k$ is the state vector containing the observed film-height information at time step $k$. 
The parameters $\acute{A}(\mathbf{s}_k)$, $\acute{f}(\mathbf{s}_k)$, and $\acute{\phi}(\mathbf{s}_k)$ are determined by the PPO policy as functions of the instantaneous system state. 
This formulation, applied only to single–actuator configurations (either a jet or an electromagnet), was introduced to bias the learning process toward harmonic actuation while still allowing the policy to decide whether the forcing should remain strictly periodic. 
Suppose the learned parameters converge to nearly constant values. In that case, the resulting actuation corresponds to a classical phase–opposition control law, where the actuator oscillates at the dominant wave frequency but in opposite phase to attenuate the free–surface deformation. 
If, instead, the policy continuously modulates $\acute{A}$, $\acute{f}$, or $\acute{\phi}$, the control signal departs from a pure sinusoid and adapts its waveform to mitigate the incoming, not perfectly periodic, wave pattern.

In the second formulation, referred to as the \emph{non-harmonic policy}, the PPO agent directly outputs the actuator command at each time step as a function of the instantaneous film-height observations. 
This corresponds to a fully closed-loop feedback control law in which the policy learns a mapping from the system state to the continuous actuation signal. 
The non-harmonic policy, therefore, represents a generalisation of the previous case, enabling non-periodic and adaptive control strategies that can respond to arbitrary disturbances in real time.

\begin{figure}
    \centering
    \includegraphics[width=\linewidth]{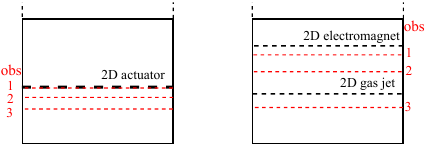}
    \caption{Schematic showing the numbering of observations and actuators in the 3D undulation control environment.}
    \label{fig:schme_numbering_spectral}
\end{figure}

Figure~\ref{fig:schme_numbering_spectral} illustrates the numbering and placement of the three liquid film observations (red dashed lines) and the actuators (black dashed lines) for the case with a single (left) and double (right) actuators.

The three liquid film observations are given by the average of the liquid film thickness sampled at four points along $\hat{z}$ at three locations along $\hat{x}$. For a single 2D actuator, these are placed at  $\hat{x}= (32,31,30)(x = (5.16,5,4.83)\rm{cm}$ and $\hat{z}=(6.3,9.4,12.56,15.7)((1,1.5,2,2.5)\,\rm{cm})$. For the two 2D actuators, these are placed at  $\hat{x}=(33,31,29)((5.32,5,4.68)\,\rm{cm})$ and $\hat{z}=(6.2,9.4,12.6,15.8) ((1,1.5,2,2.5)\,\rm{cm})$.

Concerning the reward area $\Omega_r$ (yellow area in Figure \ref{undulation_3D_spectral}), in the case of a single actuator, this is defined as $\Omega_r=\hat{x}\in (10.06,31.17)((1.62,5.02)\,\rm{cm}) and \hat{z}\in (4.55,17)((0.73,2.85)\,\rm{cm})$ and in the case of two actuators $\Omega_r=\hat{x}\in (11.31,29.4)((1.82,4.74)\,\rm{cm}) and \hat{z}\in (4.55,17)((0.73,2.85)\,\rm{cm})$
\section{Results} \label{sec:results}

This section is divided into three subsections. Subsection~\ref{subsec:res_asym} reports on the asymptotic behaviour of the steady-state solution and the \textit{closure terms} for a magnetic field approaching zero. Subsection ~\ref{sec5p1} reports on the wave dynamics that arise when setting the non-linear phase speed in the leading order in the IBL model. Finally, subsection ~\ref{sec5p2} reports on the results of Reinforcement Learning control of the unstable undulation perturbations using gas jets and electromagnetic actuators, considering both standalone and tandem configurations.

\subsection{Asymptotic Behaviour}\label{subsec:res_asym}
This subsection focuses on the behaviour of the steady-state solution found in Subsection~\ref{Steady} and the \textit{closure relations} introduced in Subsection~\ref{subsec:integral_model} as the magnetic field $\hat{b}$ and the Hartmann number $\Ha$ approach zero. 
\begin{figure}
    \centering
  \begin{subfigure}[b]{\linewidth}
    \includegraphics[width=0.9\linewidth]{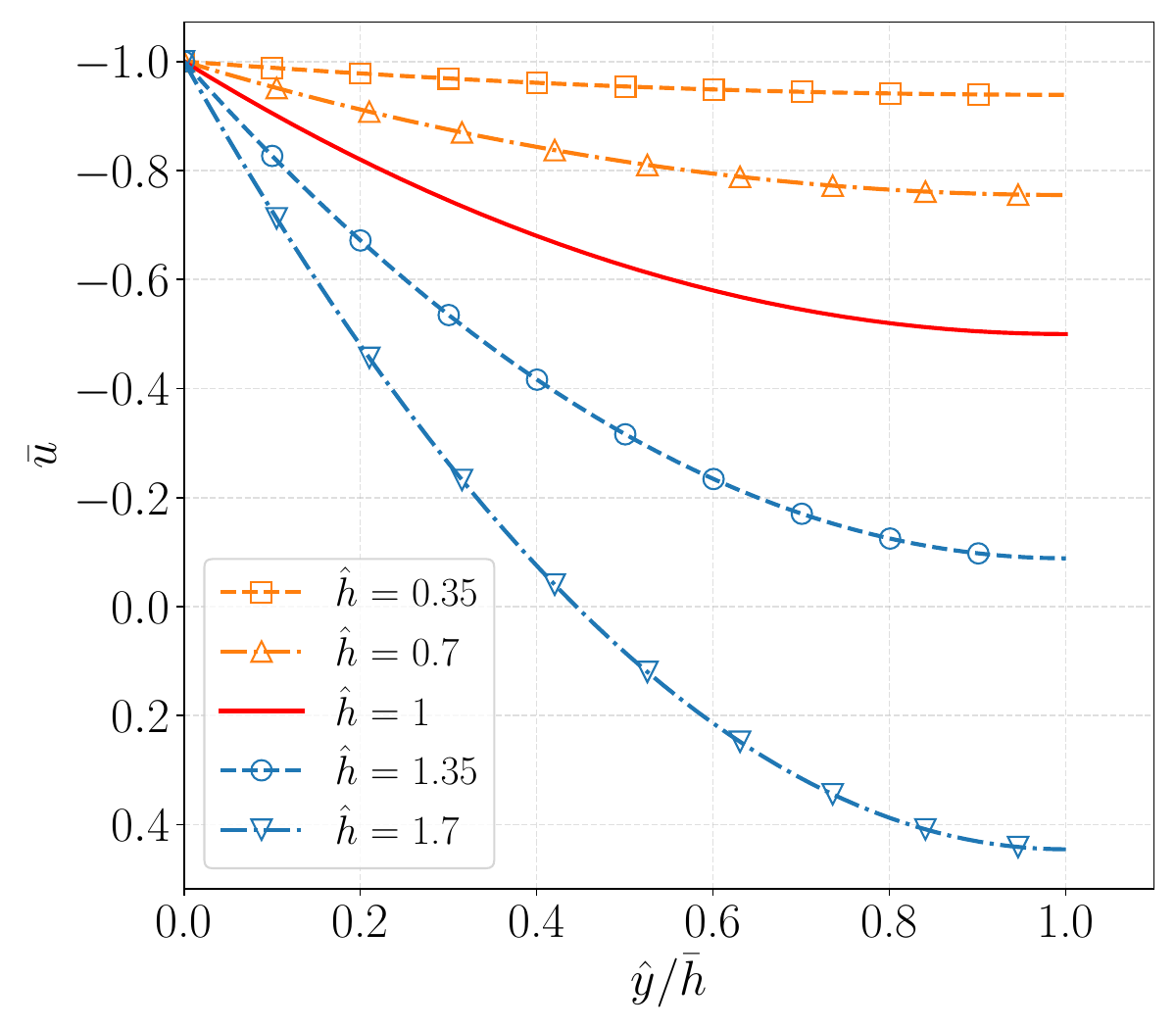}
    \caption{}
    \label{}
  \end{subfigure}
    \begin{subfigure}[b]{\linewidth}  
    \includegraphics[width=0.9\linewidth]{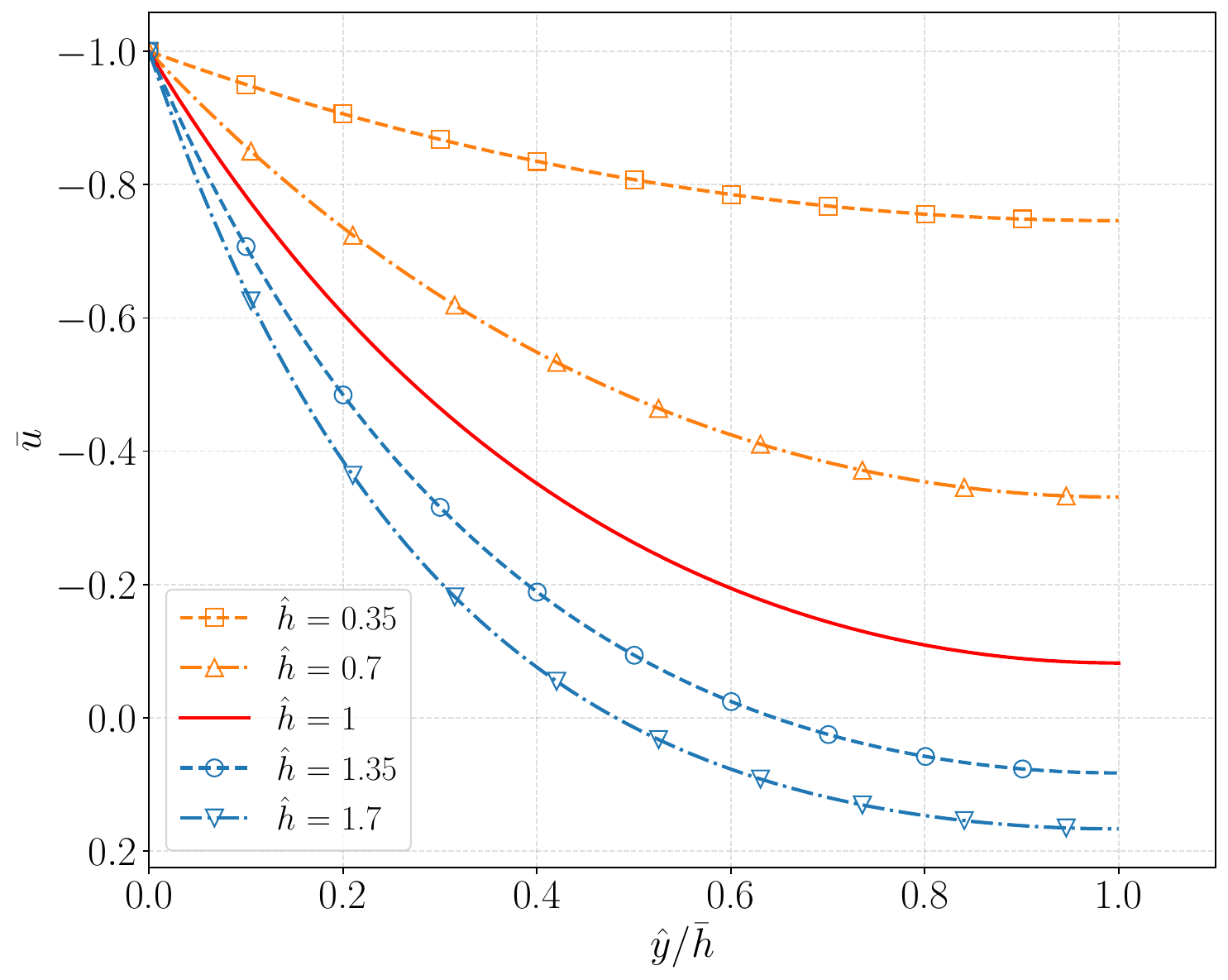}
    \caption{}
    \label{fig:scaling_fig_magnetic_b}
  \end{subfigure}
    \caption{Steady-state velocity profiles in the thin film ($\hat{h}<1$) (orange shadowed) and in the thick film ($\hat{h}>1$) (blue shadowed) for the case (a) without magnetic field and (b) with $\hat{b}=1$ and $\Ha=3$.}
    \label{fig:scaling_fig_magnetic_vel}
\end{figure}

Expanding the steady state velocity profile \eqref{eq:final_eq_velocity_magnetic} in Taylor series around ($\hat{b}=0,\Ha=0$) gives:
\begin{equation}
\label{eq:expansion_steady_vel_profile}
\begin{gathered}
    \bar{u} = \left(\hat{h}\hat{y}-\frac{\hat{y}^2}{2}-1\right)\\
    +\hat{b}^2 \Big(-\frac{1}{24}(\hat{y}(8 h^3-4 (\hat{y}^2+6)\hat{h}+\hat{y}^3\\
    +12 \hat{y})) \Ha^2+O(\Ha^4)\Big)+O(\hat{b}^4)\,.
\end{gathered}
\end{equation}
The velocity reaches a well-defined limiting value, as it recovers, at leading order, the solution that was previously derived in the case without a magnetic field by \citet{ivanova2023evolution}. This result serves as the first analytical validation of our model, confirming its asymptotic consistency.

Figure~\ref{fig:scaling_fig_magnetic_vel} presents the steady-state results for case (a) without and (b) with a magnetic field ($\Ha = 3$), for different values of $\hat{h}$. In both cases, the velocity profile is parabolic, but the variation between the different profiles is more pronounced in the absence of the magnetic field. In particular, the magnetic field exerts a stronger effect when $\hat{h} > 1$, in which case the streamwise velocity remains negative, balancing the gravitational force. In contrast, in the absence of the magnetic field, the downward gravitational force causes the velocity to become positive over most of the liquid film thickness.
\begin{figure}
\centering
    \includegraphics[width=\linewidth]{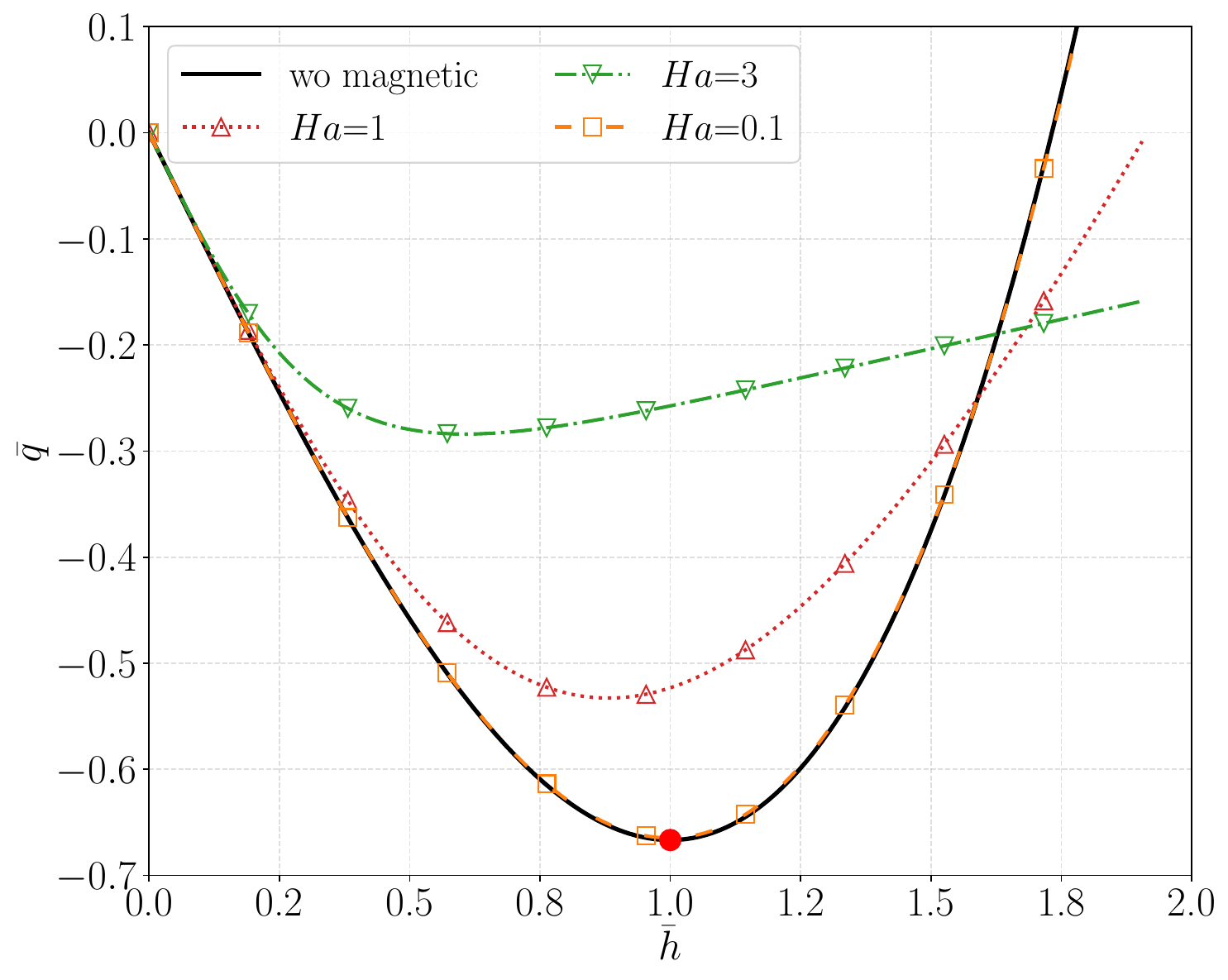}
    \label{fig:scaling_fig_magnetic_a}
    \caption{The relationship between the nondimensional liquid film thickness $\Bar{h}$ and the nondimensional flow rate $\bar{q}$: (a) without a magnetic field (black continuous line) and (b) with $\hat{b} = 1$ for different values of $\Ha$ (coloured lines with markers).}
    \label{fig:steady_state_q_magn}
\end{figure}

Moving to the steady-state flow rate \eqref{eq:final_q_func_std_magn}, Figure~\ref{fig:steady_state_q_magn} shows the relationship between $\bar{q}$ and $\Bar{h}$ for different values of $\Ha$ (lines with markers) and the case without a magnetic field (black continuous line). Like the velocity profile, $\bar{q}$ recovers the case without a magnetic field as $\Ha \to 0$. In the absence of the magnetic field, $\bar{q}$ is equal to zero for $\hat{h}=0$ and $\hat{h}=\sqrt{3}$ and it has two solution branches: \textit{thin film} ($\hat{h} < 1$) and \textit{thick film} ($\sqrt{3} \geq \hat{h} > 1$), with a minimum at $\bar{h} = 1$, corresponding to the flat film solution of \citet{derjaguin_thickdipcoating}. The branch with $\bar{q} > 0$ in the range $\hat{h} > \sqrt{3}$ is attainable solely by withdrawing the substrate from the bath and, at the same time, feeding the liquid film from above. 

When the magnetic field is applied, the minimum shifts to lower flow rates and thinner film thicknesses as $\Ha$ increases. In addition, the position of the zero net flow rate changes to higher values of $\Bar{h}$ as $\Ha$ increases. This indicates that the locus of possible steady-state solutions, attainable by simply extracting the liquid film from a bath, expands with the application of the magnetic field.

The same asymptotic behaviour observed for the steady-state velocity profile in \eqref{eq:expansion_steady_vel_profile} also applies to the flux terms and the wall shear stress \textit{closure relations}. The Taylor expansions of \eqref{eq:closure_flux} and \eqref{eq:wall_shear_stress_magn} around the point ($\hat{b}=0, \Ha=0$) are as follows:

\begin{subequations}
\begin{equation}
\begin{split}
F_{12} \approx& \frac{\hat{h}^4 \hat{\tau}_{gx}^2+24 \left(\hat{h}^2+2 \hat{h} \hat{q}_x+6 \hat{q}_x^2\right)}{120\hat{h}}\\& +\frac{6 \hat{h}^2 \hat{\tau}_{gx} (\hat{h}+\hat{q}_x)}{120\hat{h}}
\end{split}
\end{equation}\\
\begin{equation}
\begin{split}
F_{13} \approx &\frac{\hat{h}^4 \hat{\tau}_{gx} \hat{\tau}_{gz}+3\hat{h}^2 \hat{\tau}_{gz}(\hat{h}+\hat{q}_x)}{120 \hat{h}}+\\&+ \frac{3 \hat{h} \hat{q}_z(\hat{h}\hat{\tau}_{gx}+8)}{120 \hat{h}} + \frac{144\hat{q}_x\hat{q}_z}{120 \hat{h}}\,,
\end{split}
\end{equation}\\ 
\begin{equation}
\begin{split}
F_{23} \approx& \frac{\hat{h}^4\hat{\tau}_{gz}^2+6 \hat{h}^2 \hat{q}_z\hat{\tau}_{gz}+144\hat{q}_z^2}{120\hat{h}}\,,
\end{split}
\end{equation}\\
\begin{equation}
\begin{split}
\hat{\tau}_{w,x} \approx& \left(\frac{\hat{\tau}_{g,x}}{2}-\frac{3(\hat{h}+\hat{q}_x)}{\hat{h}^2}\right)\,,
\end{split}
\end{equation}\\
\begin{equation}
\begin{split}
\hat{\tau}_{w,z} \approx& \left(\frac{\hat{\tau}_{g,x}}{2}-\frac{3 \hat{q}_z}{\hat{h}^2}\right)\,.
\end{split}
\end{equation}
\end{subequations}

The closure relation exhibits a regular limit, recovering at leading order the values obtained in the absence of the magnetic field found by \citep{ivanova2023evolution}. This provides an additional analytical validation of our model, showcasing the consistency of the term entering the unsteady equations.

\subsection{Leading-Oder convective velocity}\label{sec5p1}
Having established the consistency of the IBL model with the non-magnetic case under both steady and unsteady conditions, this section investigates the effect of magnetic forces on the phase speed of surface waves. To this end, we neglect all terms of order $\varepsilon$ in \eqref{eq:model_gen}, assuming the pressure gradient terms of order $\varepsilon$, and $\hat{b}=1$ everywhere in the domain. Under these assumptions, the solution to \eqref{eq:model_gen} is given by the steady-state solution \eqref{eq:final_q_func_std_magn} for the flow rate, with zero flow rate along the $z$-axis. Substituting this into the integral form of the continuity equation yields a convection equation for the liquid film thickness, which is given by:
\begin{equation}
    \partial_t\hat{h} + c(\hat{h},\Ha)\partial_x\hat{h} = 0\,,
\end{equation}
where $c$ is the nonlinear phase speed, reading:
\begin{equation}
\label{eq:leading_order_adv}
    c(\hat{h},\Ha) = \frac{1-\left(\Ha^2+1\right) \text{sech}(\Ha \hat{h})^2}{\Ha^2}\,.
\end{equation}

The Taylor expansion of \eqref{eq:leading_order_adv} around $\Ha=0$, reads:
\begin{equation}
    \begin{aligned}
    c = (\hat{h}^2-1)+\Ha^2\left(\hat{h}^2-\frac{2}{3} \hat{h}^4\right)+O\left(\Ha^4\right).
    \end{aligned}
\end{equation}

As for the cases treated in subsection~\ref{subsec:res_asym}, the leading-order term is consistent with the case without a magnetic field \cite{ivanova2023evolution}.

\begin{figure}
\centering
    \includegraphics[width=\linewidth]{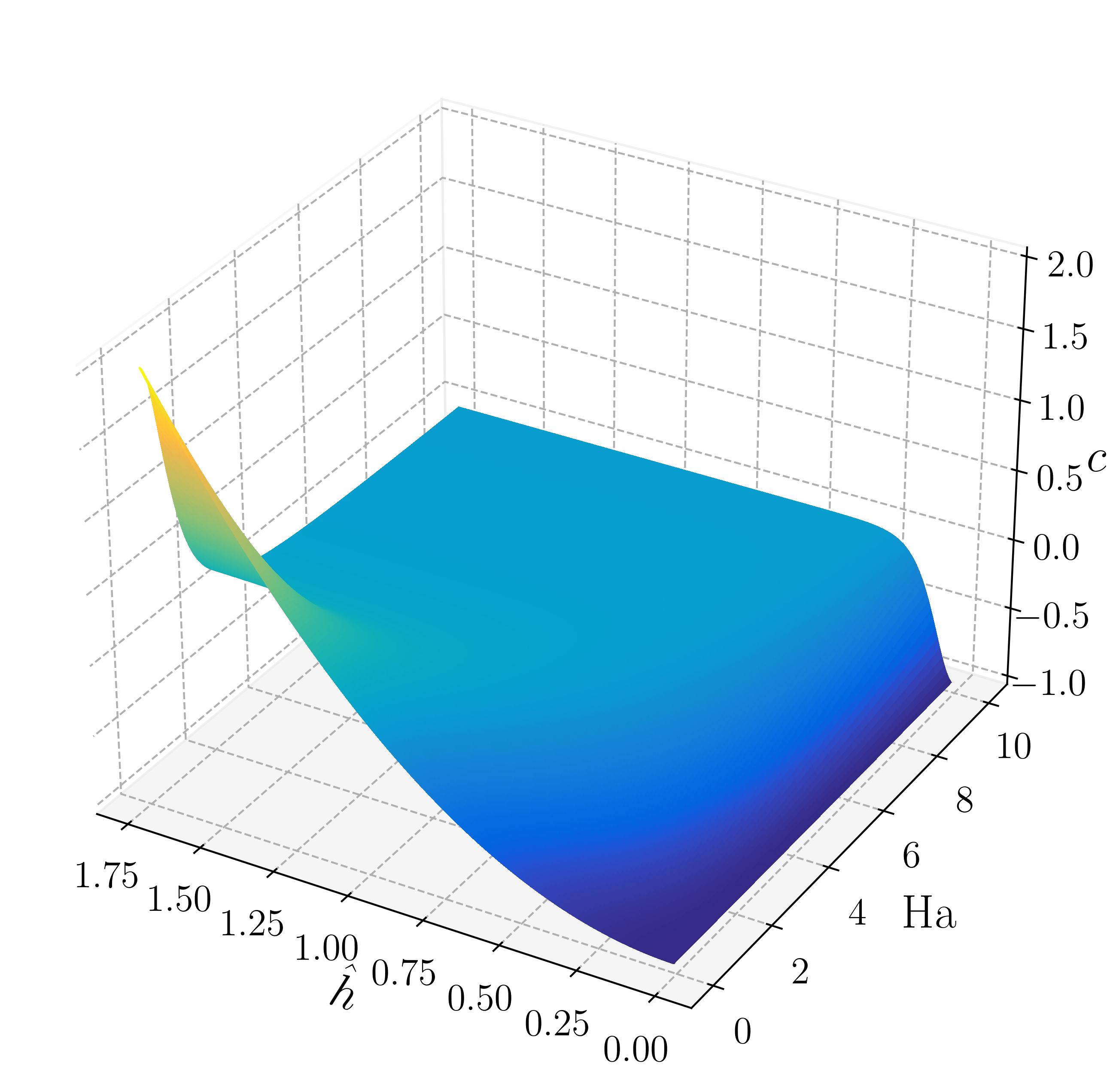}
  \caption{Variation of the leading order phase speed $c$ as a function of the liquid film nondimensional thickness $\hat{h}$ and the Hartmann number $\Ha$.}
  \label{fig:leading_order_vel}
\end{figure}
Figure~\ref{fig:leading_order_vel} illustrates the phase speed $c$ \eqref{eq:leading_order_adv} as a function of the non-dimensional thickness $\hat{h}$ and the Hartmann number $\Ha$. For small values of $\Ha$, $c$ has a quadratic dependence on $\hat{h}$, with waves propagating upstream ($c < 0$) for $\hat{h} < 1$ and downstream ($c > 0$) for $\hat{h} > 1$. For $\Ha\geq 2$, $c$ reaches a plateau at approximately $c \approx 0.1$ for $\hat{h} > 0.3$. As we have already seen in figure~\ref{fig:scaling_fig_magnetic_b}, the magnetic effects for these conditions counterbalance the gravitational effects, resulting in an equilibrium solution where waves always propagate in the direction of substrate movement. This highlights how the magnetic force may affect the absolute/convective instability properties of these flat-film solutions relative to the case without a magnetic field \citep{pino2024linear}.

\subsection{Undulation control using Reinforcement Learning}\label{sec5p2}
This section presents the results of the undulation control problem introduced in subsection~\ref{subsec:test_cases}, using the PPO algorithm described in subsection~\ref{subsec:proximal_policy}. Three scenarios are considered: two with a single actuator (either a gas jet or an electromagnetic actuator) and one with both actuators operating simultaneously. The performance of the trained agent is assessed by analysing the evolution of reward during training episodes (learning curves) and by evaluating the agent's performance over 10 test episodes at the end of training. The results of this evaluation are reported as the mean and standard deviation of episode rewards, which are compared with those from the uncontrolled case. To provide further insight into the control mechanism, we also present the evolution of the control function and snapshots of the liquid film along the centreline during one evaluation episode.

\subsubsection{Control with single gas-jets}
The first case involves a single gas-jet actuator with a prescribed harmonic control function, and the agent is free to take any action at any time.
\begin{figure}
\centering
    \includegraphics[width=\linewidth]{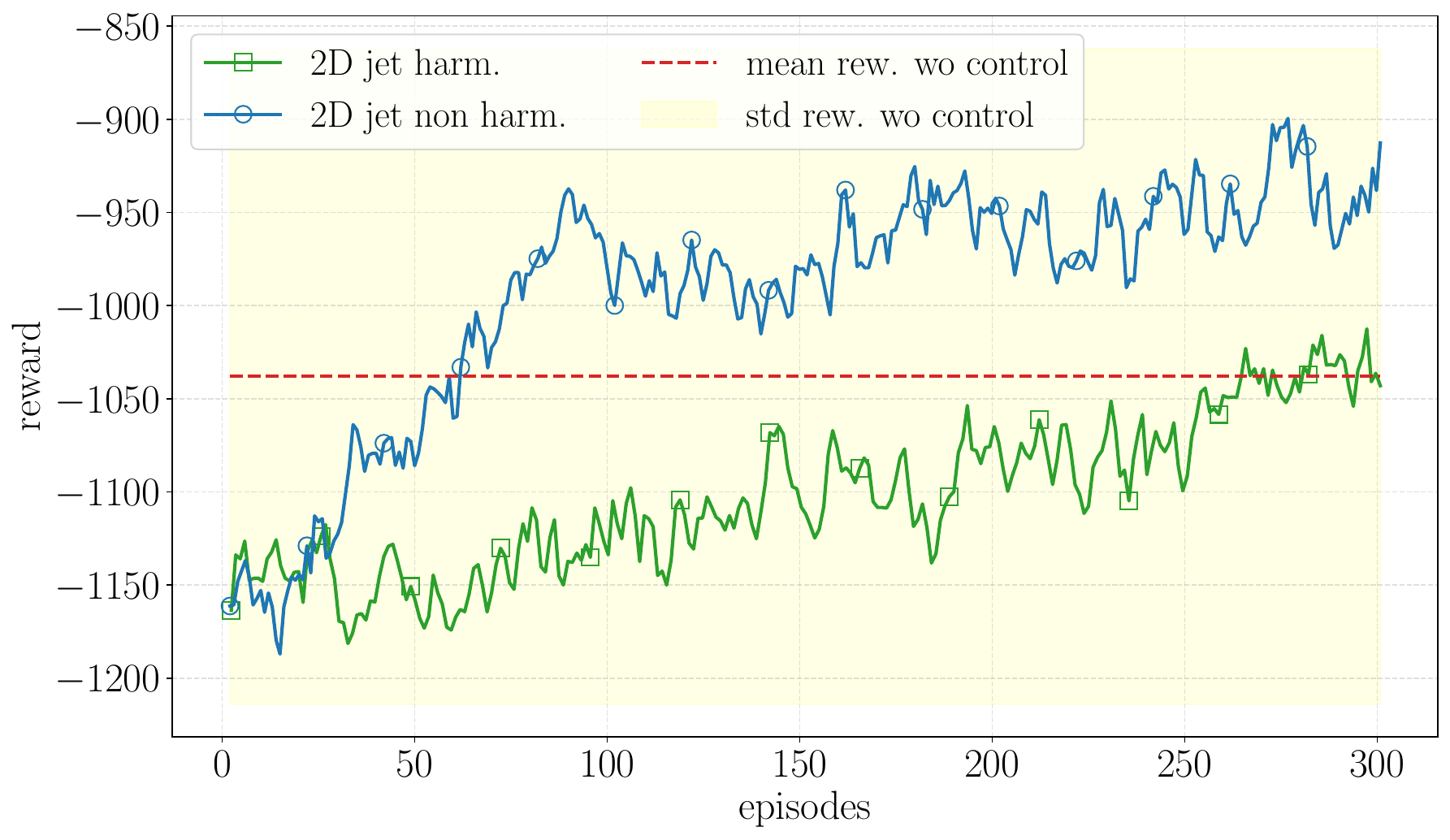}
  \caption{Learning curves for the harmonic (green line with squares) and non-harmonic (blue line with circles) control functions with a gas jet.}
  \label{fig:learning_curve_2D_jet_spectral}
\end{figure}

Figure~\ref{fig:learning_curve_2D_jet_spectral} shows the learning curves for the case with a single 2D gas jet, comparing the harmonic control function (green curve with squares) and the nonharmonic control function (blue curve with circles) against the mean (red dashed line) and the standard deviation (yellow shaded area) of the uncontrolled case. Both control functions achieve mean rewards at the end of the training that exceed those of the uncontrolled case.

\begin{table}
\centering
\caption{Total reward means and standard deviation for a single gas jet in the evaluation episodes.}
\label{tab:mean_std_eval_2D_magn}
\begin{tabular}{c@{\hspace{0.4cm}}c@{\hspace{0.3cm}}c@{\hspace{0.3cm}}c@{\hspace{0.3cm}}}
\toprule
 \textbf{Reward} & \textbf{wo Control} & \textbf{Harmonic} & \textbf{Non-harmonic} \\ 
\midrule
\begin{tabular}[c]{@{}c@{}}Mean \end{tabular}& 
\textbf{-1038} & -826 & -790 \\ \bottomrule
\begin{tabular}[c]{@{}c@{}}Standard \\ Deviation\end{tabular}& 
\textbf{176} & 153 & 136  \\ \bottomrule
\end{tabular}
\end{table}
Table~\ref{tab:mean_std_eval_2D_magn} summarises the mean and standard deviation of the evaluation episodes for controlled and uncontrolled cases. As in the learning phase, both control functions yield a higher mean reward than in the uncontrolled case. Furthermore, in the controlled case, the reward variability around the mean is reduced, as indicated by a smaller standard deviation. This shows that the PPO identified an effective control action that reduced undulation amplitude in the reward area.

\begin{figure}
  \begin{subfigure}[b]{\linewidth}
    \includegraphics[width=0.9\linewidth]{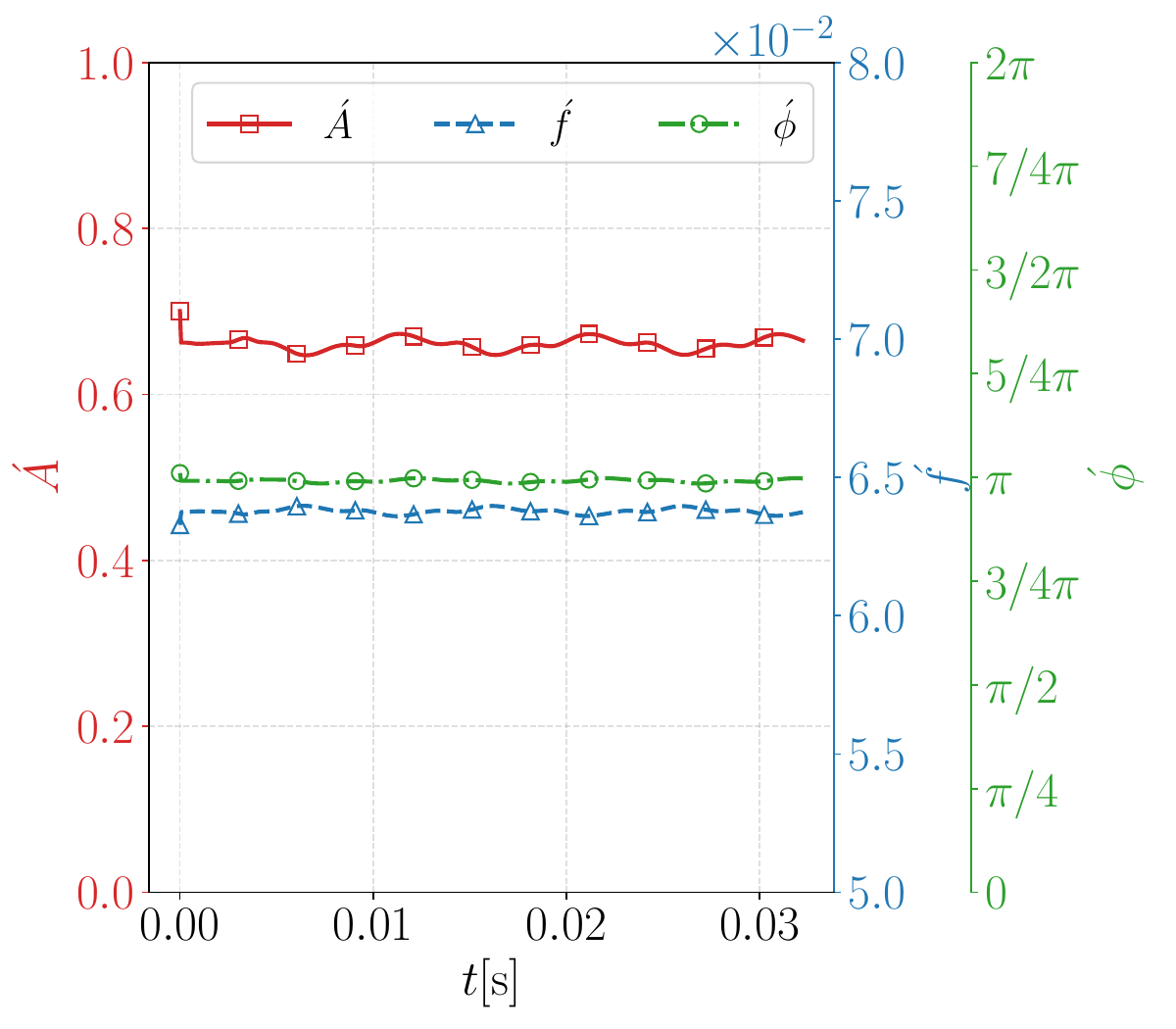}
    \caption{}
\label{fig:results_harmonic_2D_jet_3D_undulation_control_spectral_a}
  \end{subfigure}
  \hfill
  \begin{subfigure}[b]{\linewidth}
    \includegraphics[width=0.9\linewidth]{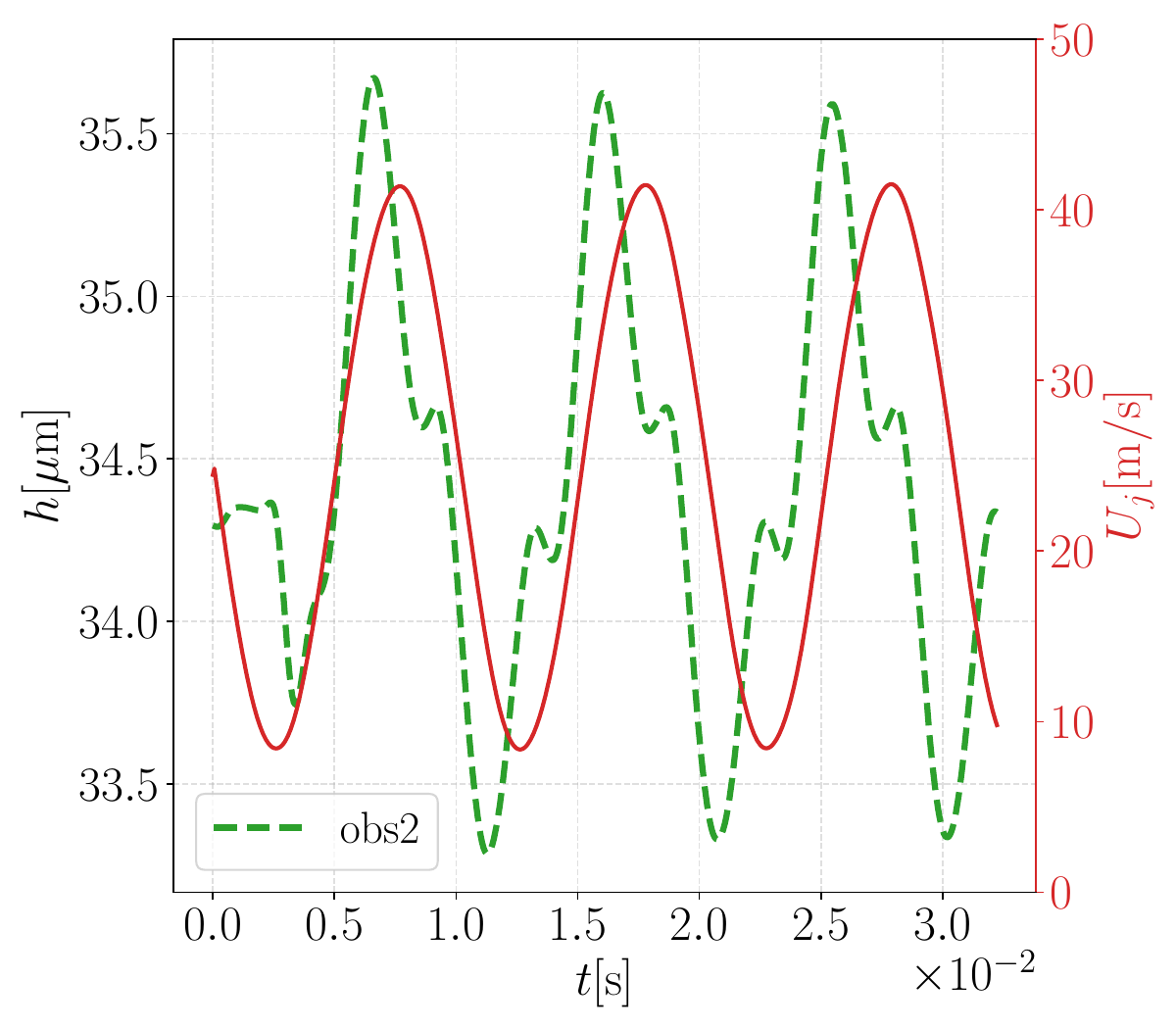}
    \caption{}
\label{fig:results_harmonic_2D_jet_3D_undulation_control_spectral_b}
  \end{subfigure}
  \caption{Evolution of the amplitude ($\acute{A}$), frequency ($\acute{f}$) and phase shift ($\acute{\phi}$) (a), the obs2 and the control function $U_j$ (b) during an evaluation episode for the harmonic 2D gas jet control.}
\label{fig:results_harmonic_2D_jet_3D_undulation_control_spectral}
\end{figure}

Focusing on the harmonic case, the evolution of the optimal control function during an evaluation episode is reported in Figure~\ref{fig:results_harmonic_2D_jet_3D_undulation_control_spectral}. Figure~\ref{fig:results_harmonic_2D_jet_3D_undulation_control_spectral_a} shows the amplitude (solid red line with squares), the nondimensional frequency (blue dashed line with triangles), and the phase shift (dotted dashed line with circles) of the harmonic control function during an evaluation episode. These three parameters remain nearly constant, with an amplitude of $\acute{A} \approx 0.68$, a frequency of $\acute{f} \approx 0.065$, and a phase shift of $\acute{\phi} \approx \pi$. The frequency aligns with the average frequency of the undulation waves, which ranges between $0.05$ and $0.08$. This shows how a single harmonic function is sufficient to significantly reduce the amplitude of the undulation instability.

\begin{figure}
  \begin{subfigure}[b]{\linewidth}
    \includegraphics[width=0.9\linewidth]{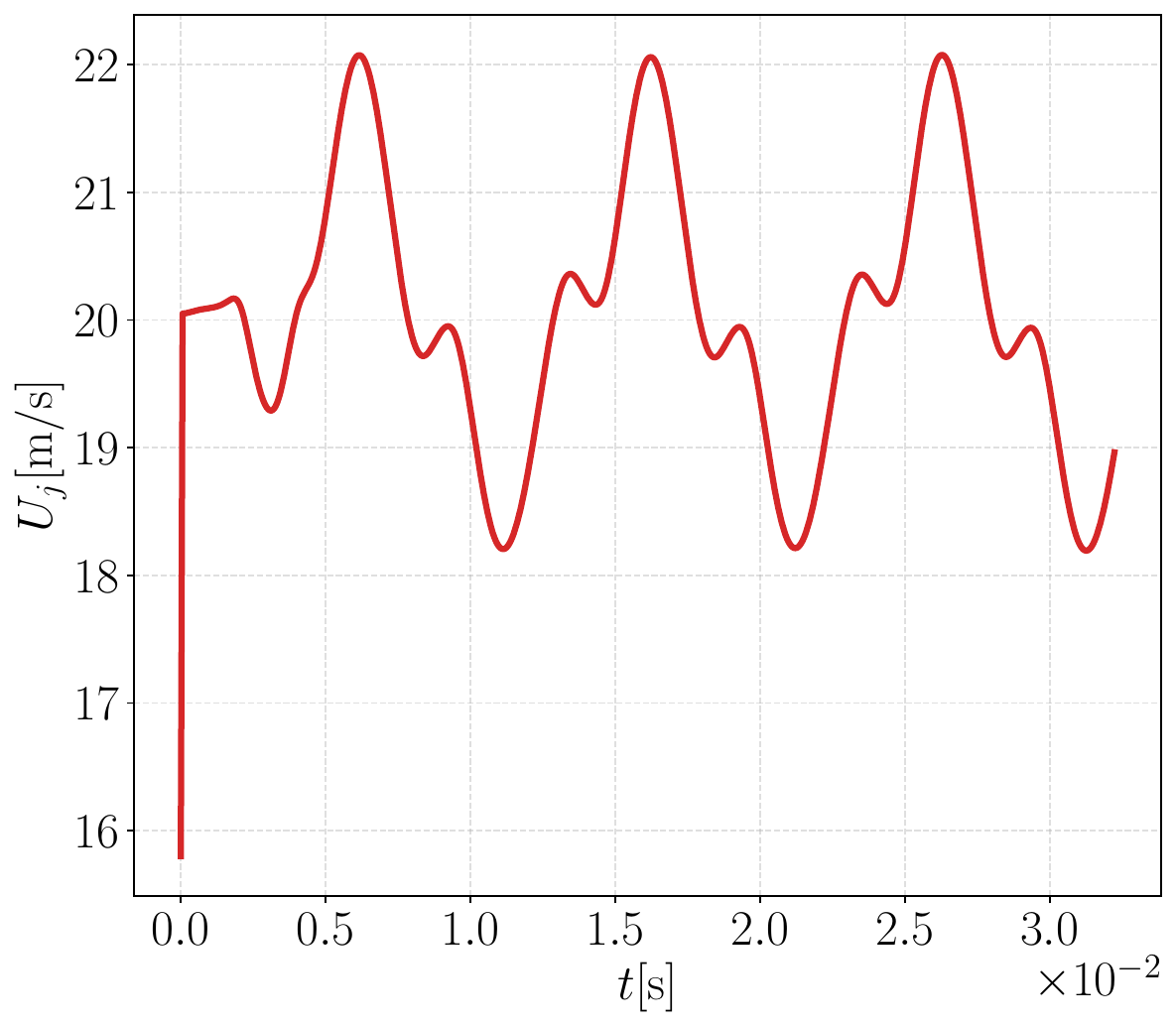}
    \caption{}
    \label{}
  \end{subfigure}
  \hfill
  \begin{subfigure}[b]{\linewidth}
    \includegraphics[width=0.9\linewidth]{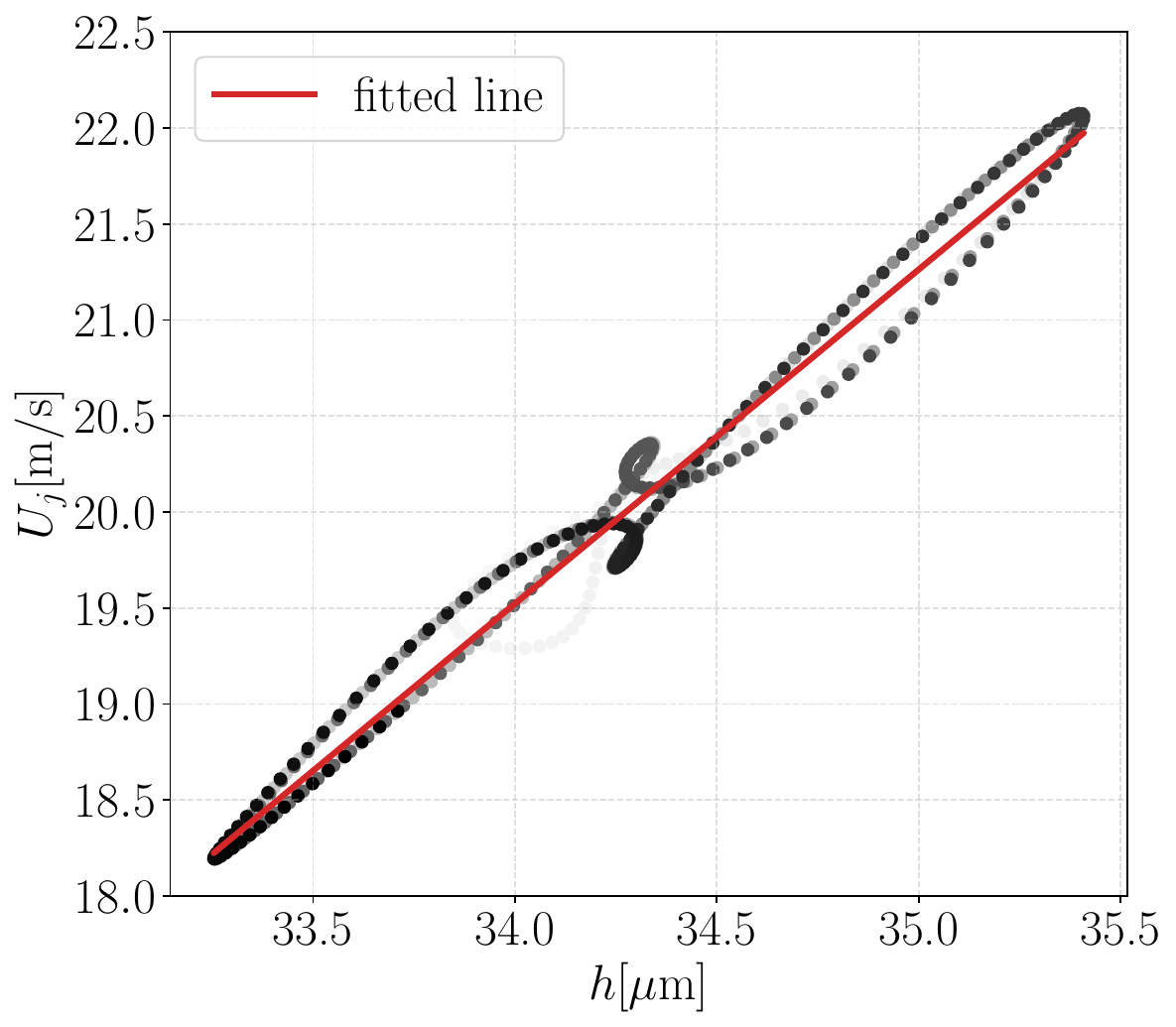}
    \caption{}
    \label{fig:results_harmonic_2D_jet_3D_undulation_control_spectral_f}
  \end{subfigure}
  \caption{Evolution of the optimal control action as a function of (a) time and (b) obs1 with a single gas-jet.}
  \label{fig:obs_actions_2D_non_harm_jet_zinc}
\end{figure}
To gain a better understanding of the control mechanism, we examine the evolution of the control function alongside observations of the liquid film collected at one of the three observation locations. Figure~\ref{fig:results_harmonic_2D_jet_3D_undulation_control_spectral_b} shows the evolution of the nozzle exit velocity (solid red line) and the observation at obs2 (dashed-green line). The positions of the crests and valleys of the two functions are very similar, with only a slight phase shift due to the observation point being located ahead of the control actuator. This shows that the jet primarily flattens the crests with only a minor effect on the valleys. However, because the nozzle exit velocity is always nonzero, this indicates that the optimal control law involves thinning the liquid film and a harmonic component nearly in phase with the instability wave.
\begin{figure*}
  \begin{subfigure}[b]{0.325\textwidth}
    \includegraphics[width=\textwidth]{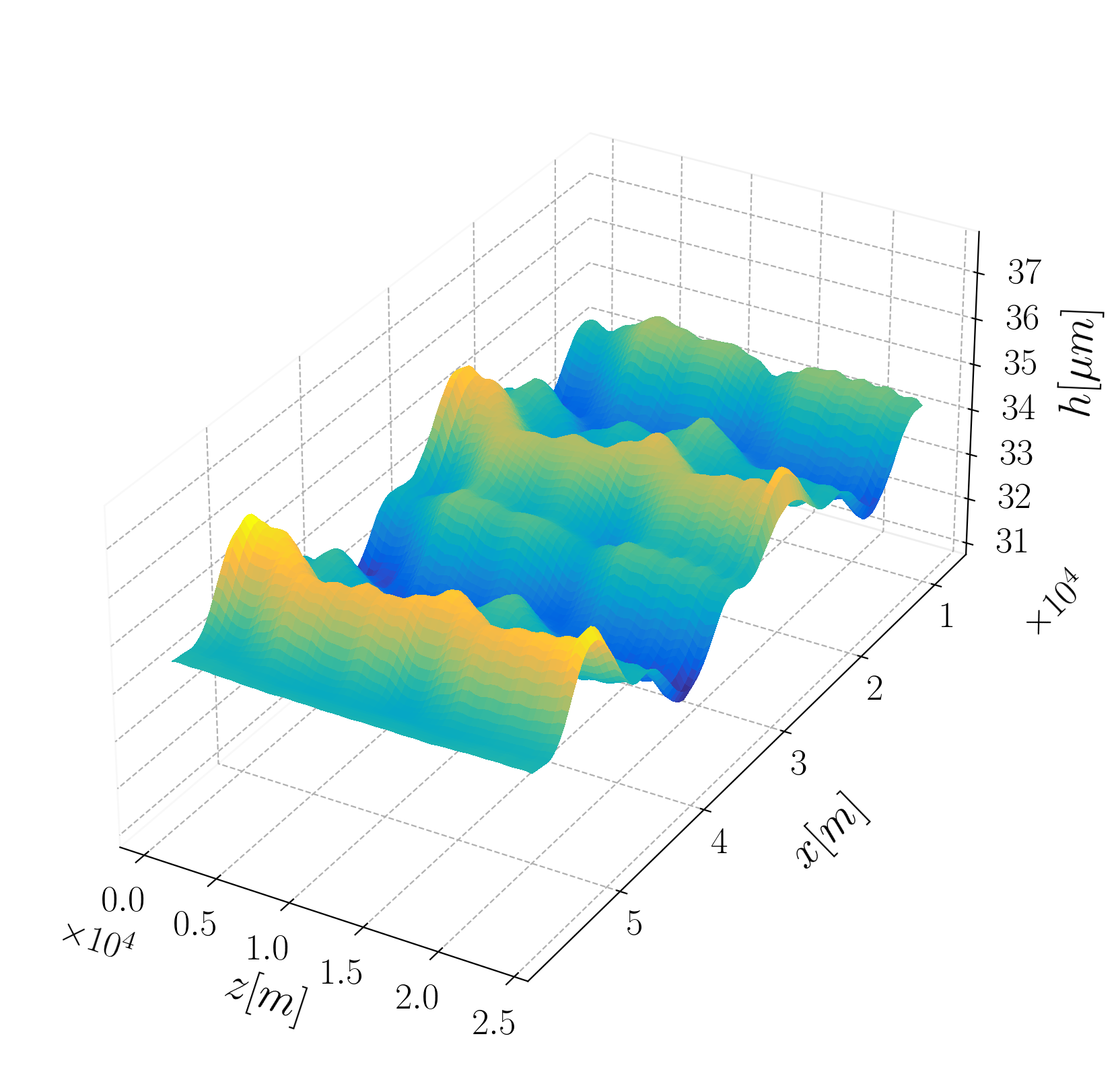}
    \caption{}
  \end{subfigure}
  \hfill
  \begin{subfigure}[b]{0.325\textwidth}
    \includegraphics[width=\textwidth]{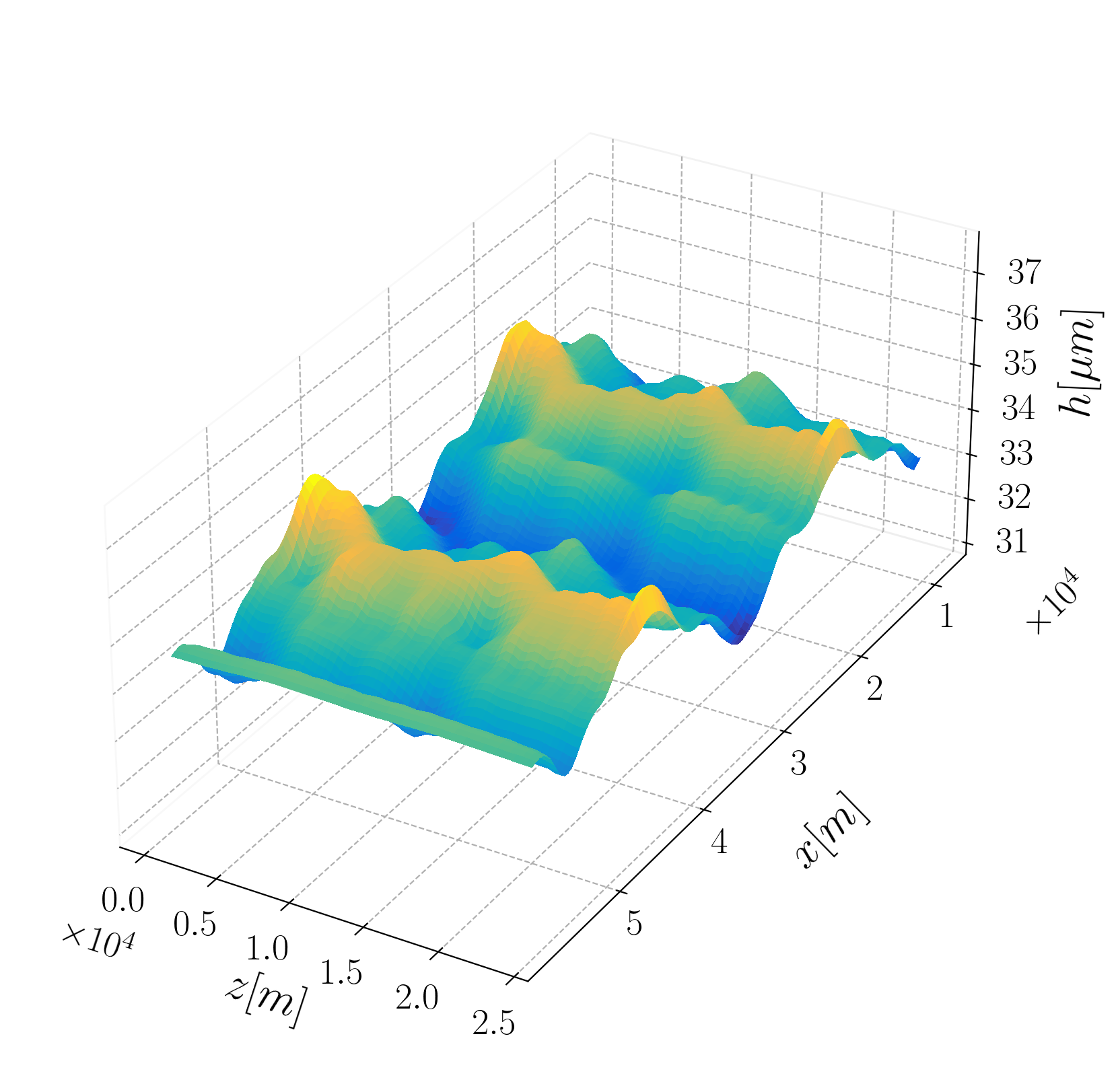}
    \caption{}
  \end{subfigure}
  \hfill
  \begin{subfigure}[b]{0.325\textwidth}
    \includegraphics[width=\textwidth]{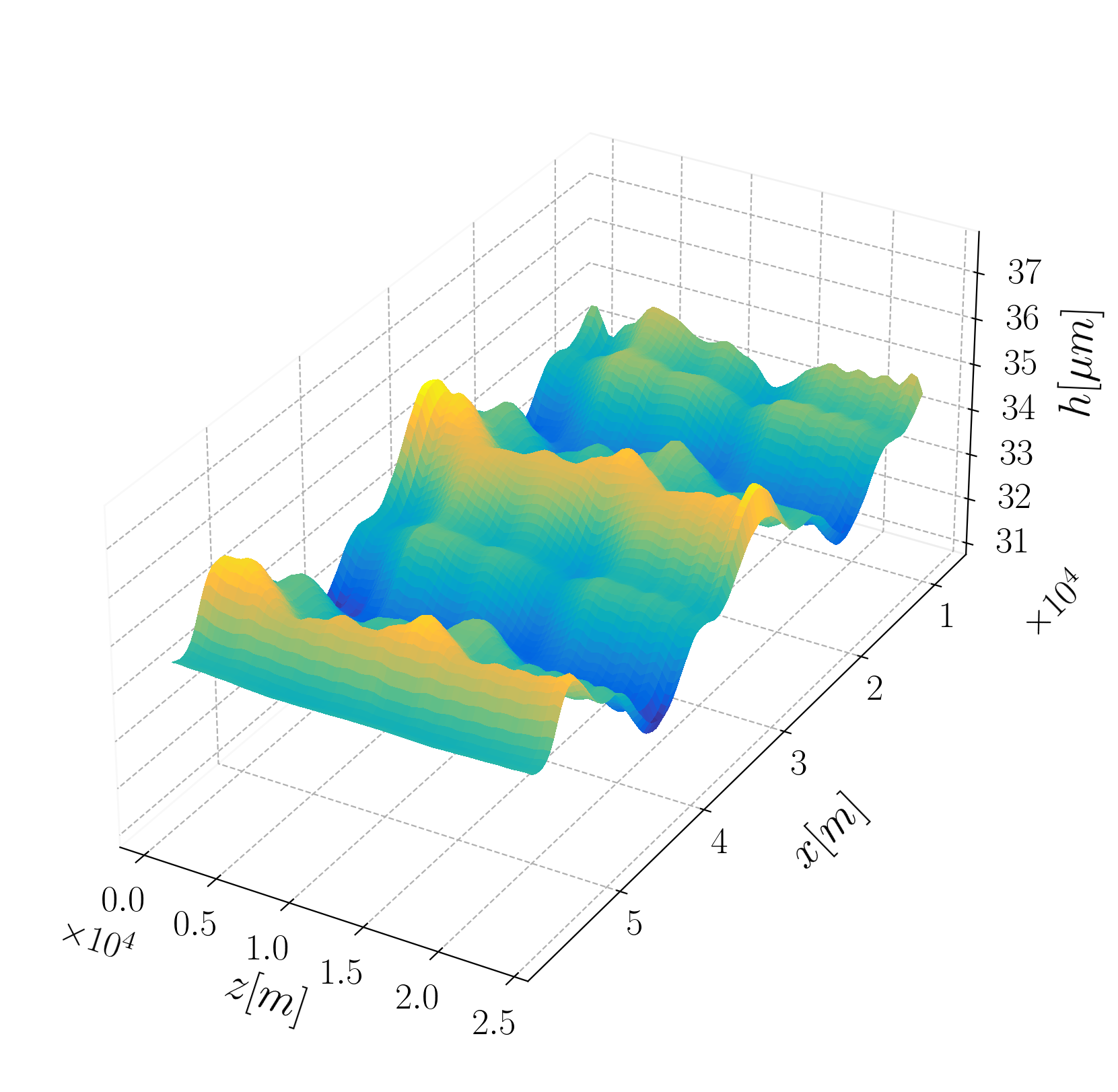}
    \caption{}
  \end{subfigure}
  \begin{subfigure}[b]{0.325\textwidth}
    \includegraphics[width=\textwidth]{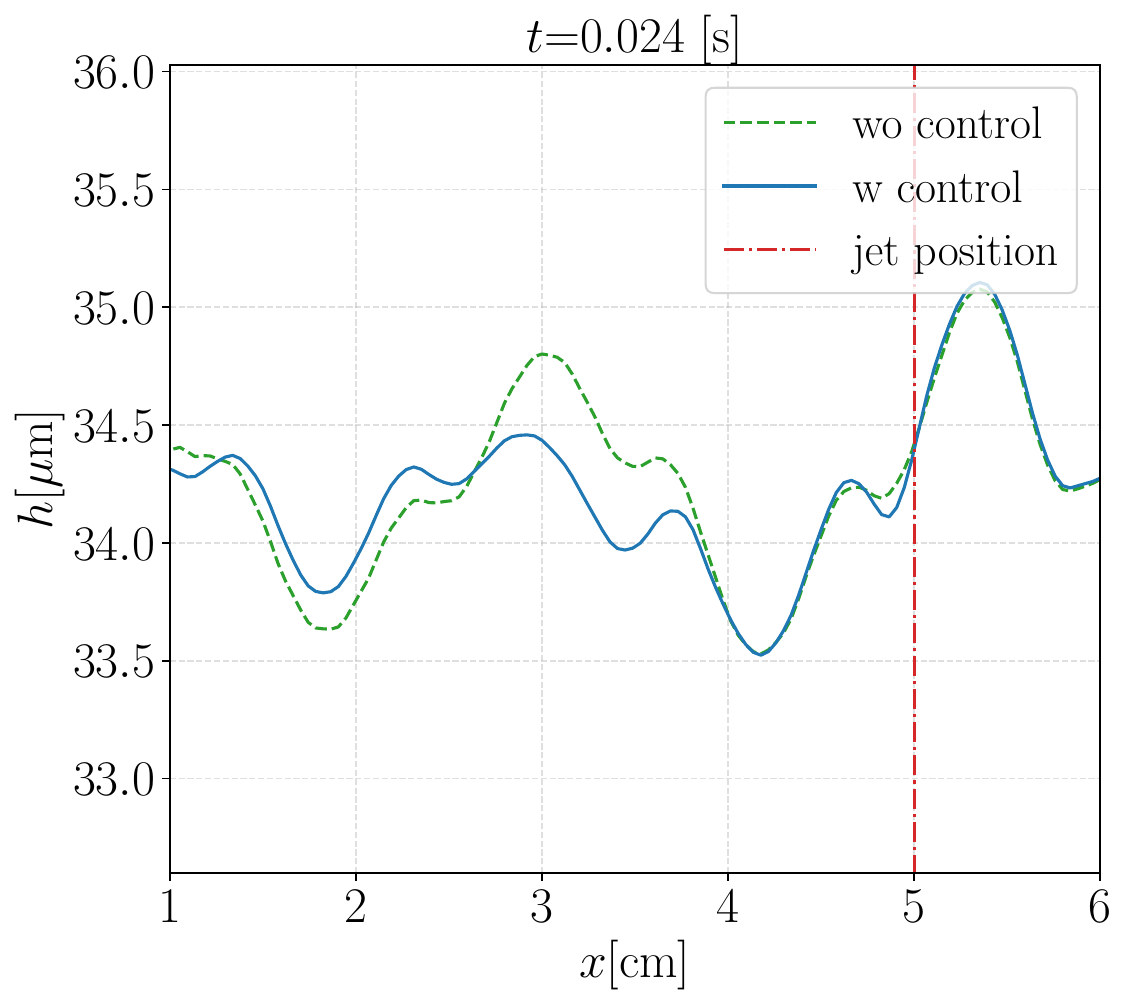}
    \caption{}
  \end{subfigure}
  \hfill
  \begin{subfigure}[b]{0.325\textwidth}
    \includegraphics[width=\textwidth]{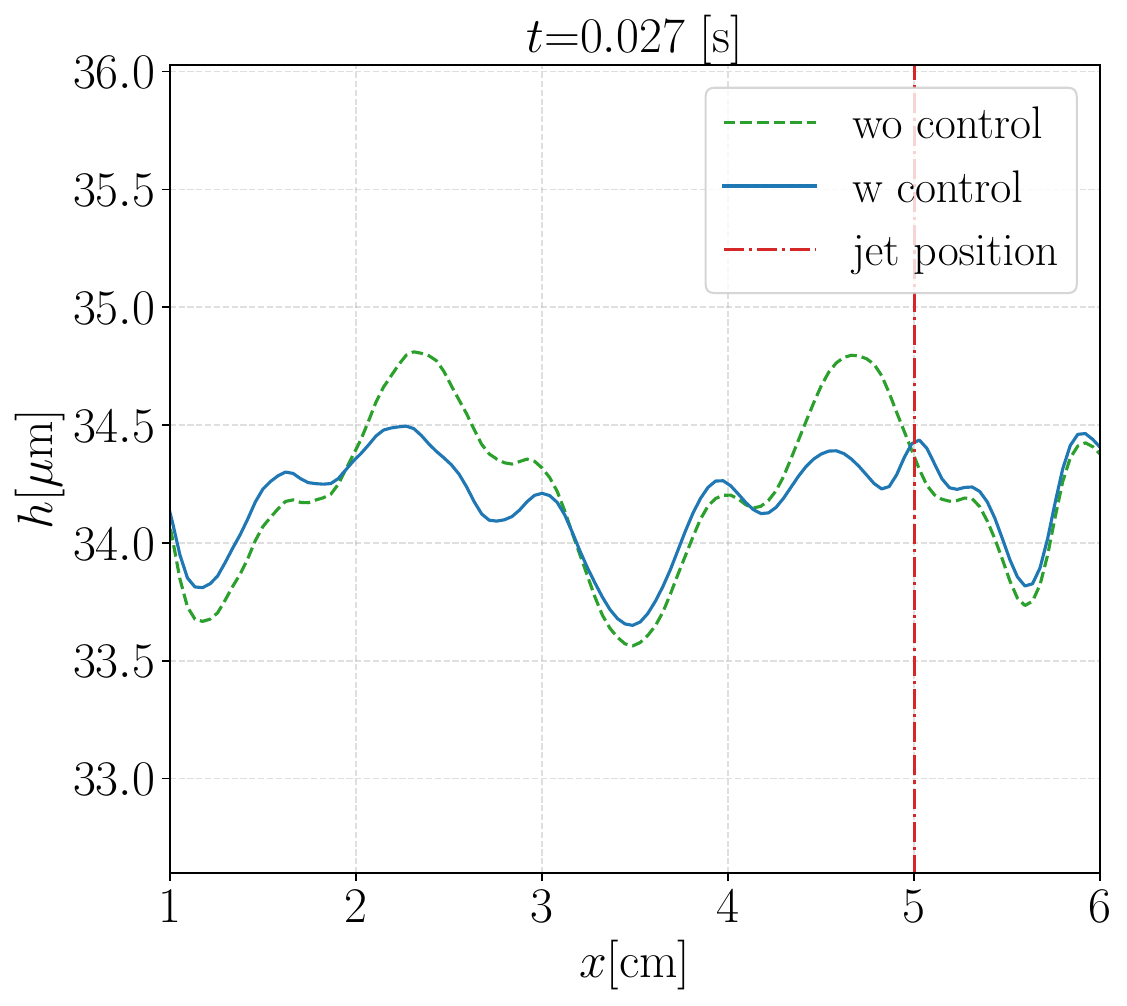}
    \caption{}
  \end{subfigure}
  \hfill
  \begin{subfigure}[b]{0.325\textwidth}
    \includegraphics[width=\textwidth]{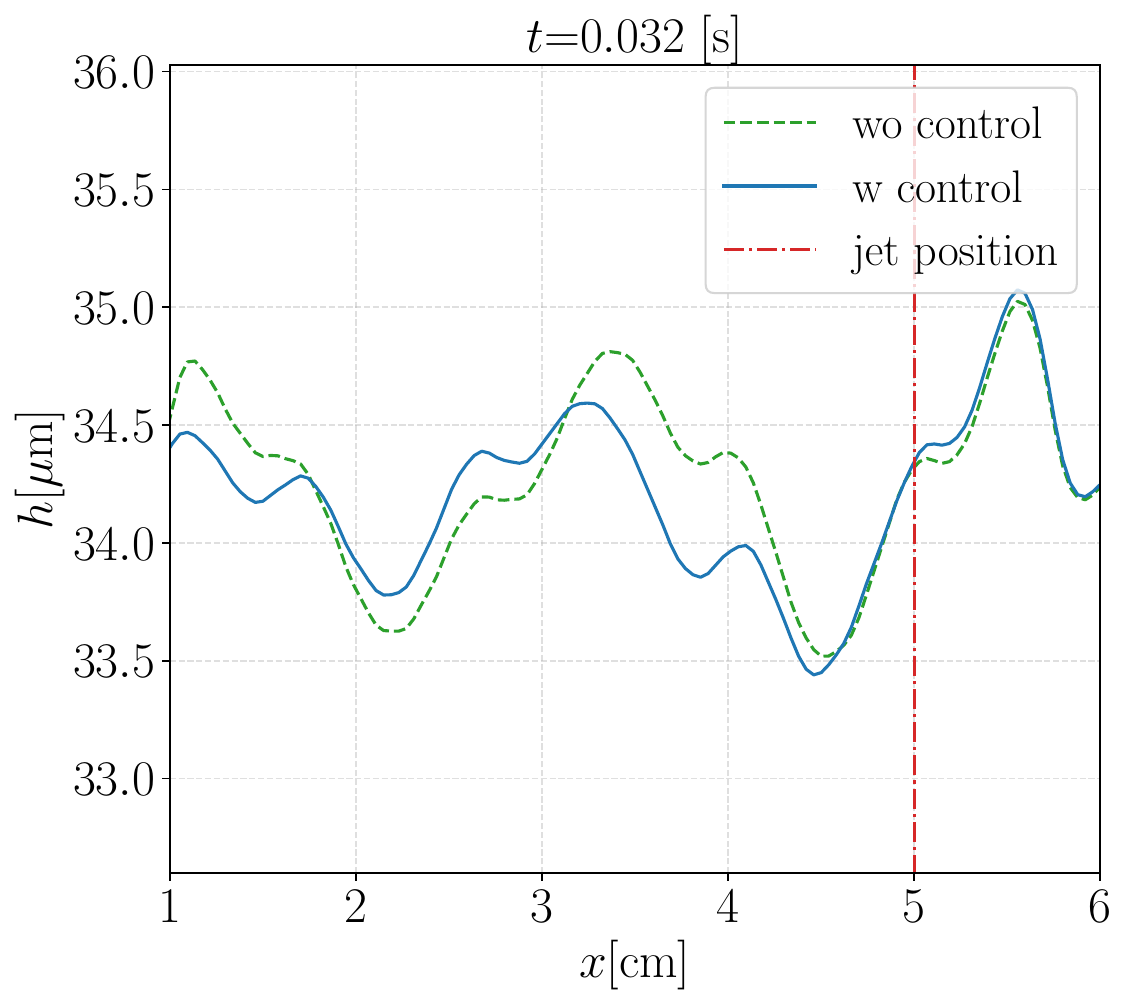}
    \caption{}
  \end{subfigure}
  \caption{Evolution of (a, b and c) the controlled 3D liquid film at different time steps and (d,e and f) the corresponding film profiles along the streamwise direction $x$ at $z = 1\,\text{cm}$ for the controlled (solid blue line) and uncontrolled (dashed green line) cases using a single gas jet actuator.}
\label{fig:evolution_3D_liquid_film_non_harm_2D_gas_jet_zinc}
\end{figure*} 

Moving to the case where the PPO is free to take any actions, Figure~\ref{fig:obs_actions_2D_non_harm_jet_zinc} shows the evolution of (a) the control actions and the relation between the control actions and (b) the obs1 with the fitted line (red curve) for the non-harmonic control function. The control action is proportional to the mean liquid-film thickness at the impingement point, with a small phase lag.

Looking at the plot along a line in the streamwise direction, Figure~\ref{fig:evolution_3D_liquid_film_non_harm_2D_gas_jet_zinc} shows the evolution of the controlled three-dimensional liquid film (a, b, and c) and the comparison between the uncontrolled (green dashed line) and the controlled (continuous blue line) along $x$ at $z=1\rm{cm}$ (d, e, and f). In the harmonic case, the controller reduces the amplitude of the crests with small effects on the valleys.

\subsubsection{Control with single electromagnets}
\begin{figure*}
\centering
    \includegraphics[width=0.58\linewidth]{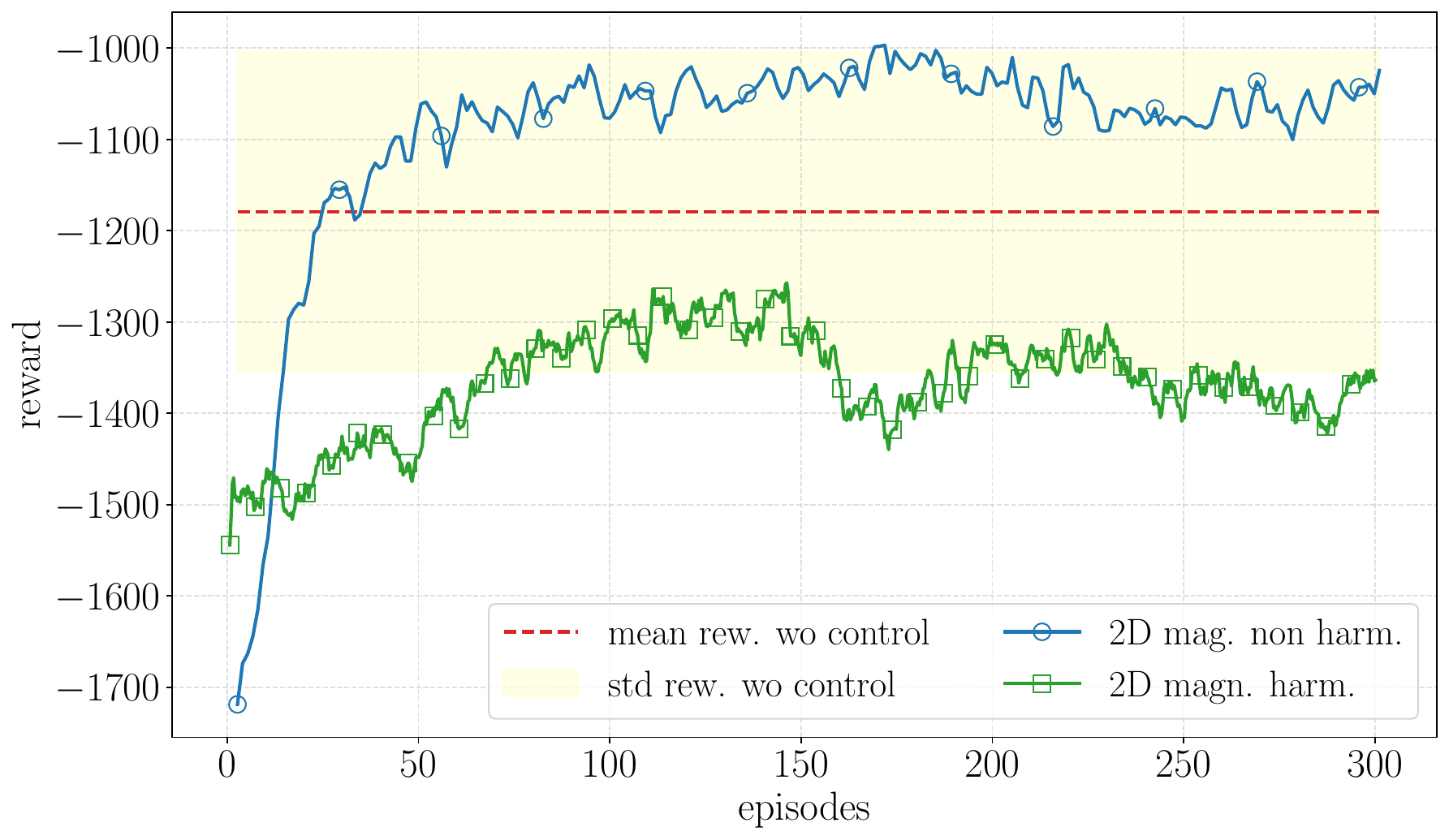}  \caption{Learning curve for single electromagnet with harmonic (green curve with squares) and non-harmonic (blue line with circles)control against mean (red dashed line) and standard deviation (yellow area) without control.}
  \label{fig:lc_magnetic_2D_alone}
\end{figure*}

Turning to the control with a single electromagnetic actuator, Figure~\ref{fig:lc_magnetic_2D_alone} shows the learning curve (continuous blue line with circles and continuous green line with squares), the mean value (red dashed line) and the standard deviation (yellow shaded area) of the reward in the absence of control. The case with a prescribed harmonic control function fails to produce good results, with the average reward well below that of the uncontrolled case, even falling outside the uncertainty region of the no-control case. This suggests that the agent was unable to find an efficient control law and therefore continued to test random action combinations, resulting in increased undulation amplitude in the reward area. In contrast, the agent discovered an optimal non-harmonic control law without a prescribed control function shape at the 40th learning episode. This episode marks the point at which the mean reward exceeded that of the uncontrolled case. Furthermore, the mean value approaches the upper bound of the uncertainty region in the uncontrolled case, underscoring the algorithm's robustness.

A similar behaviour is observed when examining performance during the evaluation episodes. Table~\ref{tab:mean_std_controlled_uncontrolled_2D_magnetic_zinc} shows the mean and standard deviation of the liquid film with and without control, using both harmonic and non-harmonic control functions with the 2D electromagnetic actuator. The harmonic control function yielded a mean reward 14\% lower than in the uncontrolled case, with a larger standard deviation. In contrast, the non-harmonic control function yielded a reward 8\% higher than the uncontrolled case, with a smaller standard deviation.
\begin{table}
\centering
\caption{Reward means and standard deviations in the evaluation episode using a 2D electromagnet.}
\label{tab:mean_std_controlled_uncontrolled_2D_magnetic_zinc}
\begin{tabular}{c@{\hspace{0.4cm}}c@{\hspace{0.3cm}}c@{\hspace{0.3cm}}c@{\hspace{0.3cm}}}
\toprule
 \textbf{Reward} & \textbf{wo Control} & \textbf{Harmonic} & \textbf{Non-harmonic} \\ 
\midrule
\begin{tabular}[c]{@{}c@{}}Mean \end{tabular}& 
\textbf{-1179} & -1345 & -1075 \\ \bottomrule
\begin{tabular}[c]{@{}c@{}}Standard \\ Deviation\end{tabular}& 
\textbf{175} & 189 & 165 \\ \bottomrule
\end{tabular}
\end{table} 

\begin{figure}
  \begin{subfigure}[b]{\linewidth}
    \includegraphics[width=\linewidth]{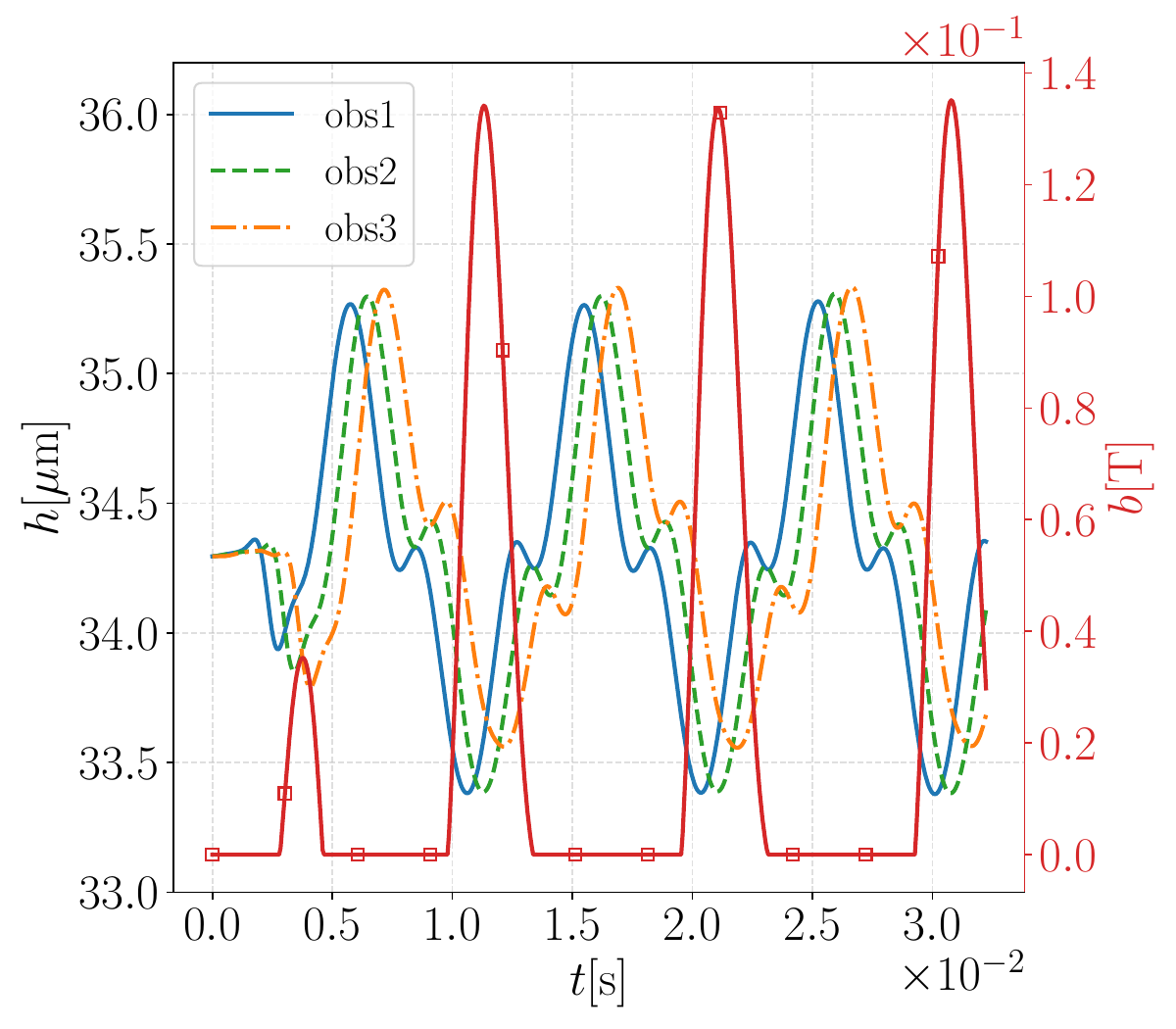}
    \caption{}
  \end{subfigure}
  \hfill
  \begin{subfigure}[b]{\linewidth}
    \includegraphics[width=\linewidth]{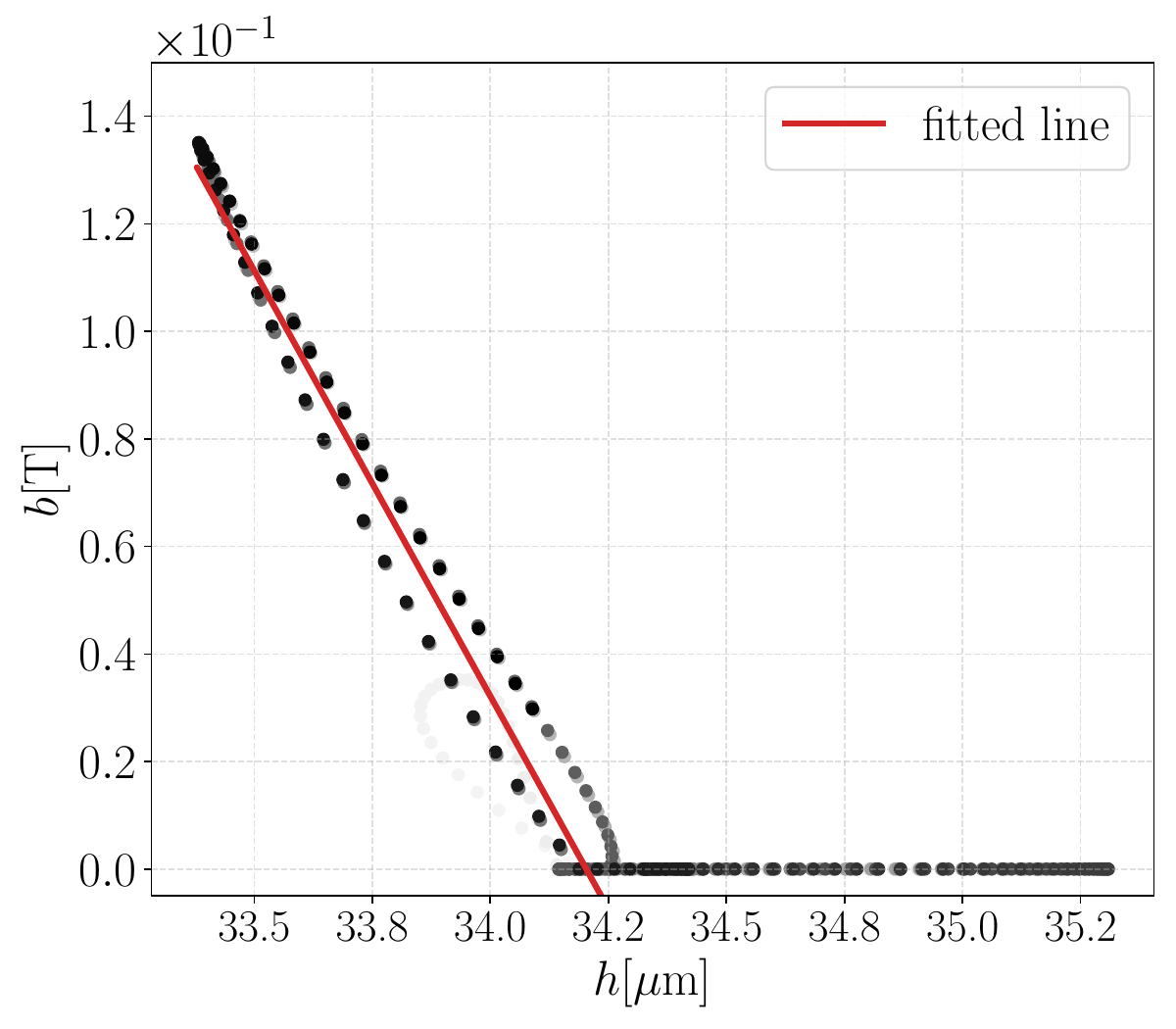}
    \caption{}
  \end{subfigure}
  \caption{Evolution of the 3D undulation control using 2D electromagnets: (a) observations (coloured curves without markers) and magnetic field intensity (red solid curve with square markers); (b) relationship between the magnetic field intensity and the second observation (black dots, with shading indicating simulation time from light at the beginning to dark at the end), together with the fitted curve (red solid line).}
  \label{fig:action_obs_magnetic_2D}
\end{figure}

The control strategy obtained with the electromagnets differs significantly from that obtained with a single 2D gas jet. Figure~\ref{fig:action_obs_magnetic_2D} shows (a) the evolution of the control function (solid red line with squares) and the observations (coloured lines without markers) and (b) the relationship between the control function and obs1 during an evaluation episode. The electromagnetic control exhibits a bang-bang-type behaviour, acting only near the valleys with a maximum intensity of $0.14 \, \rm{T}$. The actions are inversely proportional to obs1 when they exceed the flat state of the target, $\bar{h} = 0.1 \, (34.2 \, \mu\rm{m})$.

\begin{figure*}
  \begin{subfigure}[b]{0.325\textwidth}
    \includegraphics[width=\textwidth]{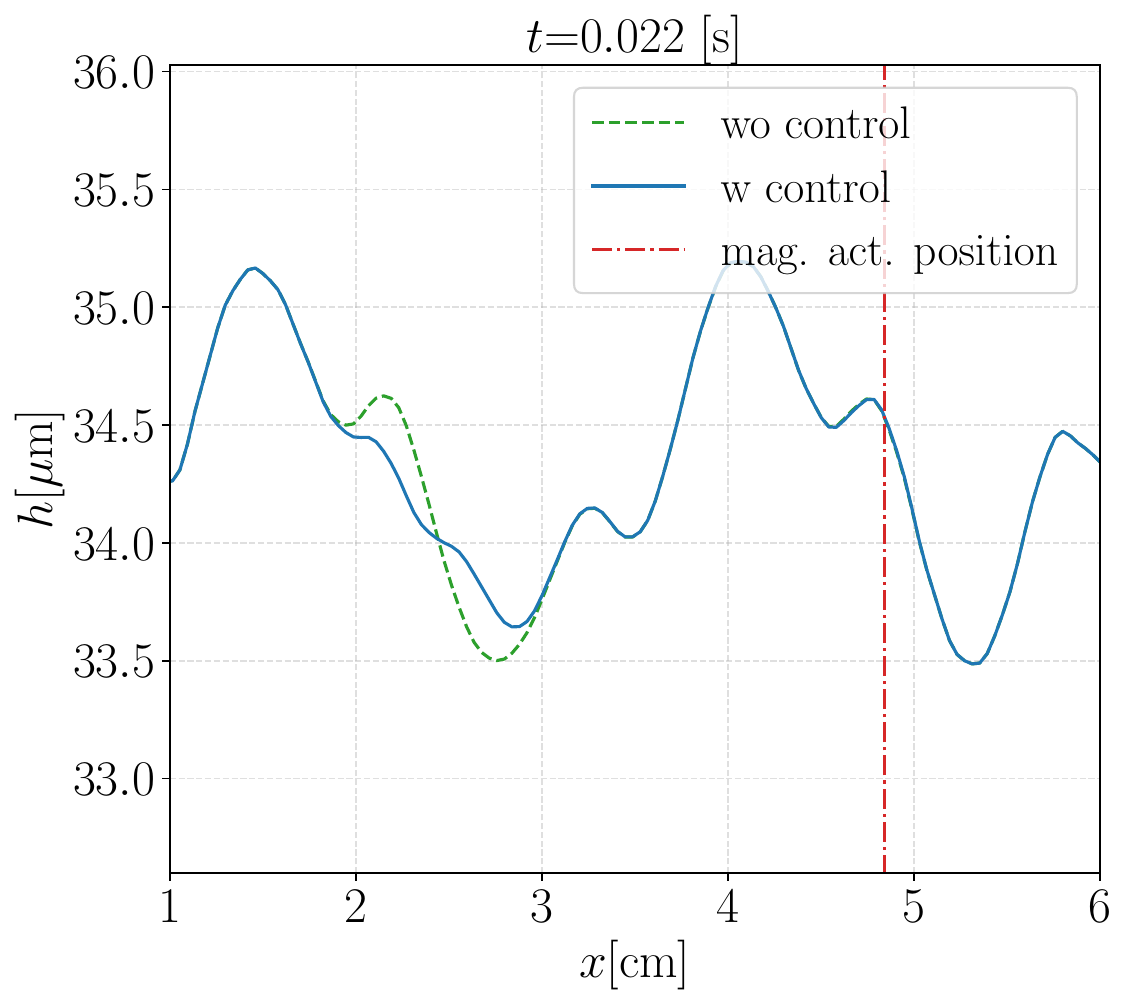}
    \caption{}
  \end{subfigure}
  \hfill
  \begin{subfigure}[b]{0.325\textwidth}
    \includegraphics[width=\textwidth]{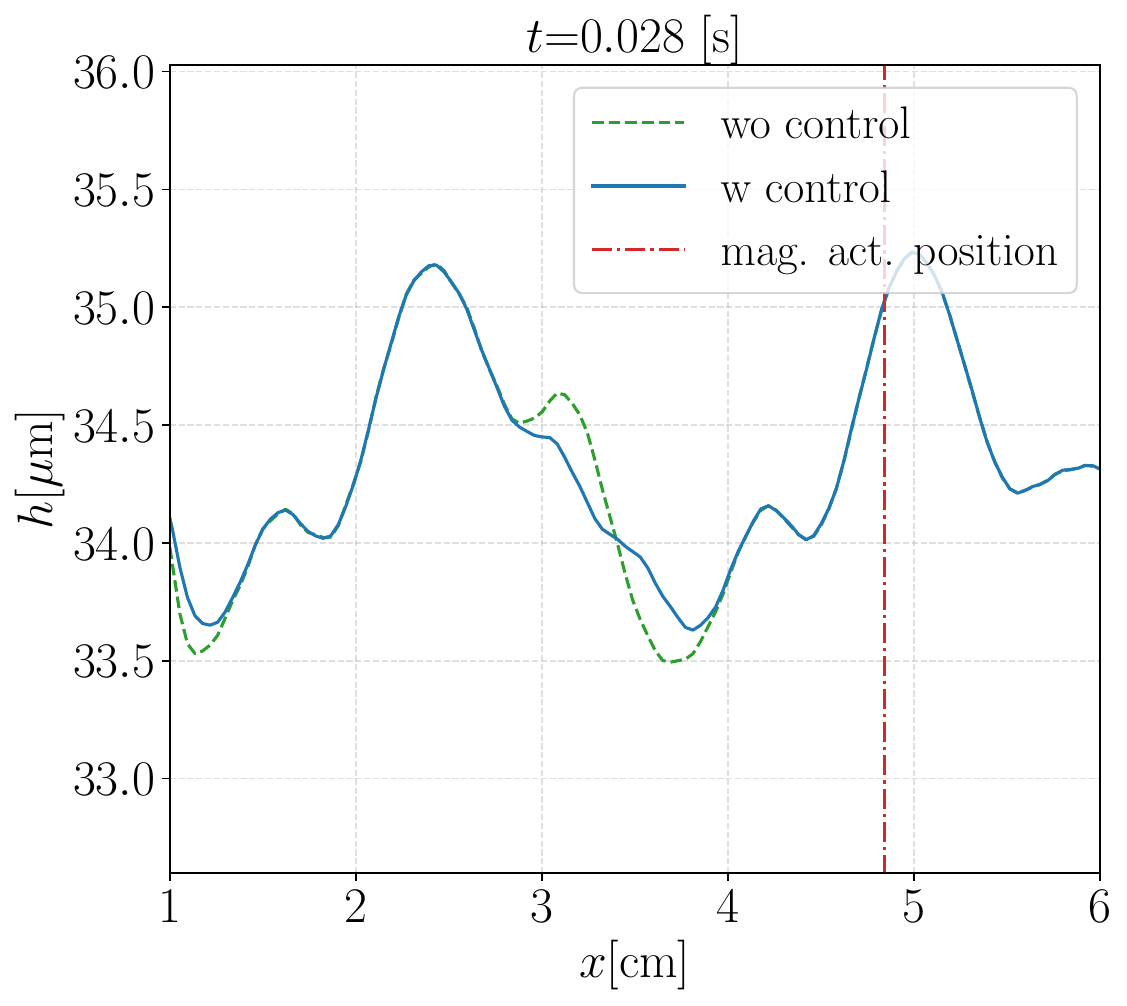}
    \caption{}
  \end{subfigure}
  \begin{subfigure}[b]{0.325\textwidth}
    \includegraphics[width=\textwidth]{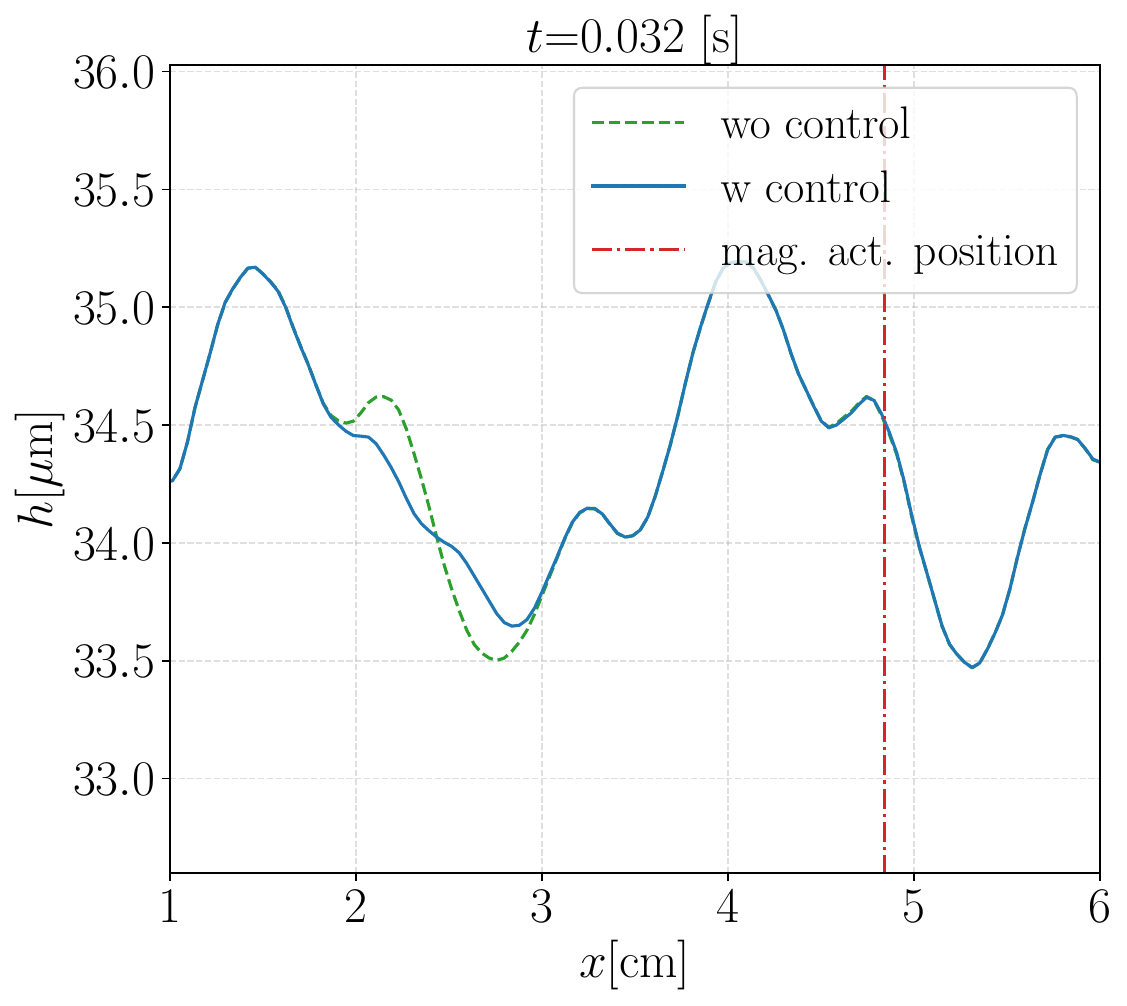}
    \caption{}
  \end{subfigure}
  \caption{Evolution of the controlled (blue continuous line) and the uncontrolled (green dashed line) along $x$ at $z=1 \rm{cm}$ for the control of the 3D undulation with a 2D electromagnetic actuator.}
  \label{fig:plot_line_magnetic_2D_spectral}
\end{figure*}

The optimal control law mechanism arises from the resistive effects of the Lorentz force on the liquid film. The Lorenz forces push the liquid in the positive $x$ direction, creating a positive/negative gradient in flow rate across the electromagnet's centerline. This results in an increase/decrease in the liquid film thickness upstream and a decrease downstream. This behaviour is evident in the comparison of wave evolution between the controlled and uncontrolled cases. Figure~\ref{fig:plot_line_magnetic_2D_spectral} shows the $ x$-profile of the liquid film with (solid blue line) and without (dashed green line) control. The action of the electromagnet raises and deforms the valleys, which reduces the amplitude of the nearby peaks.

This control mechanism is constrained by the mismatch between the characteristic time scale of the film, $t_{\mathrm{ref}}$, and that of the electromagnetic effects, $t_{\mathrm{ref,M}}$, quantified by a small Stuart number ($\N = 1.9 \times 10^{-2}$). 
For the magnetic field to exert a significant influence, its characteristic time scale should be comparable to that of the film, i.e., $\N = O(1)$. However, given the Reynolds and Capillary numbers considered in this study, this condition would require a very large Hartmann number, expressed as:
\begin{equation}
    \Ha = \sqrt{\R}\,\Ca^{1/6} = 16,
\end{equation}
which presents a major challenge for industrial implementation due to the high cost, strong thermal gradients induced by the Joule effect, and the potential destabilising impact on the strip.

\subsubsection{Control with gas jets and electromagnets}
In the previous test cases, we found that optimal control actions for the gas jet and electromagnetic actuators are complementary. The gas jet pushes the crests, whereas the electromagnets raise the valleys. In this case, we test the control of undulations with both actuators operating in tandem. Additionally, to simplify the control problem, the agent has access to only three observations of liquid film thickness, rather than six.
\begin{figure*}
\centering
    \includegraphics[width=0.58\linewidth]{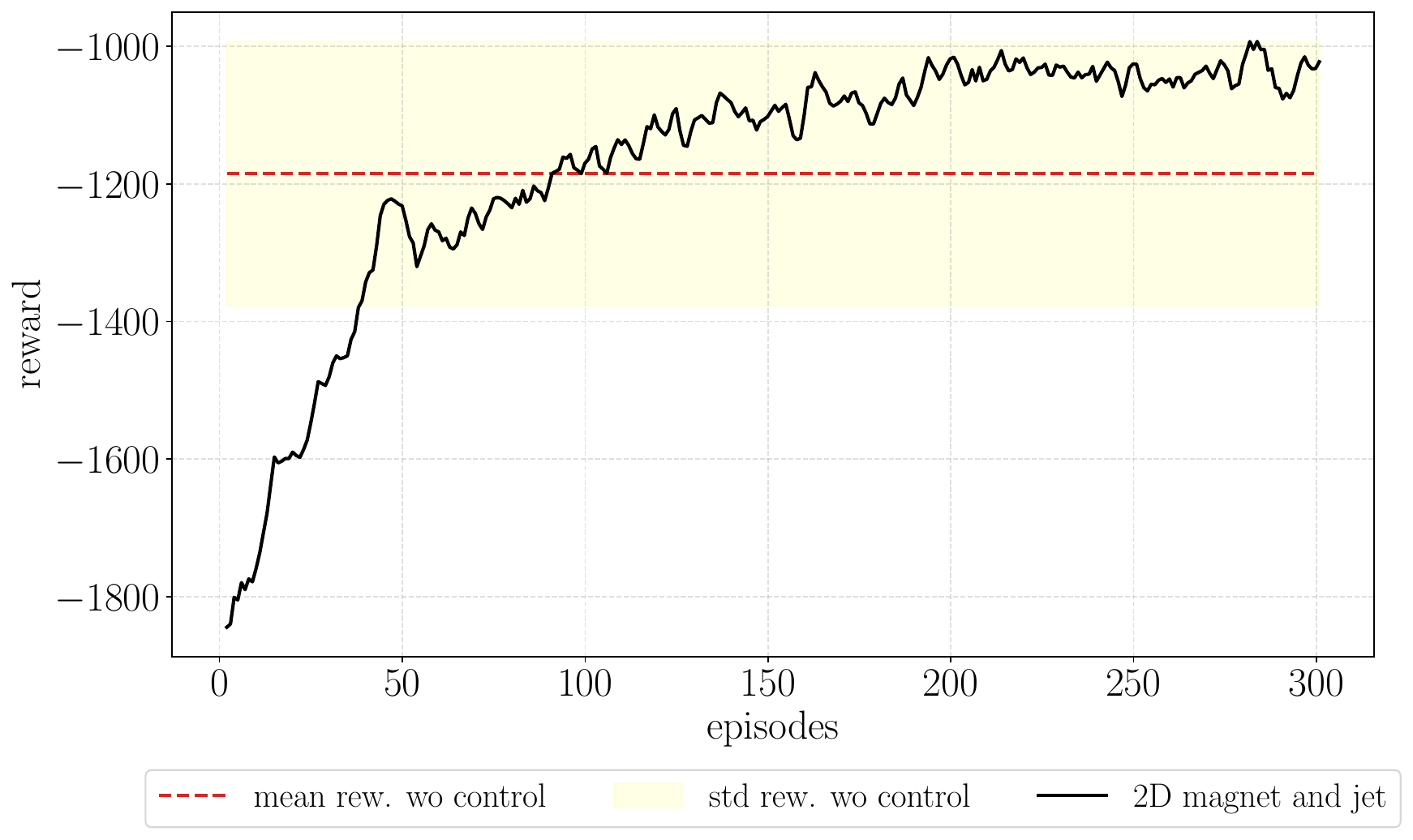}
  \caption{Learning curve for the control with 2D gas jet and electromagnetic actuators.}
  \label{fig:lc_hybrid}
\end{figure*}

Figure~\ref{fig:lc_hybrid} shows the learning curve along with the mean (red dashed line) and standard deviation (yellow shaded area) of the reward obtained without control. The PPO algorithm finds an optimal control law that outperforms the uncontrolled case after 100 training episodes. Furthermore, as observed in the single electromagnet case, the learning curve approaches the upper bound of the uncertainty region, demonstrating the robustness of the discovered control action. Similar results are evident in terms of evaluation episodes. 
\begin{table}
\centering
\caption{Mean and standard deviation in the evaluation episodes with and without control function using 2D gas jet and electromagnetic actuators.}
\label{tab:res_hybrid}
\begin{tabular}{c@{\hspace{0.6cm}}c@{\hspace{0.5cm}}c@{\hspace{0.5cm}}}
\toprule
 \textbf{Reward} & \textbf{wo Control} & \textbf{w Control} \\ 
\midrule
\begin{tabular}[c]{@{}c@{}}Mean \end{tabular}& 
\textbf{-1185} & -1036 \\ \bottomrule
\begin{tabular}[c]{@{}c@{}}Standard \\ Deviation\end{tabular}& 
\textbf{192} & 155  \\ \bottomrule
\end{tabular}
\end{table} 

Table~\ref{tab:res_hybrid} shows the mean and standard deviation of the reward obtained during the evaluation episodes, both with and without control. The optimal control law yields a mean reward 13\% higher than in the uncontrolled case, and a 20\% reduction in the standard deviation. This indicates that, even in the worst-case evaluation test, the agent outperforms the uncontrolled case on average.

\begin{figure*}[t]
  \centering
  \begin{subfigure}[b]{0.43\linewidth}
    \includegraphics[width=\linewidth]{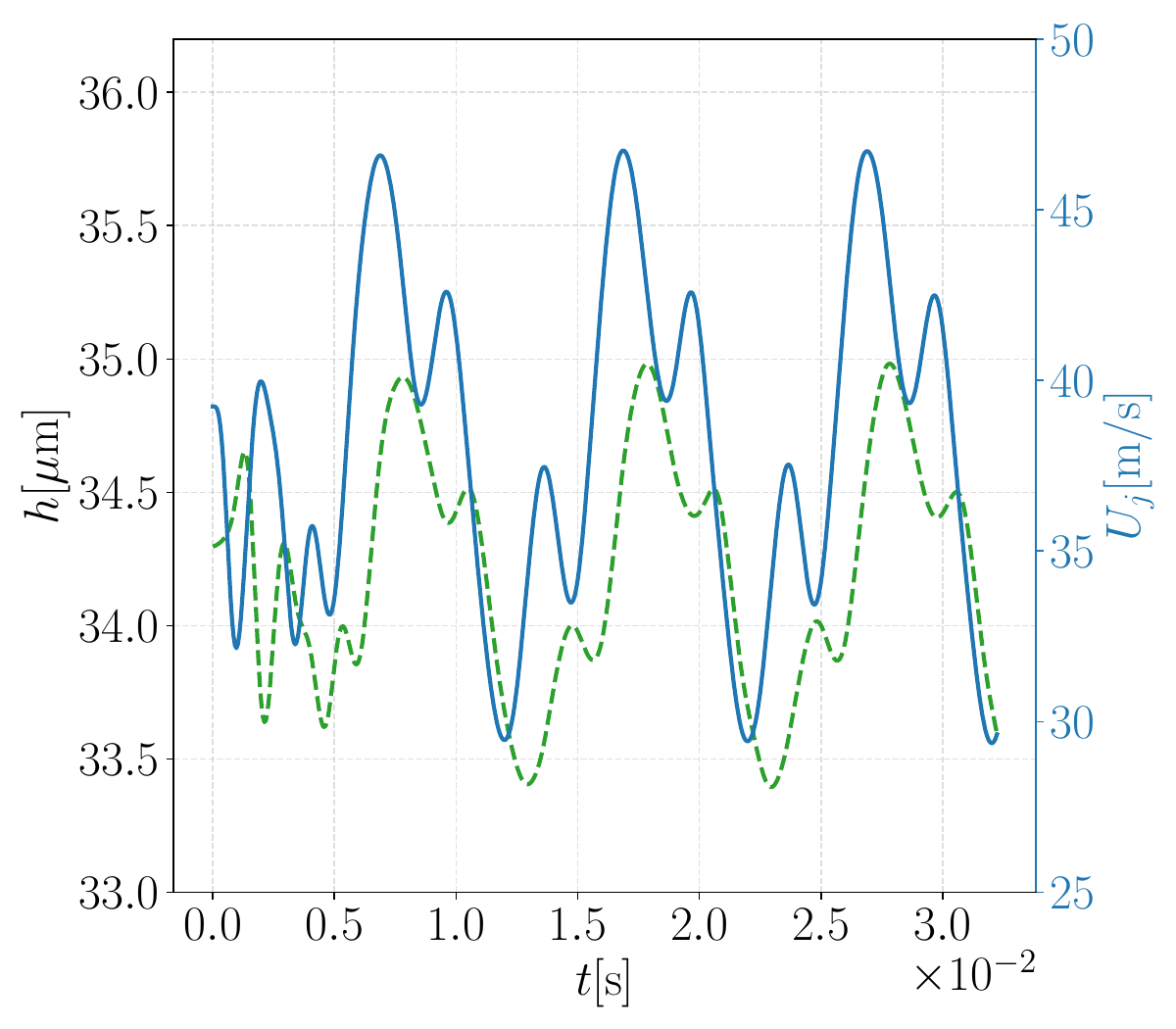}
    \caption{}
  \end{subfigure}
  \hspace{0.05\linewidth} 
  \begin{subfigure}[b]{0.43\linewidth}
    \includegraphics[width=\linewidth]{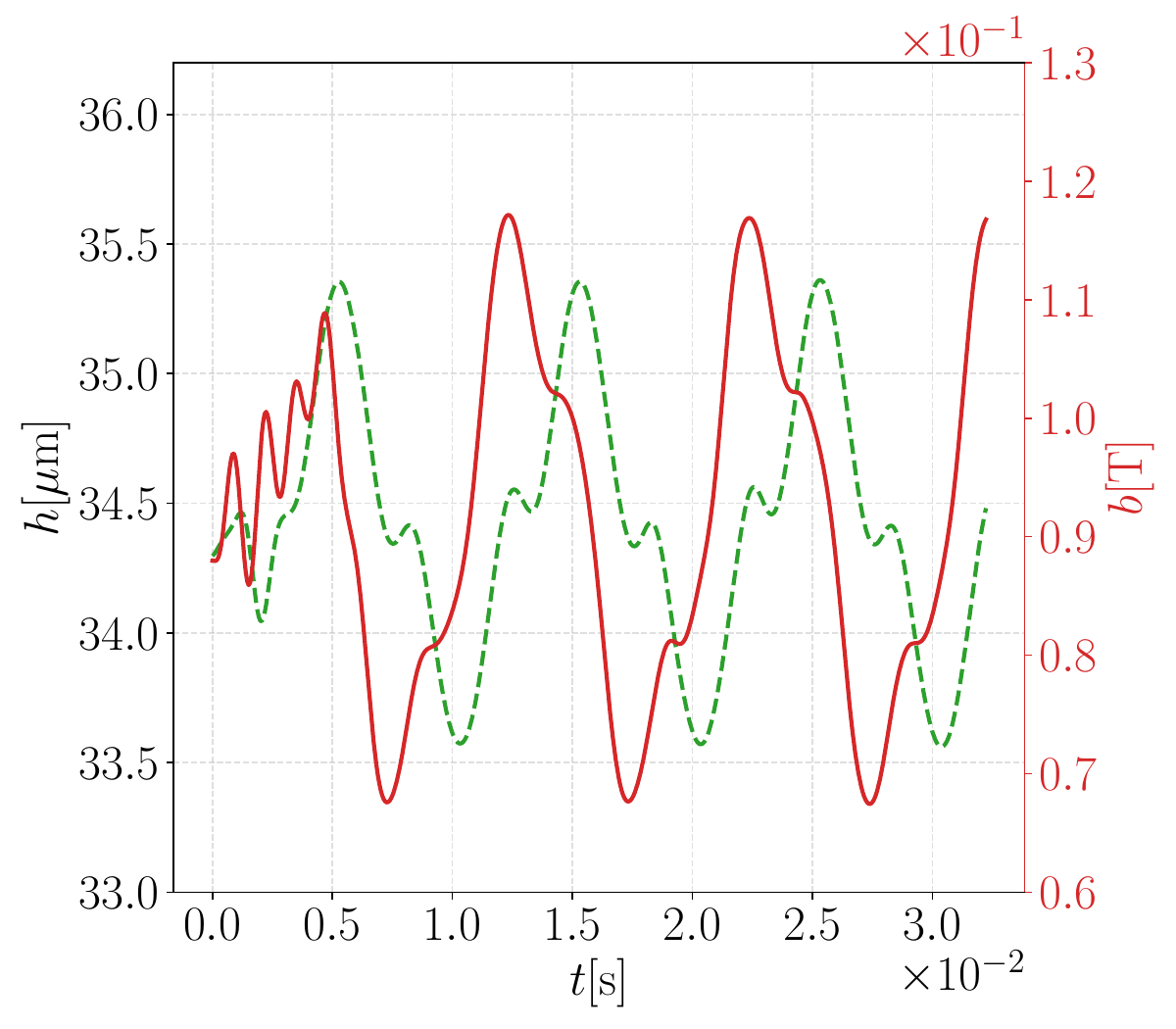}
    \caption{}
  \end{subfigure}
  \caption{Evolution of (a) the 2D jet actions (blue continuous line) and obs3 (green dashed line) and (b) the 2D electromagnet (continuous red line) and obs2 (green dashed line).}
  \label{fig:actions_actuators}
\end{figure*}

Moving to the evolution of optimal control functions during an evaluation episode, Figure~\ref{fig:actions_actuators} shows (a) the actions of the 2D gas jet (blue solid line) and obs3 (green dashed line) and (b) the actions of the 2D electromagnet (red solid line) and obs2 (green dashed line). As in the case of a single gas jet, the optimal control law is nearly proportional to the liquid film observation, with a small phase shift due to the distance between the observation point and the actuator. In contrast, the magnetic control action differs significantly from that in the standalone case. Here, the magnetic field is never zero and varies nonproportionally relative to the observation, unlike the gas jet's upstream action, which alters the free-surface topography and forces the electromagnet to adopt a different control strategy. Additionally, since both actuators are always nonzero, the optimal control action implies further thinning of the liquid film.

\begin{figure*}
\centering
  \begin{subfigure}[b]{0.325\textwidth}
  \centering
    \includegraphics[width=\textwidth]{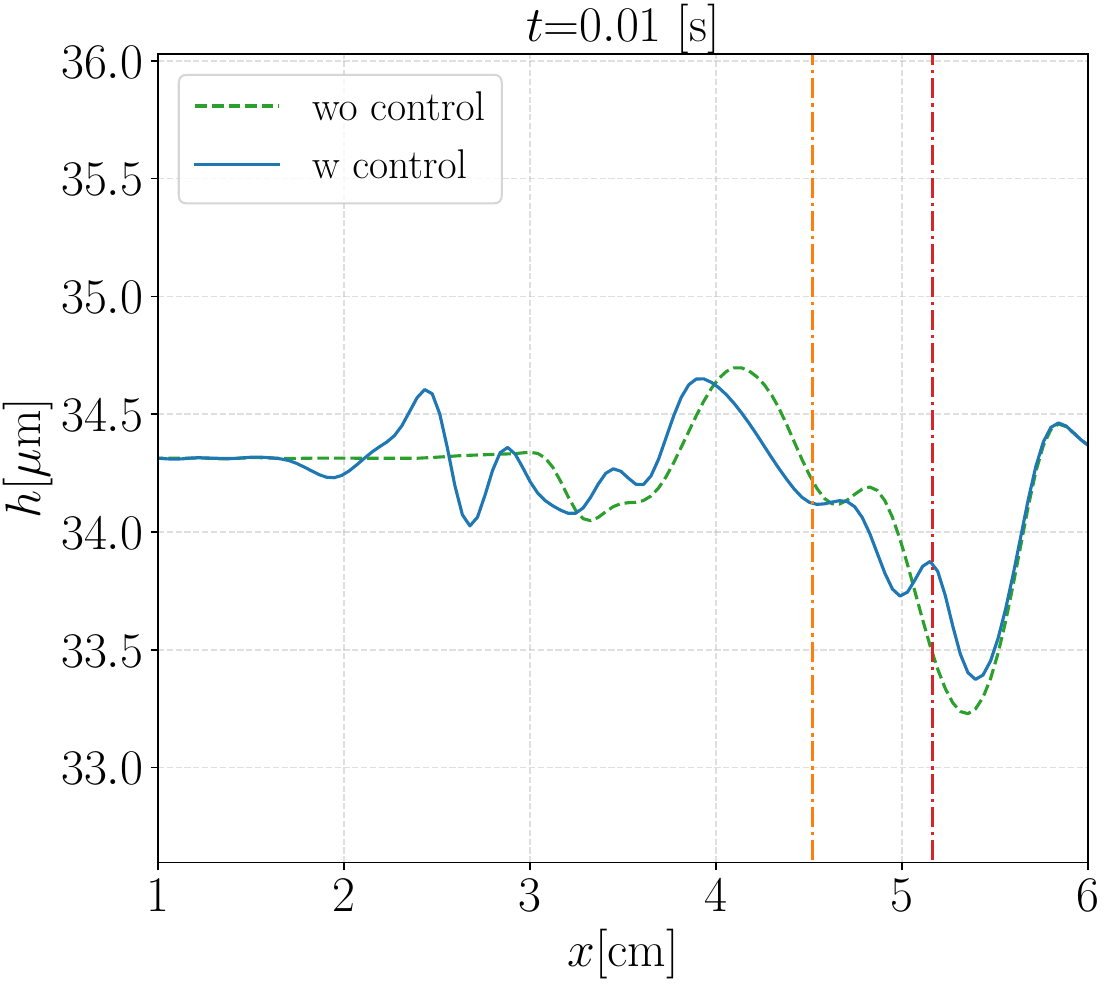}
    \caption{}
  \end{subfigure}
  \hfill
  \begin{subfigure}[b]{0.325\textwidth}
    \centering
    \includegraphics[width=\textwidth]{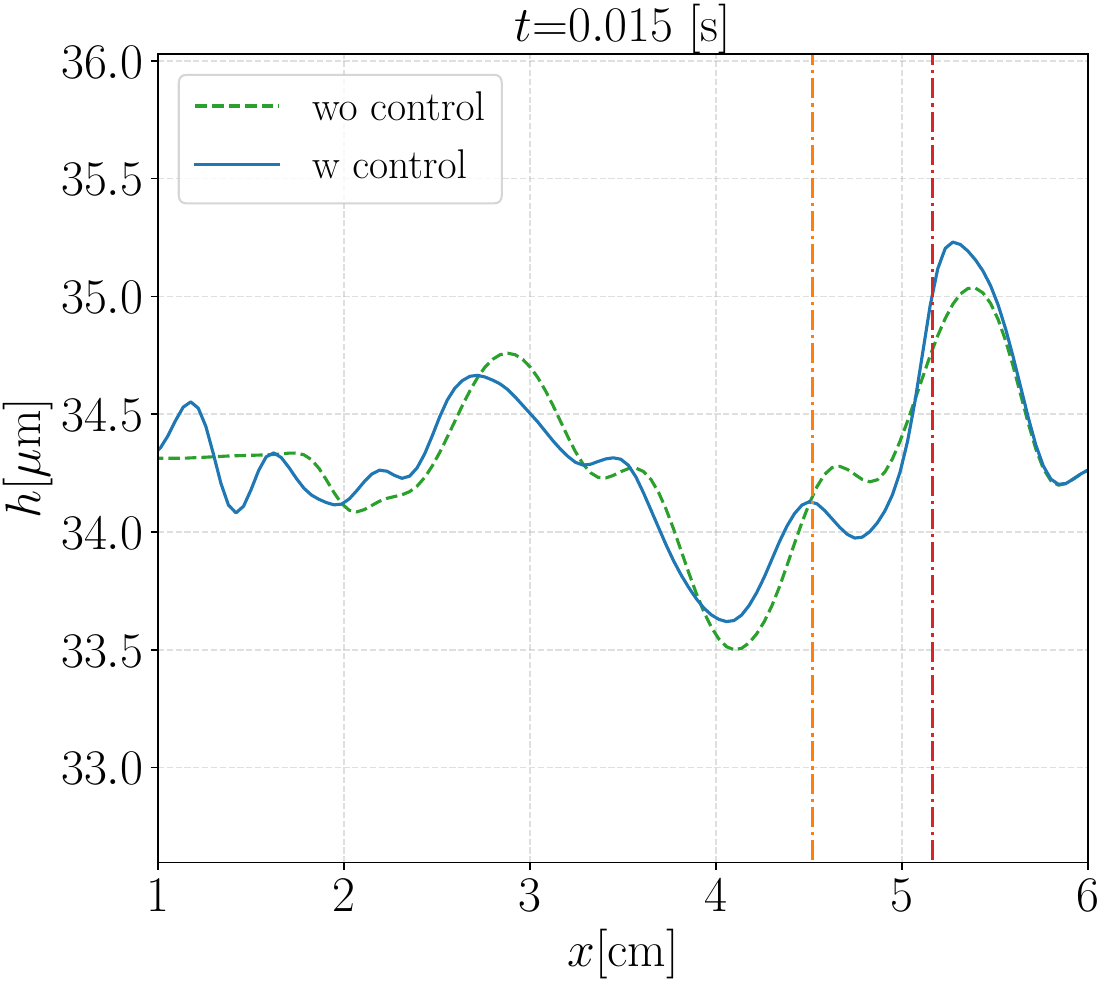}
    \caption{}
  \end{subfigure}
  \begin{subfigure}[b]{0.325\textwidth}
    \centering
    \includegraphics[width=\textwidth]{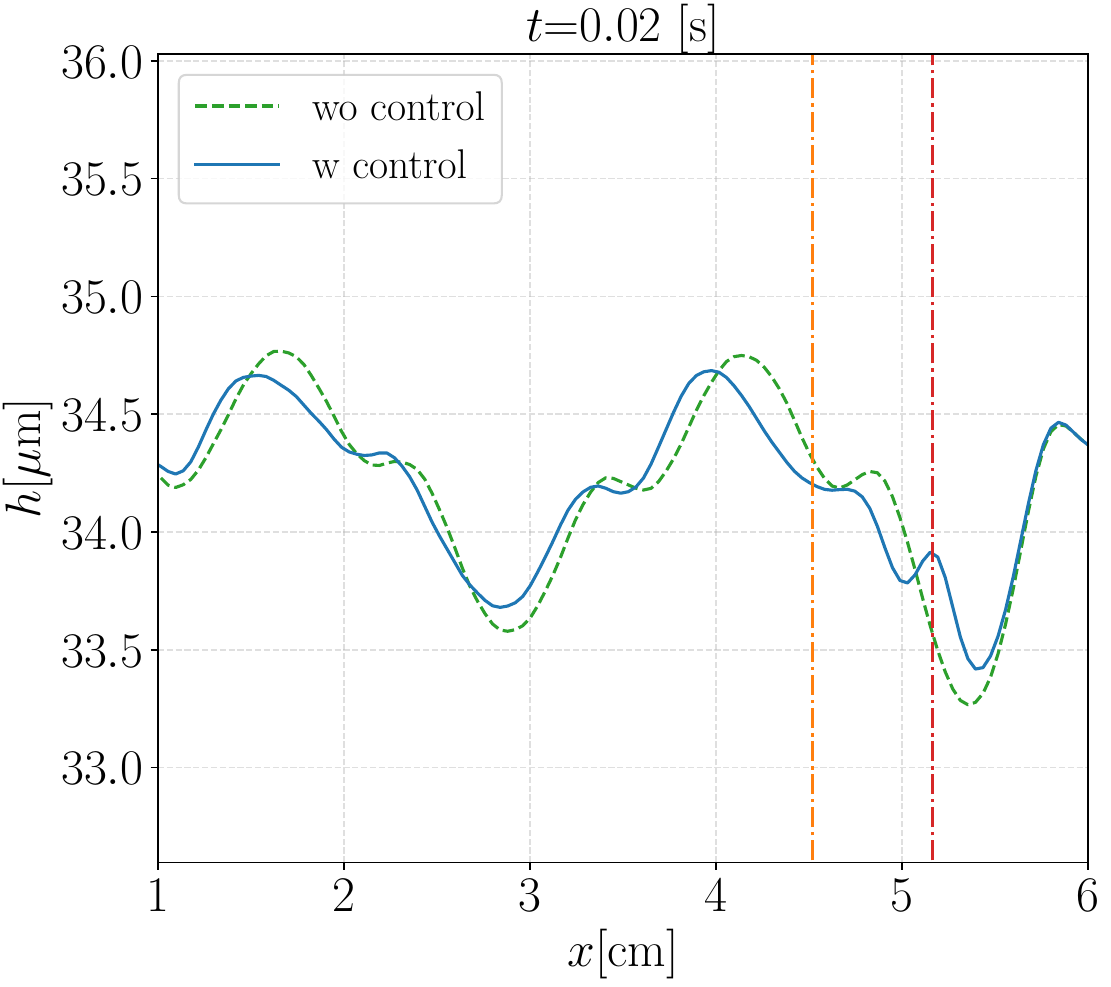}
    \caption{}
  \end{subfigure}
\hfill
  \begin{subfigure}[b]{0.325\textwidth}
    \centering
    \includegraphics[width=\textwidth]{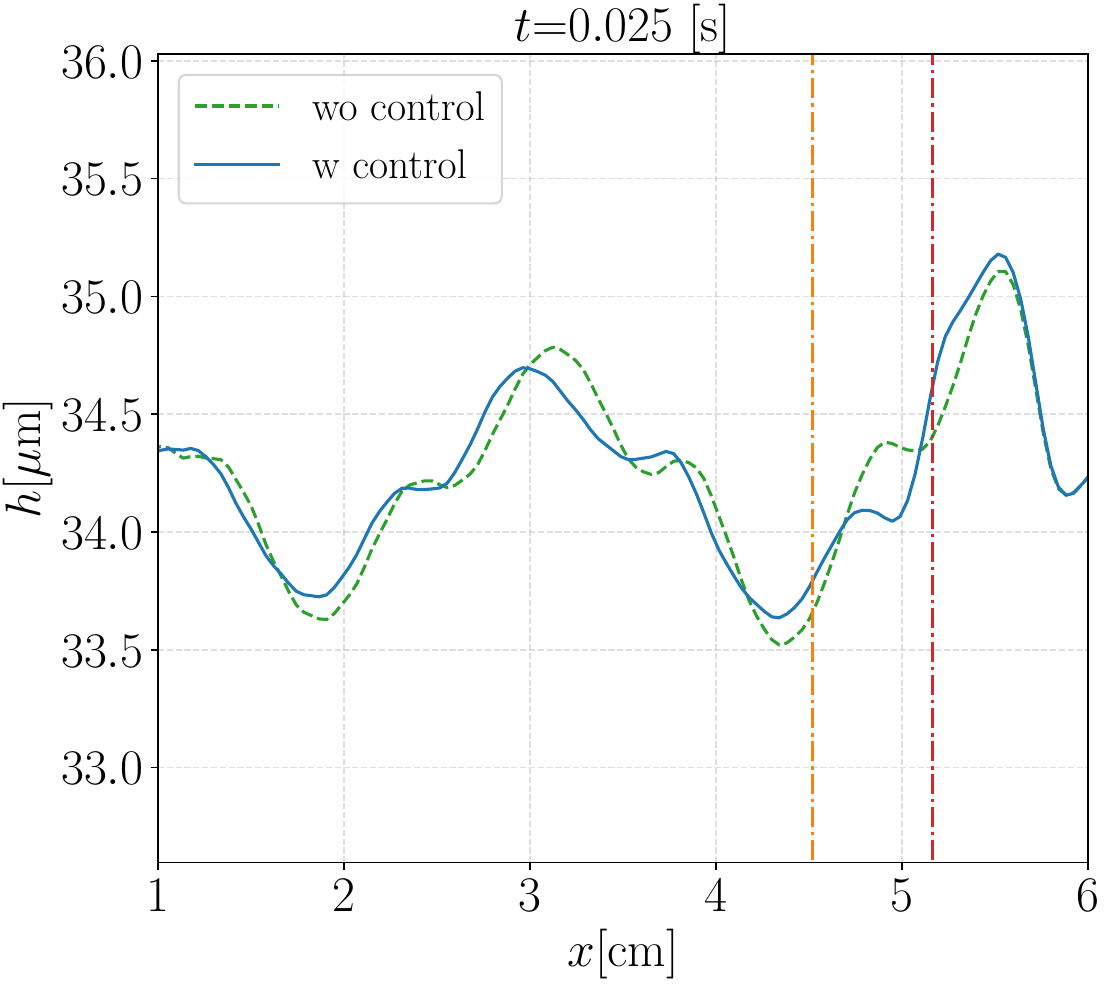}
    \caption{}
  \end{subfigure}
  \begin{subfigure}[b]{0.325\textwidth}
    \centering
    \includegraphics[width=\textwidth]{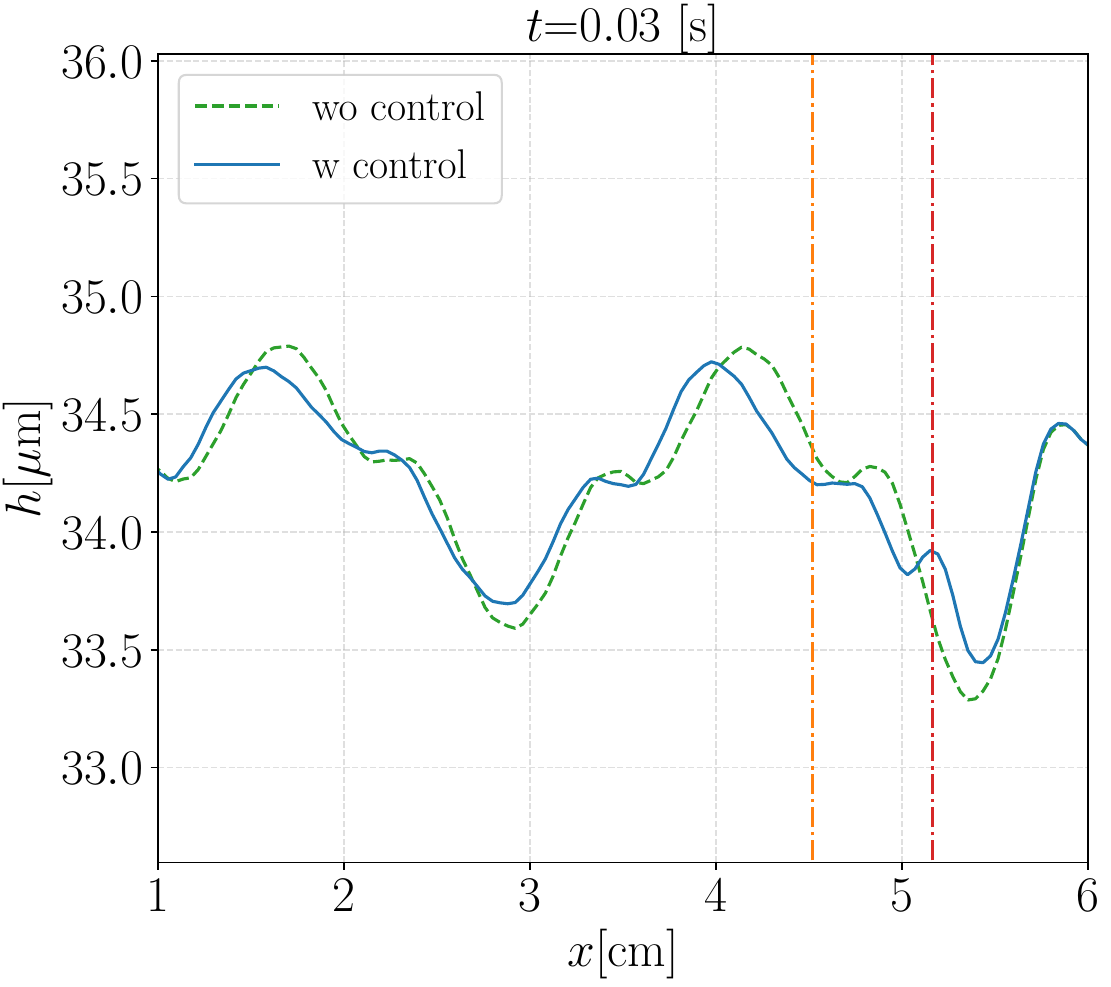}
    \caption{}
  \end{subfigure}
  \hfill
  \begin{subfigure}[b]{0.325\textwidth}
    \centering
    \includegraphics[width=\linewidth]{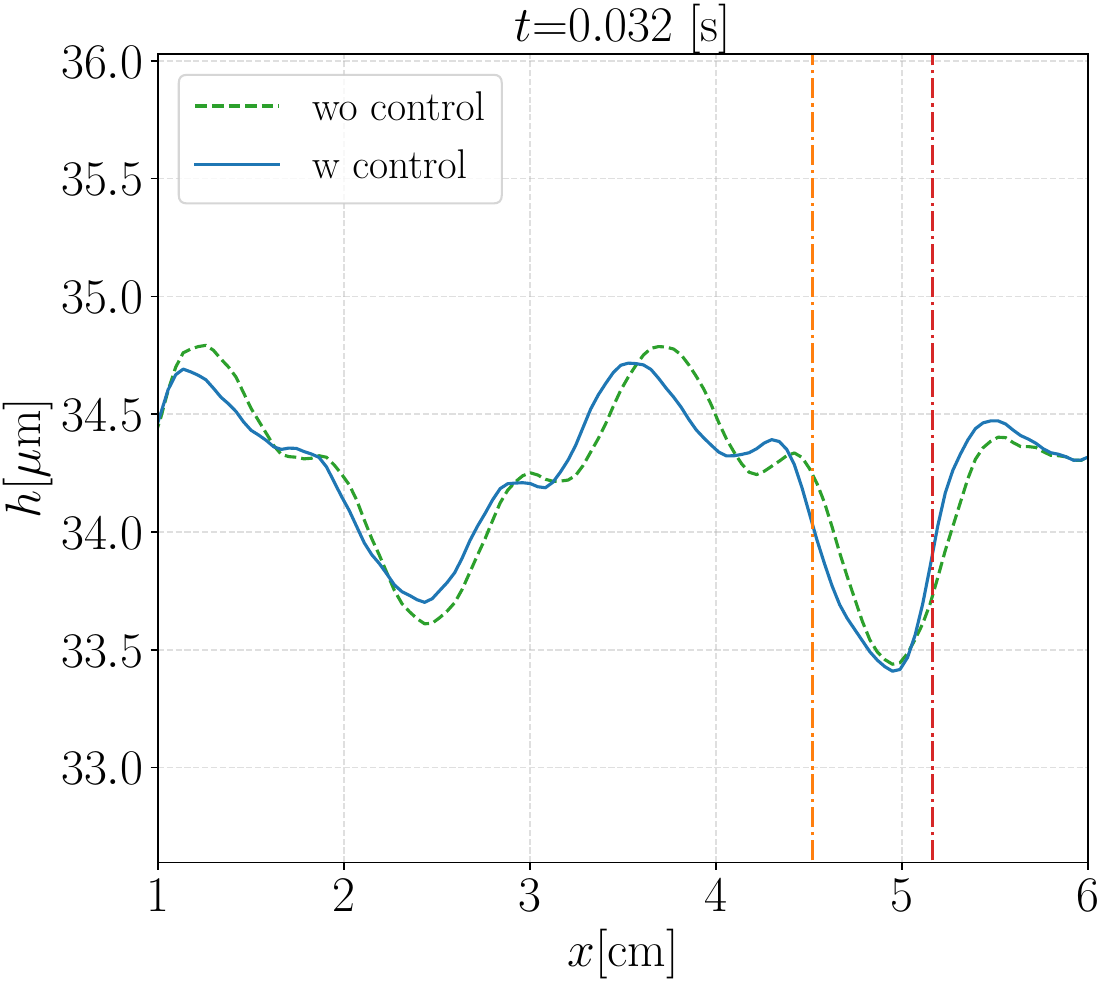}
    \caption{}
  \end{subfigure}
  \caption{Evolution of the dimensional liquid film with (continuous blue line) and without (green dashed line) control using a 2D gas jet (red dash-dotted line) and a 2D electromagnet (orange dash-dotted line).}
  \label{fig:plot_line_hybrid_sim}
\end{figure*}

The control mechanism, which combines pushing crests and raising valleys, is most apparent when examining the evolution of the liquid film along a line in the streamwise direction. Figure~\ref{fig:plot_line_hybrid_sim} shows the evolution of the liquid film with control (solid blue line) and without control (dashed green line) along $x$ at $z = 1 \, [\rm{cm}]$. For a single actuator, the gas jets depress the crests through additional wiping. However, this is counteracted by the electromagnet, resulting in waves with a smaller peak-to-valley amplitude in the controlled case than in the uncontrolled case.


\section{Conclusion and Perspectives}\label{sec:conclusion}
In this study, we developed a 3D Integral Boundary-Layer model to describe a liquid film on a moving substrate subjected to a transverse magnetic field in the context of hot-dip galvanising. The model was implemented in a numerical environment using the Fourier pseudo-spectral method. To reproduce open-domain conditions, we applied a linear, perfectly matched layer around the boundaries to absorb the incoming waves. Leveraging this numerical environment, we addressed a control problem involving unstable waves interacting with gas-jet and electromagnetic actuators. The optimal control law for the actuators was determined using the proximal policy optimisation (PPO) reinforcement learning algorithm. 

The Integral Boundary Layer (IBL) model is consistent with the model without magnetic forces: as the magnetic field intensity decreases, the flat-film steady-state solution and the closure relations for the flux terms and wall shear stress converge to their well-known counterparts in the absence of magneto-hydrodynamic effects. It should be noted, however, that the investigated conditions at a reduced Reynolds number $\delta = 266$, representative of galvanising flows, lie beyond the strict asymptotic range in which the model is formally derived. Yet, since previous studies have demonstrated that such models are predictive well beyond this limit, we believe that the present formulation still captures the essential coupling between hydrodynamic and electromagnetic effects. Accordingly, this work should be regarded not as an industrially deployable solution, but as a proof of concept demonstrating the feasibility of combining reduced-order modelling and reinforcement learning for the control of magnetohydrodynamic coating flows.

Our analysis of the unsteady model at the leading order revealed that the convective velocity is always negative for $\Ha > 2$. This highlights the strong influence of the magnetic field on the absolute and convective nature of steady-state solutions, and further investigation could provide valuable insight into whether small defects on the free surface affect industrial production processes. In addition, our study suggests that an effective magnetic field requires $\Ha = 16$ for galvanising conditions, which may be impractical for real-world applications due to high thermal gradients caused by the Joule effect and the destabilising magnetic effects on the strip's dynamics.

Regarding the control problem, this study demonstrates that an electromagnetic feedback regulator can effectively reduce the amplitude of unwanted waves by raising their valleys. This effect is achieved through a control mechanism driven by the resistive influence of the Lorentz force. From a practical implementation perspective, control performance against undulation perturbations can be enhanced by arranging multiple pairs in series along the streamwise direction, thereby leveraging the translational invariance of the governing equations.

These findings open new possibilities for extending this approach to more complex scenarios, particularly when incorporating the thermal effects generated by the electromagnetic field, which may help stabilise the liquid film via Marangoni forces. A notable feature of our model is the absence of a hysteresis term in the magnetic field. In reality, the magnetic field does not respond instantaneously to changes in $b_t$; its response is determined by the magnetic properties of the solenoid’s core. While this simplification is adequate for the accuracy requirements of this study, a more refined model would need to include hysteresis effects. Another promising avenue for future research is exploring the impact of high-frequency magnetic fields, which can produce a `skin effect' that confines the induced electric current to a thin layer near the free surface.

This investigation provides valuable insights into the dynamics of the coating process, capturing the intricate multiscale interactions between the liquid zinc and control actuators at a computational cost lower than that of Large Eddy Simulations (LES). Beyond its immediate applications, the model can be adapted to other engineering problems. This opens new avenues for modelling, experimental investigations, and control design, ranging from `liquid walls' used to confine plasma in tokamak nuclear fusion reactors to the shaping of versatile circuitry for next-generation electronic devices.

\appendix
\section{Magnetohydrodynamic governing equations and boundary conditions}\label{sec_apx:governing_eq}
The liquid film is governed by the 3D incompressible Navier-Stokes equations given by:
\begin{subequations}
\label{eq:Navier-Stokes}   
\begin{gather}
  \nabla\cdot\mathbf{u}=0,\label{eq:continuity}\\
  \rho(\partial_t\mathbf{u} + (\mathbf{u}\cdot\nabla)\mathbf{u})=-\nabla p + \rho\nu\nabla^2\mathbf{u} - \rho g \mathbf{i} + \mathbf{f}_L,
\end{gather}
\end{subequations}
where $\mathbf{f}_L$ is the Lorentz force given by the interaction between the magnetic field in the liquid and the induced current. The relative motion of the liquid zinc with respect to the external magnetic field generates an induced current in the bulk $\mathbf{j}=(j_x,j_y,j_z)^T$, given by Faraday's law of induction, neglecting the electric potential difference \cite{Dumont2011d}:
\begin{equation}
    \mathbf{j} = \sigma_M(\mathbf{u}\times\mathbf{b}) = (-\sigma_Mbw,0,-\sigma_Mbu)^T.
\end{equation} 
The interaction of the induced current with $\mathbf{b}$ results in a Lorentz force $\mathbf{f}_L$ given by the cross product of $\mathbf{j}$ and $\mathbf{b}$:
\begin{equation}
    \mathbf{f}_L = \mathbf{j}\times\mathbf{b} = (-\sigma_Mb^2u, 0, -\sigma_Mb^2w)^T.
\end{equation}
At the strip surface ($y=0$), we enforce the non-slip condition:
\begin{eqnarray}
\label{eq:bc_strip}
	\mathbf{u} \big|_{y=0} = (-U_p, 0, 0)^T.
	\label{eq:bc-kinematic-wall}
\end{eqnarray}

At the free surface ($y=h(x,z,t)$), we impose the kinematic boundary condition:
\begin{equation}
	v|_{y=h} = \partial_th + u\partial_xh + w \partial_zh,
	\label{eq:bc-kinematic-interface}
\end{equation}
along with the dynamic condition:
\begin{equation}
\label{eq:dynamics_bs}
    [\bm{\tau}\mathbf{n} -  \bm{\tau}_m\mathbf{n}]^{\rm l}_{\rm g} = 2\sigma K(h),
\end{equation}
where $K(h)$ is the free surface mean curvature \citep[Chapter~2]{kalliadasis2011falling}, $[f]^{\rm l}_{\rm g}=f - f_{\rm g}$ defines the jump at the interface between the liquid and the gas phases, $\bm{\tau}$ is the viscous stress tensor, and $\bm{\tau}_m$ is the Maxwell stress tensor \citep[Section~3.9]{Davidson_2001} defined as: 
\begin{equation}
    \bm{\tau}_m = \frac{\mathbf{b}\mathbf{b}^T}{\mu_m} - \frac{\left\lVert\mathbf{b}\right\lVert^2\mathbf{I}}{2\mu_m}.
\end{equation}

By introducing the following notations:
\begin{subequations}
    \begin{gather}
    p_g - \mathbf{n}^T(2\mu_g\mathbf{E}_g\mathbf{n})=p_g (x,z,t),\\        
     \mathbf{t}_x^T(2\mu_g\mathbf{E}_g\mathbf{n})=\tau_{g,x}(x,z,t),\\   
     \mathbf{t}_z^T(2\mu_g\mathbf{E}_g\mathbf{n})=\tau_{g,z}(x,z,t),   
    \end{gather}
\end{subequations}
where $\mathbf{E}$ is the strain-rate tensor, we obtain the three scalar equations representing the projections of the force balance \eqref{eq:dynamics_bs} in the local Cartesian reference frame (O;$n,\tau_x,\tau_z$) (defined in \eqref{eq:local_ref_frame}):

\begin{subequations}
\label{eq:forces-balances}
\begin{gather}
\begin{split}
p_g&-p +\mathbf{n}^T(2\mu\mathbf{E}\mathbf{n}) - \mathbf{n}^T(\bm{\tau}_{\rm{ml}}\mathbf{n})\\&+\mathbf{n}^T(\bm{\tau}_{\rm{mg}}\mathbf{n}) = 2\sigma K(h)\mathbf{n}, 
    \end{split}\\
    \begin{split}
    &\mathbf{t}_x^T(2\mu\mathbf{E}\mathbf{n}) + \mathbf{t}_x^T(\bm{\tau}_{\rm{ml}}\mathbf{n}) - \mathbf{t}_x^T(\bm{\tau}_{\rm{mg}}\mathbf{n})- \tau_{g,x} = 0, 
    \end{split}\\
    \begin{split}
    &\mathbf{t}_z^T(2\mu\mathbf{E}\mathbf{n}) + \mathbf{t}_z^T(\bm{\tau}_{\rm{ml}}\mathbf{n})-\mathbf{t}_z^T(\bm{\tau}_{\rm{mg}}\mathbf{n}) - \tau_{g,z} = 0.
    \end{split}
\end{gather}
\end{subequations}

The 3D liquid film on a moving substrate is therefore represented by the governing equations \eqref{eq:Navier-Stokes} and the boundary conditions \eqref{eq:bc_strip}, \eqref{eq:bc-kinematic-interface}, and \eqref{eq:forces-balances}.
Expanding \eqref{eq:forces-balances} gives:
\begin{subequations}
\label{eq:forces-balances-exp}
\begin{equation}
    \begin{gathered}
    \frac{2\mu[\partial_zh\left(\partial_xh \left(\partial_z u + \partial_x w\right)-\partial_z v-\partial_y w\right)}{(\partial_z h)^2+(\partial_x h)^2+1}\\- \frac{\partial_xh\left(\partial_yu+\partial_xv\right)+(\partial_x h)^2 \partial_x u]+(\partial_z h)^2 \partial_z w+\partial_y v)}{(\partial_z h)^2+(\partial_x h)^2+1}\\
     +p_g=\frac{\sigma  \left[\partial_{zz}h \left((\partial_x h)^2+1\right)-2 \partial_z h \partial_x h \partial_{xz}h\right]}{\left[(\partial_z h)^2+(\partial_x h)^2+1\right]^{3/2}}\\
    +\frac{\sigma  \left(\left((\partial_z h^2+1\right) \partial_{xx}h\right)}{\left((\partial_z h)^2+(\partial_x h)^2+1\right)^{3/2}}+p\\ - \frac{b^2 (\chi_g-\chi) \left[(\partial_z h)^2+(\partial_x h)^2-1\right]}{2\mu_{M}(\chi_g+1) \left[(\partial_zh)^2+(\partial_xh)^2+1\right]}\,,
    \end{gathered}
\end{equation}
\begin{equation}
    \begin{gathered}
        \tau_{g,x}+\frac{\mu \left(\partial_z h \left(\partial_xh \left(\partial_zv+\partial_yw\right)+\partial_zu+\partial_xw\right)\right)}{\sqrt{(\partial_xh)^2+1} \sqrt{(\partial_zh)^2+(\partial_xh)^2+1}}\\+
        \frac{\mu \left(2 \partial_xh\left(\partial_xu-\partial_yv\right)\right)}{\sqrt{(\partial_xh)^2+1} \sqrt{(\partial_zh)^2+(\partial_xh)^2+1}}\\+
        \frac{\mu  \left(\left((\partial_xh)^2-1\right) \partial_yu+\left((\partial_xh)^2-1\right) \partial_xv\right)}{\sqrt{(\partial_xh)^2+1} \sqrt{(\partial_zh)^2+(\partial_xh)^2+1}}        
        \\ + \frac{b^2(\chi-\chi_g)\partial_xh}{\mu(\chi_g+1) \sqrt{(\partial_xh)^2+1} \sqrt{(\partial_zh)^2+(\partial_xh)^2+1}}=0\,,
    \end{gathered}
\end{equation}
\begin{equation}
    \begin{gathered}
        \tau_{gz}+\frac{\mu \left(\partial_xh \left(\partial_zh \left(\partial_yu+\partial_xv\right)+\partial_zu+\partial_xw\right)\right)}{\sqrt{(\partial_zh)^2+1} \sqrt{(\partial_zh)^2+(\partial_xh)^2+1}}\\
        +\frac{\mu \left(2\partial_zh\left(\partial_zw-\partial_yv\right)\right)}{\sqrt{(\partial_zh)^2+1} \sqrt{(\partial_zh)^2+(\partial_xh)^2+1}}\\+\frac{\mu \left(\left((\partial_zh)^2-1\right) \partial_zv+\left((\partial_zh)^2-1\right) \partial_yw\right)}{\sqrt{(\partial_zh)^2+1} \sqrt{(\partial_zh)^2+(\partial_xh)^2+1}}\\+ \frac{b^2(\chi-\chi_g)\partial_xh}{\mu(\chi_g+1) \sqrt{(\partial_zh)^2+1}\sqrt{(\partial_zh)^2+(\partial_xh)^2+1}}=0\,.
    \end{gathered}
\end{equation}
\end{subequations}

The full magnetohydrodynamic model is given by \eqref{eq:Navier-Stokes} with boundary conditions \eqref{eq:bc_strip}, \eqref{eq:bc-kinematic-interface} and \eqref{eq:forces-balances-exp}.

\section{Approximated magnetic field of a circular finite-length solenoid}
\label{sec:approx_solenoid}
This section details the assumptions behind the derivation of the approximated Gaussian magnetic field \eqref{eq:approx_magnetic_field_11} presented in Subsection~\ref{subsec:model_magnet}.

We consider a circular solenoid with radius $R_s$ and length $L_s$, with a cylindrical reference frame $\mathcal{O}(r_s,z_s,\varphi_s)$ centred at its midpoint. The analytical expression for the axial $b_{za}$ and radial $b_{ra}$ components of the generated magnetic field read \cite{hampton2020closed}:
\begin{subequations}
\label{eq:magn_field_sol}
    \begin{equation}
    \label{eq:magn_field_sol_axial}
        \begin{gathered}
            b_{ra} = \frac{b_0}{\pi}\sqrt{\frac{R_s}{r_sm_{+}}}\Big(E(m_{+})+\Big(1-\frac{m_{+}}{2}\Big)K(m_{+})\Big)\\-\frac{b_0}{\pi}\sqrt{\frac{R_s}{rm_{-}}}\Big(E(m_{-})+\Big(1-\frac{m_{-}}{2}\Big)K(m_{-})\Big),
        \end{gathered}
    \end{equation}
    \begin{equation}
    \label{eq:magn_field_sol_radial}
        \begin{gathered}
            b_{za} = \frac{b_0\zeta^{+}}{4\pi}\sqrt{\frac{m_{+}}{R_sr_s}}\Big(K(m_{+})+\Big(\frac{R_s-r_s}{R_s+r_s}\Big)\Pi(u,m_{+})\Big)\\-\frac{b_0\zeta^{-}}{4\pi}\sqrt{\frac{m_{-}}{R_sr_s}}\Big(K(m_{-})+\Big(\frac{R_s-r_s}{R_s+r_s}\Big)\Pi(u,m_{-})\Big),
        \end{gathered}
    \end{equation}
\end{subequations}
where $E$, $K$, and $\Pi$ are the complete elliptic integrals of the first, second, and third kinds, respectively, $b_0$ is the magnetic field at the solenoid's core in the limit of $L_s \to \infty$ given by:
\begin{equation}
\label{eq:def_core_magn}
    b_0 = \mu_m I n,
\end{equation}
where $\mu_m$ is the magnetic permeability of the core material, $n$ is the number of coils and $I$ is the electric current flowing through the coils of the solenoid. The variables $u$, $m_{+}$, and $m_{-}$ are expressed in terms of $r_s$ and $z_s$ by the following expressions:
\begin{subequations}
    \begin{equation}
        u = \frac{4R_sr_s}{(R_s+r_s)^2},
    \end{equation}
    \begin{equation}
        m_{+}=\frac{4R_sr_s}{(R_s+r_s)^2+\zeta^{2}_{-}}, 
    \end{equation}
    \begin{equation}
         m_{-}=\frac{4R_sr_s}{(R_s+r_s)^2+\zeta^{2}_{+}},
    \end{equation}
\end{subequations}
with $\zeta_{+}$ and $\zeta_{-}$ reading:
\begin{equation}
    \zeta_{+} = z_s + \frac{L_s}{2}, \quad\quad \zeta_{-} = z_s - \frac{L_s}{2}.
\end{equation}
\begin{figure}
      \centering
    \includegraphics[width=0.48\textwidth]{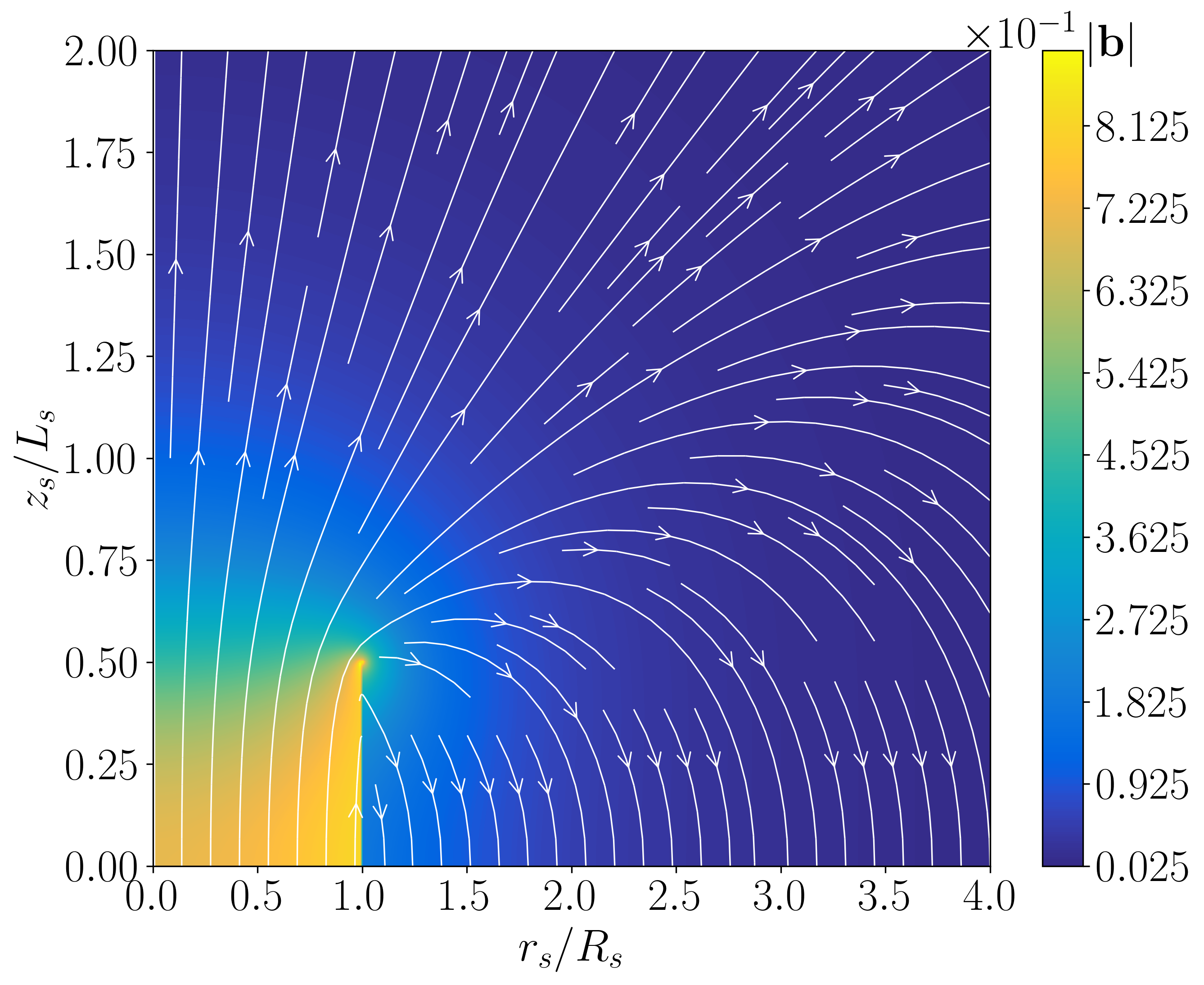}
    \caption{Normalised solenoid magnetic field $|\mathbf{b}|/b_0$ with magnetic lines (white continuous lines) in an axial-radial plane scaled with the solenoid radius $R_s$ and lengths $L_s$.}
    \label{fig:magnetic_modelling_tot_0}
\end{figure}
\begin{figure}
  \begin{subfigure}[b]{0.48\textwidth}
      \centering
    \includegraphics[width=\textwidth]{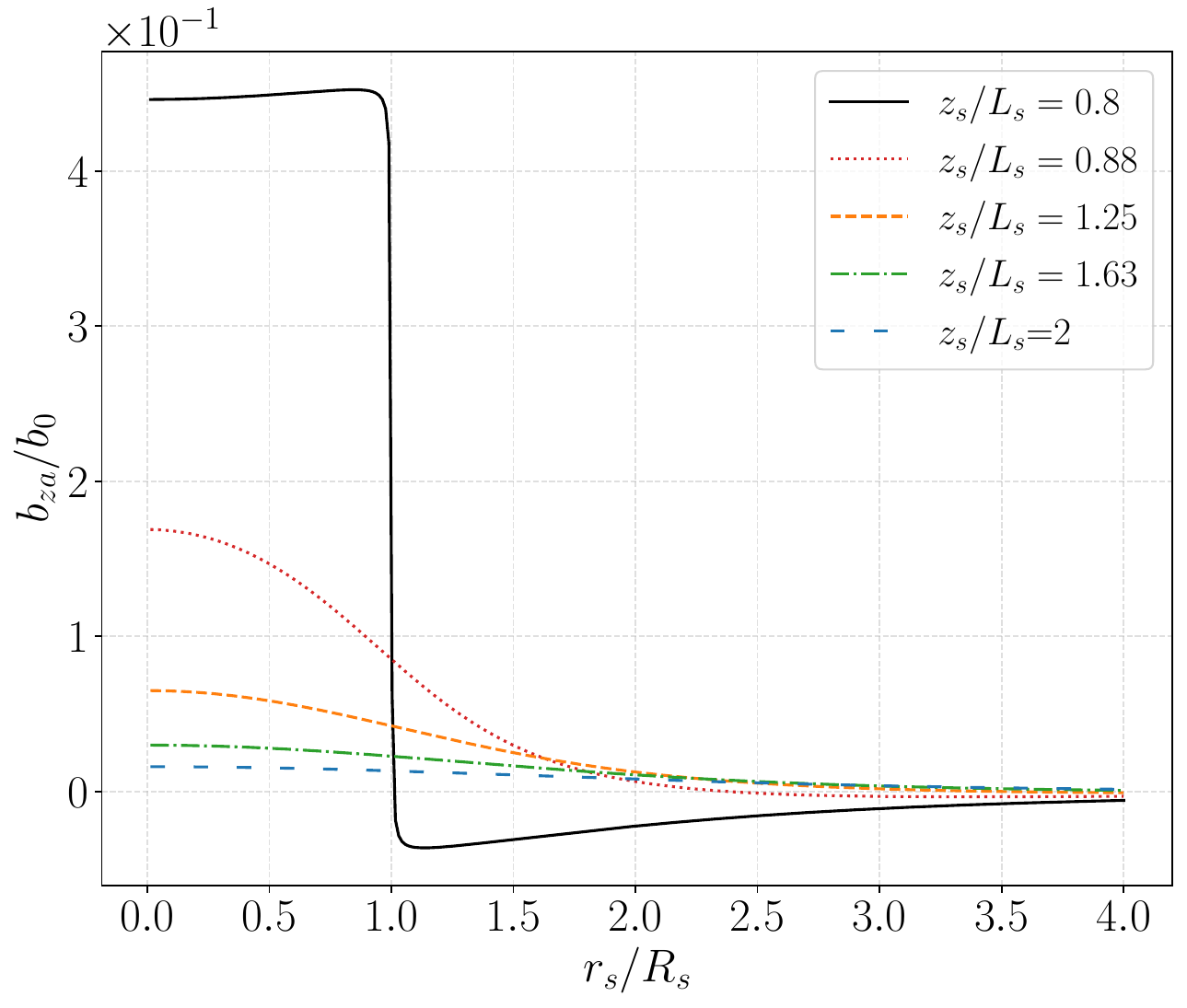}
    \caption{}
    \label{fig:magnetic_modelling_3}
  \end{subfigure}
  \hfill
  \begin{subfigure}[b]{0.48\textwidth}
    \centering
    \includegraphics[width=\textwidth]{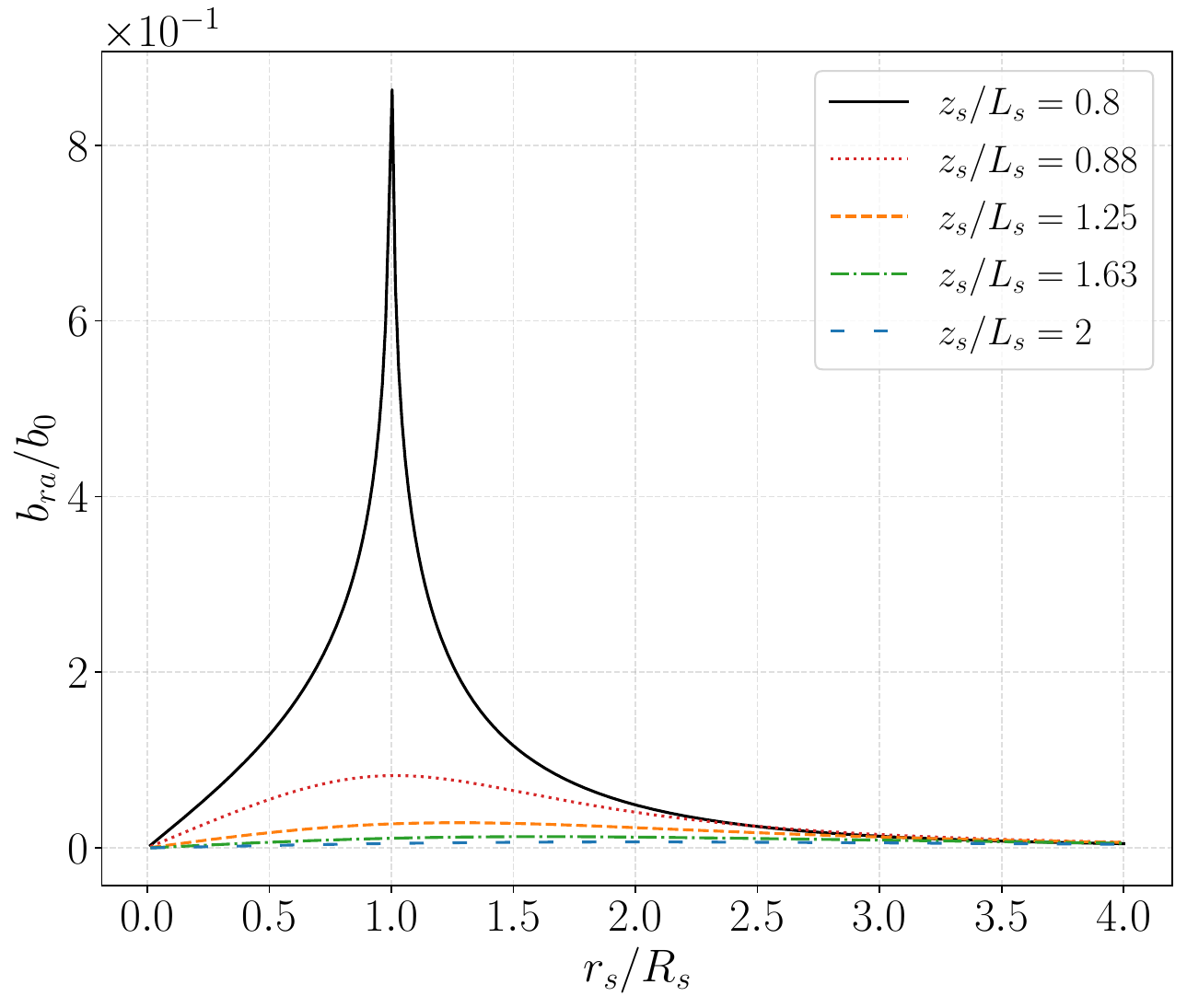}
    \caption{}
    \label{fig:magnetic_modelling_4}
  \end{subfigure}
    \caption{(a) Axial} and (b) radial components of the magnetic induction field in the reduced radial direction $r_s/R_s$ at different reduced axial locations $z_s/L_s$.
    \label{fig:magnetic_modelling_tot}
\end{figure}

Figure~\ref{fig:magnetic_modelling_tot_0} shows the normalized magnetic field (colour map) and magnetic field lines (white continuous lines), and figure and \ref{fig:magnetic_modelling_tot} shows the radial distributions of (a) the axial $b_{za}$ and (b) radial $b_{ra}$ components of the magnetic field normalized with $b_0$ at different axial positions, $z_s/L_s$, outside the solenoid. The field intensity drops sharply outside the solenoid, falling below $3\%$ of $b_0$ for $r_s/R_s > 1.5$ and $z_s/L_s > 1$. Near the solenoid exit ($z_s/L_s = 0.8$), $b_{za}$ remains nearly constant along $r_s$, while $b_{ra}$ exhibits a peak at the edge ($r_s/R_s = 1$), where the field lines are most curved. As $z_s$ increases, both $b_{za}$ and $b_{ra}$ become smoother, adopting a bell-shaped profile in $r_s$, with the axial component dominating, being one order of magnitude larger than the radial component.

Based on these observations, assuming that the electromagnetic actuator is positioned far enough from the moving substrate, we neglected the radial component $b_{ra}$ of the magnetic field and approximated its axial component $b_{za}$ as a Gaussian function in space, modulated by a time-varying amplitude $b(t)$, reading: 
\begin{equation}
\label{eq:approx_magn_0}
    b_{za} \approx b_{zs} = b(t)\,\exp\left(\frac{-r_s^2}{2\gamma^2}\right),
\end{equation}

To define a relationship between the Gaussian's standard deviation $\gamma$, the amplitude $b(t)$ and the geometrical and physical characteristics of the solenoid $R_s$, $L_s$, $I$ and $n$, we assume that $r_s \gg R_s$ or $r \to 0$, where $u \to 0$. Thereby, the complete elliptic integral of the third kind $\Pi$ reduces to $\Pi(0,m_{-}) = K(m_{-})$. This simplification leads to a more tractable expression for the axial magnetic field \eqref{eq:magn_field_sol_axial}, which reads: 
\begin{equation}
\label{eq:ax_sim_magnetic_1}
\begin{gathered}
    b_{za} = \frac{B0}{4\pi}\frac{2R_sL_s}{\sqrt{R_sr_s}(R_s + r_s)}\\\times\left[\zeta_{+}\sqrt{m_{+}}K(m_{+})-\zeta_{-}\sqrt{m_{-}}K(m_{-})\right].
\end{gathered}
\end{equation}

At this point, we further assume that $R_s + r_s \gg \zeta_{+}$ and $R_s + r_s \gg \zeta_{-}$ which results in the following conditions for $r\rightarrow 0$:
\begin{equation}
    z\ll R-L/2, \qquad\qquad R>L
\end{equation}
and for $r\gg R$:
\begin{equation}
    z\ll r -L/2, \qquad\qquad L>2R.
\end{equation}

Leveraging this assumption, we can approximate $m_{+}\sim u$ and $m_{-}\sim u$ which simplify \eqref{eq:ax_sim_magnetic_1} into:
\begin{equation}
\label{eq:simplified_magn_1}
    b_{za} = \frac{B_0 RL}{2\pi}\frac{K(u)}{(r_s + R_s)^2}.
\end{equation}

The difference in terms of absolute value between \eqref{eq:simplified_magn_1} with $K$ approximated by its first two terms in its power series expansion and the Taylor expansion of the approximated magnetic field \eqref{eq:approx_magn_0} truncated at its fifth order around $r_s=0$ gives the error function $e(r_s)$, which reads:
\begin{equation}
\label{eq:error_approx_Gaussian}
\begin{gathered}
    e(r)=\bigg|\left(\frac{b_0L_s}{4R_s}-b(t)\right)-\frac{r_s\left(b_0L_s\right)}{2 R_s^2}\\+r_s^2 \left(\frac{b(t)}{2 \gamma^2}+\frac{7b_0L_s}{4R_s^3}\right)-\frac{7 r_s^3\left(b_0L_s\right)}{R_s^4}\\+r_s^4\left(\frac{153b_0L_s}{4R_s^5}-\frac{b(t)}{8 \gamma^4}\right)\\-\frac{435r_s^5\left(b_0L_s\right)}{2R_s^6}+O\left(r_s^6\right)\bigg|\,.
\end{gathered}
\end{equation}

To minimize $e(r_s)$, the coefficients at $O(1)$ and $O(r_s^4)$ are set to zero, yielding the following expressions for $b(t)$ and $\gamma$:
\begin{equation}
\label{eq:appx_approx_amp_std}
    b(t) = \frac{b_0L_s}{4R_s}, \qquad\qquad \gamma=\frac{R_s}{2^{3/4} \sqrt{3} \sqrt[4]{17}},
\end{equation}
where their nondimensional forms are reported in subsection~\ref{subsec:model_magnet}.

Replacing $b_0$ in \eqref{eq:appx_approx_amp_std} with its definition \eqref{eq:def_core_magn} gives a relation between $b(t)$ and the evolution of the electric current in the solenoid in time:
\begin{equation}
    b(t) = \frac{\mu_m I(t) n L_s}{4R_s}.
\end{equation}

Concerning the magnetic field in the radial direction $b_{ra}$ in \eqref{eq:magn_field_sol_radial}, it is straightforward to observe that this can be considered negligible with the approximation reported above.

\section{Governing and auxiliary equations implemented in the spectral BLEW 3D environment}
\label{appx:PML}
Here, we report the modified governing and auxiliary differential equations arising from the perfectly matched layer transformation. The modified 3D IBL equations, without considering Maxwell stresses, with the auxiliary variable arising from the PML formulation reads: 
\begin{subequations}
\begin{equation}
        \partial_{\hat{t}}\hat{h} + \zeta_1 + \zeta_2 = 0,
    \end{equation}
    \begin{equation}
    \begin{split}
        \partial_{\hat{t}}\hat{q}_{\hat{x}} + \zeta_3 + \zeta_4 = \delta^{-1}\Big[\hat{h}(-\zeta_5 + \zeta_8 + \zeta_{11} + 1)\\ - H_a^2\hat{b}^2\hat{q_x} + \Delta\tau_{\hat{x}}\Big]
    \end{split}
    \end{equation}
    \begin{equation}
    \begin{split}
        \partial_{\hat{t}}\hat{q}_{\hat{z}} + \zeta_{12} + \zeta_{13} = \delta^{-1}\Big[\hat{h}(-\zeta_{14} + \zeta_{15} + \zeta_{16})\\ -  H_a^2\hat{b}^2\hat{q_z} + \Delta\tau_{\hat{z}}\Big]
    \end{split}
    \end{equation}
\end{subequations}
with the new set of auxiliary differential equations given by:
\begin{equation}
\begin{aligned}
        &\partial_{\hat{t}}(\zeta_1 - \partial_{\hat{x}}\hat{q}_{\hat{x}})=-\sigma_x\zeta_1
        &&\partial_{\hat{t}}(\zeta_2 - \partial_{\hat{z}}\hat{q}_{\hat{z}})=-\sigma_z\zeta_2
        \\        
        &\partial_{\hat{t}}(\zeta_3 - \partial_{\hat{x}}F_{12})=-\sigma_x\zeta_3&
        &\partial_{\hat{t}}(\zeta_4 - \partial_{\hat{z}}F_{22})= -\sigma_z\zeta_4 
        \\
        &\partial_{\hat{t}}(\zeta_5 - \partial_{\hat{x}}\hat{p}_{g})=-\sigma_x\zeta_5& 
        &\partial_{\hat{t}}(\zeta_6 - \partial_{\hat{x}}\hat{h})=-\sigma_x\zeta_6
        \\ 
        &\partial_{\hat{t}}(\zeta_7 - \partial_{\hat{x}}\zeta_6)=-\sigma_x\zeta_7& 
        &\partial_{\hat{t}}(\zeta_8 - \partial_{\hat{x}}\zeta_7)=-\sigma_x\zeta_8
        \\
        &\partial_{\hat{t}}(\zeta_9 - \partial_{\hat{z}}\hat{h})=-\sigma_z\zeta_9&
        &\partial_{\hat{t}}(\zeta_{10} - \partial_{\hat{z}}\zeta_9)=-\sigma_z\zeta_{10}
        \\
        &\partial_{\hat{t}}(\zeta_{11} - \partial_{\hat{x}}\zeta_{10})=-\sigma_x\zeta_{11}&
        &\partial_{\hat{t}}(\zeta_{12} - \partial_{\hat{x}}F_{13})=-\sigma_x\zeta_{12}
        \\
        &\partial_{\hat{t}}(\zeta_{13} - \partial_{\hat{z}}F_{23})=-\sigma_z\zeta_{13}&
        &\partial_{\hat{t}}(\zeta_{14} - \partial_{\hat{z}}\hat{p}_g)=-\sigma_z\zeta_{14}
        \\
        &\partial_{\hat{t}}(\zeta_{15} - \partial_{\hat{z}}\zeta_{7})=-\sigma_z\zeta_{15}& &\partial_{\hat{t}}(\zeta_{16}-\partial_{\hat{z}}\zeta_{10})=-\sigma_z\zeta_{16}
    \end{aligned}
\end{equation}

\begin{acknowledgments}
F. Pino was supported by an F.R.S.-FNRS FRIA grant, and E. Fracchia was supported by the Short Training Program scholarship provided by the Von Karman Institute for Fluid Dynamics. This project was funded by Arcelor-Mittal Research. B. Scheid is Research Director at F.R.S.-FNRS.
\end{acknowledgments}

\bibliography{apssamp}
\end{document}